\documentclass[apj]{emulateapj}
\usepackage{amssymb}
\usepackage{multirow}

\renewcommand{\r}{\mathbf{r}}

\renewcommand{\k}{\mathbf{k}}
\newcommand{\msun}{$\,{\rm M}_\odot$}
\newcommand{\Lsun}{$\,{\rm L}_\odot$}

\newcommand{\mcrit}{$M_{\rm crit}$}
\newcommand{\fcrit}{$f_{\rm crit}$}
\newcommand{\fbary}{$f_{\rm b}$}
\newcommand{\mtot}{$M_{\rm tot}$}
\newcommand{\mvir}{$M_{\rm vir}$}

\newcommand{\mbary}{$M_{\rm b}$}
\newcommand{\rvir}{$r_{\rm vir}$}
\newcommand{\rs}{$r_{\rm s}$}

\newcommand{\kms}{\ km s$^{-1}$}

\newcommand{\comments}[1]{}

\begin{document}

\shorttitle{Lowest Mass Galaxies}
\shortauthors{Bland-Hawthorn, Sutherland \& Webster}

\title{Ultrafaint dwarf galaxies $-$ the lowest mass relics from before reionization}

\author{ }
\affil{ }

\author{Joss Bland-Hawthorn}
\affil{Sydney Institute for Astronomy, School of Physics, University of Sydney, NSW 2006, Australia}
\email{jbh@physics.usyd.edu.au}

\author{Ralph Sutherland}
\affil{Research School of Astronomy \& Astrophysics, Australian National University, Cotter Rd,
Weston, ACT 2611, Australia}

\author{David Webster}
\affil{Sydney Institute for Astronomy, School of Physics, University of Sydney, NSW 2006, Australia}

\begin{abstract}
New observations suggest that ultrafaint dwarf galaxies (UFD) -- the least luminous systems bound by dark matter halos ($\lesssim 10^5$\Lsun) -- may
have formed before reionization. The extrapolated virial masses {\it today} are 
uncertain with estimates ranging from $10^8$\msun\ to as high as 
$10^9$\msun\ depending on the assumed form of the underlying potential.
We show that the progenitor halo masses of UFDs can,
in principle, be as low as \mvir\ $\approx$ $10^7$\msun. Under the right conditions, such a halo can survive the energy input
of a supernova (SN) and its radiative progenitor. A clumpy (fractal) medium is 
much less susceptible to both internal and external injections of energy. It is less 
prone to SN sweeping (particularly if it is off-centred) because the coupling 
efficiency of the explosive energy is much lower than for a diffuse interstellar
medium. With the aid of the 3D hydro/ionization code {\em Fyris}, we
show that sufficient baryons are retained to form stars following a single supernova event
in dark matter halos down to \mvir\ $\approx$ $10^7$\msun\ in the presence of radiative cooling.
In these models, the gas survives the SN explosion, is enriched with the specific 
abundance yields of the discrete events, and reaches surface densities where low
mass stars can form. Our highest resolution simulations reveal why cooling is so
effective in retaining gas compared to any other factor. In the early stages, 
the super-hot metal-enriched SN ejecta exhibit strong cooling, leading to much of the
explosive energy being lost. Consistent with earlier work, the baryons do {\it not} survive in 
smooth or adiabatic models in the event of a supernova. 
The smallest galaxies may not contribute a large fraction of matter to the formation of galaxies,
but they carry signatures of the earliest epochs of star formation, as we show. 
These signatures may allow us to distinguish a small primordial galaxy
from one that was stripped down to its present size through tidal
interaction. We discuss these results in the context of local ultra-faint dwarfs and 
damped Ly$\alpha$ systems ($z\sim 2$) at very
low metallicity ([Fe/H] $\sim$ $-3$). We show that both classes of objects are consistent
with primordial low-mass systems that have experienced only a few enrichment events.
\end{abstract}

\keywords{dwarf galaxies -- stellar populations -- star clusters -- elemental abundances}

\section{Introduction}
\label{s:intro}

One of the most important questions in astrophysics is what
constituted the earliest baryonic systems, and whether relics of these
survive to the present day. Primordial objects are important to identify at any redshift because 
these retain chemical signatures of the first and second generations of stars
\citep[hereafter K13]{karlsson13}. Furthermore, their existence and distribution provide
important constraints on warm dark matter models \citep{kennedy14} and how reionization
propagated through the local universe \citep{busha10, ocvirk13}.

The earliest baryonic systems must have contained dark matter 
to allow gas cooling to proceed in order to form the first star \citep{abel02}.
Gas cooling and accretion
may have started as early as $z\sim 30$ in a dark matter halo with a mass of roughly
$10^{5-6}$ M$_\odot$. The first stars are thought to have been very massive, short lived, and
unique to their time \citep{yoshida06,gao07}. 
There is an emerging consensus on the sequence of events involved in forming the first 
stars because the hydrodynamic processes appear to be relatively well defined \citep{bromm11}. 
The same cannot be said about their chemical yields (K13). Some of these stars may have seeded 
the first black holes in their cores. An unknown combination of metals is expected to fall back onto 
the black hole seed \citep{podsiadlowski02,umeda03}, thus complicating any attempt to
deduce the first stellar yields.

While many questions remain concerning the formation of the first stars
\citep{clark11, greif11, oshea11, norman11}, 
the processes leading to `second stars' and subsequent generations 
are even less well understood \citep{ritter12,jeon14}. Whether the first stars inhibit or
aid the formation of the second generation is a complex problem \citep{ahn2007}. In particular,
the emergence of long-lived, low mass stars
depends in part on whether dust formed from the ejected material of the first stars
\citep{clark2008}. If dust did not form in appreciable amounts, the cosmic microwave background 
inhibits gas cooling, in which case low mass stars may not have emerged until $z\lesssim 10$ 
when mean metallicities reached $\sim 10^{-4}$ Z$_\odot$ \citep{schneider10, bromm11}. 

\smallskip
\noindent{\sl The first baryonic systems.}
So what were the first baryonic systems?\footnote{How globular clusters fit into the early 
history of galaxy formation, and 
what they may tell us about the first stellar generations, remains a mystery.
These ancient systems, which are relatively metal-enriched 
([Fe/H]$\gtrsim$-2.5) and free of dark matter, were long considered to be the oldest baryonic relics of the early universe \citep{peebles1968}. } Were these exclusively the domain of
the most massive stars, or did they include low to intermediate mass stars
\citep{tsuribe2008, clark2008}, and do any of these stars survive to
the present day \citep{okrochkov2010}? There are few if any reliable
observational constraints thus far on the initial mass function (IMF) of the 
earliest galaxies. In this paper, our goal is to explore astrophysical 
situations in which very low mass galaxies,
i.e. baryons confined by dark matter, have survived from ancient times to the present day.
These objects are of great interest because they must have
preserved chemical signatures of early star formation and therefore could provide direct
constraints on the properties of the earliest generations of stars (K13).

One of the most interesting developments in recent years is the discovery of ultra-faint
dwarf galaxies in the Galactic halo 
\citep{willman05,belokurov06,zucker06,belokurov07,walsh07}.
These dark-matter dominated objects 
have very low baryonic mass ($\lesssim 10^4$\msun), are exceedingly faint ($\lesssim 10^5$ L$_\odot$), and have stellar metallicities below [Fe/H]=-2 \citep{martin08}. Remarkably, some
UFDs are less luminous than globular clusters, but the former have
dark matter whereas the latter do not, and hence must be considered as galaxies.
The characteristic halo mass of UFDs today is highly uncertain with estimates ranging
from $10^8$ to $10^9$ M$_\odot$
\citep{simon07,strigari08,martin08,bovill09,wolf10,frebel14}.

\smallskip\noindent{\sl The reionization epoch.}
UV radiation from a dispersed population of `first stars' can 
dissociate H$_2$, the primary coolant responsible for cooling within dark haloes. The diffuse ionised gas is then too warm to stay bound to low-mass haloes. 
In order for gas to be confined, the dark-halo mass must have exceeded 
\mcrit\ $\approx$ $10^{8.5}$ M$_\odot$ if the gas is to cool and eventually
form stars after reionization has taken hold \citep{rees1986, ikeuchi1986, efstathiou1992, barkana1999}. 
These simple models based on ionization balance and heating are supported by 
3D numerical simulations \citep{okamoto2008}. Thus  knowing the characteristic halo mass of UFDs
-- at the time of formation --  is important for ascertaining whether they formed before or after the
onset of reionization.

A more nuanced approach to reionization considers not only the impact
of positive feedback, but also the local environment
of the dwarf population. In particular, reionization must have been very patchy in the 
local universe due to the clumpy distribution of the Local Group, nearby groups and 
the Virgo cluster. Just how the expanding Str\"{o}mgren spheres from the early galaxies
evolved and overlapped is difficult to establish and presumably awaits future observations
of the distribution of neutral gas at high redshift. Numerical simulations reveal that local
reionization may have started around $z=11-14$ and have completed by $z=6-8$
\citep{busha10, ocvirk13}, a delay of order 400~Myr. A major uncertainty is the impact of 
the ionization front 
from Virgo: if the first stars formed significantly earlier, the local reionization epoch may have 
been earlier still. Regardless, much of the dwarf population
(and presumably the globular cluster population) may have acquired most of their stars before
reionization was complete \citep{ricotti05, busha10}. Indeed, their star formation may have 
been curtailed by the rising local UV background at late times.

\smallskip\noindent{\sl The origin of UFDs}.
The nature of UFDs, whether they precede or follow the epoch of reionization, is an area of
active research \citep{brown14,weisz14}. It is difficult to relate low-mass galaxies today with 
their early universe counterparts.
Some UFDs may be the dynamically stripped cores of more massive systems although
the majority, as presented here, may be relics that have preserved their halo mass 
over billions of years, with significant accretion of dark matter at later times. (It is conceivable
that the inner halo UFDs have lost the matter they accreted after $z=10$ through tidal stripping,
such that the halo masses inferred today broadly reflect their masses at the time of formation.)

The mass of the proto-Galaxy at $z=10$ would have been of roughly 
$5\times 10^9$\msun\ with a virial radius of about 5 kpc. 
This early system was surrounded by hundreds of subhalos with
masses $\lesssim 10^7$\msun.
Over cosmic time, a small fraction of halos on highly radial orbits 
would have been disrupted and absorbed by the evolving Galaxy. Their orbit distribution and 
mass ratio compared to the
progenitor halo imply that dynamical friction is relatively unimportant, thus many are 
likely to survive to the present day. 
Some fraction of these would have accreted gas \citep{susa04a,susa04b}
although the details will depend on many environmental factors \citep{benitez-llambay13}.
It is likely that the present population of a few dozen UFDs is a small fraction 
of a much bigger population to be revealed in future deep photometric surveys
\citep{bullock2010,hargis14}.

In an important new study of six UFDs \citep{brown14}, five objects are found to have formed 
75\% of their stellar mass before $z\sim 10$ (13.2 Gyr ago) and all of their stars by $z\sim 3$ 
(11.6 Gyr ago). There is a strong suggestion here of rapid quenching by a photoionising event:
it is plausible that these UFDs preceded the main epoch of local or possibly global reionization \citep[see also][]{webster15a}. In the work that follows, this is our starting assumption. We perform 
hydrodynamical simulations of early progenitors of UFDs with the aim of deriving their 
dynamical, photometric and chemical properties today. We also consider whether our models
are able to explain the new class
of very metal poor, damped Ly$\alpha$ systems at $z\sim 2$ 
which have been argued to be intrinsically low-mass systems \citep{cooke11b,cooke15}, although rather
less is known about these intriguing systems. Where our work differs from earlier
research is that we focus on the retention of baryons $-$ rather than disruption $-$ in dark matter halos that are traditionally considered to be too small to retain baryons. This
requires most of the action to take place before reionization, and in clumpy media that are 
less susceptible to supernova feedback.

\smallskip
\noindent {\sl New models.}
In our new low-mass halo models presented below, we consider the recent claim
that star formation in UFDs finished {\it before} reionization \citep{brown14} although
the isochrone ages are sufficiently uncertain that some of it may have continued
long after \citep{weisz14}. For star formation to have started so soon,
some gas must have cooled and settled into the potential well before the onset of reionization. 
We stress that complete reionization {\em does not completely exclude}
ongoing star formation for masses below $M_{\rm crit}$
\citep[e.g.][]{susa04a, susa04b, dijkstra04}. 
For it to continue to later times, the nuclear gas density must be high enough to shield the 
central regions from a global ionizing event. 

Consider the time it takes for the gas to evaporate from the low
mass halos. For a cosmic ionizing UV intensity $J_0 = 10^{-21}$ erg
cm$^{-2}$ s$^{-1}$ Hz$^{-1}$ sr$^{-1}$, with a moderate power--law form of $f_\nu \propto \nu^{-2}$, 
we obtain an ionization flux of $\sim 2\times 10^5$ photons s$^{-1}$ cm$^{-2}$.  The rate 
of evaporation of hydrogen atoms off the surface of a confined gas cloud, of radius 50~pc  
($\sim 3\times 10^{41}$~cm$^2$), assuming one ionising photon evaporates one hydrogen atom, 
is then given by 
\begin{equation}
\dot{m}_{\rm evap} \sim 1.5\times 10^{-3} (n_{\rm r}+ 1)^{-1}\; \mbox{\rm M}_\odot \, \mbox{\rm yr}^{-1}, 
\end{equation}
where the $n_{\rm r}$ term allows for the number of recombinations in the escaping wind. Thus, the
timescale for evaporation for the \mbary\ $\sim$ $10^5$ M$_\odot$ of gas in the models shown here 
($\tau_{\rm evap}\sim f_b M_{\rm b}/\dot{m}_{\rm evap}$) can exceed 100~Myr, which is long 
enough for star formation to proceed in the interior while the outer regions are evaporating.
With reference to positive feedback processes,
various authors have noted that H$_2$ formation is {\it enhanced} in the limit of a weak 
ionizing intensity reaching the dense gas core \citep{kang92, tajiri98, susa00, susa04a} which can 
further accelerate star formation. Some of these processes may assist ongoing star formation in UFDs after 
the onset of reionization \citep{brown14,weisz14}.

For dwarf galaxies in the outer halo at the time of their
formation, their internal radiation fields are likely to have dominated local 
reionization \citep{ocvirk2011}.  Here we consider the impact of UV radiation from massive
young stars born within the low mass halo, and the impact of the subsequent supernova
explosion on gas retention.  An alternative route to suppressing star formation in low mass 
systems is to remove or heat up the gas through starburst-driven winds \citep{maclow99}. 
In their 2D axisymmetric models, constant energy input corresponding to central explosions 
is found to couple efficiently to a smooth gas component 
confined by a flattened disk. Once again, these models rule out any gas surviving in a 
dark-matter halo with a mass $\lesssim$ \mcrit. But these authors did not consider discrete SN events 
and modelled the 
winds moving through a smooth medium rather than a more physical clumpy medium where the 
hot wind fluid couples less efficiently with the dense interstellar medium \citep{cooper2008}.
We consider the impact of winds in smooth and clumpy media, and winds that are off-centred
within the galaxy. Both off-centred explosions and clumpy media are likely at any epoch and
therefore both cases must be considered in detail. 

Here we investigate the conditions under which a `minimum mass' system confined 
by a dark-matter halo, specifically a relic of the early universe,
survives to the present day.
We consider feedback arising from high-mass stars due to their stellar
winds, their radiation fields, and their subsequent supernova
explosions. We find that these limiting cosmological systems may have
had total masses of $M_{\rm min} \sim 10^7$ M$_\odot$, well below
the limit of what is thought to have survived the reionization epoch.
In our follow-up papers, we perform new calculations of chemical evolution within these
`minimum mass' galaxies over cosmic time \citep{webster14, webster15a, webster15b}. 
Conceivably, one can hope to learn about the star formation before, during and immediately after the
reionization epoch from distinct chemical signatures in these low mass systems 
\citep[e.g.][]{blandhawthorn10b}. 

In our new models, the {\it initial} baryon fraction is roughly
$f_b \approx 10\%$ of the virial mass although the final baryon fraction can be 
orders of magnitude smaller. This is consistent with
observations of dark-matter dominated systems observed today in the low mass limit
\citep{mcgaugh2010}.  
As we shall see, our new models explain some ultra-faint dwarf galaxies
\citep{willman05, irwin07, simon07, martin08, strigari08}
and the peculiar chemical signatures observed in very metal poor, damped Ly$\alpha$ 
systems \citep{pettini08, penprase10, cooke11a}.
We discuss our new simulations in light of both classes of objects.

In Section 2, we describe the scope of our numerical experiments. The 
model parameters are summarised in Section 3. In Section 4, we introduce the
simulations and present our results in Section 5. In Section 6, we discuss
these results in light of new observations before giving the conclusions in
Section 7.

\medskip
\section{Statement of the problem}
\label{s:problem}

\subsection{What is a first galaxy?}

Our interest in galaxy systems with the lowest possible mass arises from
a desire to understand the chemistry of the first stars \citep[K13]{frebel12, frebel14}.
Our assumption is that the surviving remnants of the first galaxies will have 
small dark halo masses and possibly low baryon mass fractions for the
reasons outlined in \S\ref{s:intro}.
Many of these systems will have merged with larger galaxies long ago, but
some  are likely to be falling into larger systems today for the first time.

So what constitutes a first galaxy? This must retain a long-lived stellar 
system and importantly a confining dark matter halo.\footnote{We recognise
that `tidal dwarf galaxies' form a rare class of objects that may not have associated 
dark matter, but generally a galaxy is defined by the existence of a confining
dark matter halo, at least at the time of formation.} Whether
low mass stars formed alongside or immediately after the first stars is
unknown. It is likely that the process was patchy, with 
the very first stars forming in rare density peaks, but with an increasing
fraction of the universal volume becoming ionized as the first stellar generation 
took hold.

Here we make the simple assumption that the universe was enriched to a
threshold metallicity of order [Fe/H]=-4 before reionization was complete
\citep[e.g.][]{bromm11}.
Some of this gas was accreted onto low-mass dark matter halos, and 
star formation began. Our threshold metallicity allows us to consider
a `normal' initial mass function in order to describe how star formation
proceeded at these early redshifts \citep{bromm01, schneider02}.
Given the stochasticity of star formation at our low masses, there may be 
a significant amount of low-mass star formation before the first supernova
event. Some remnants of first galaxies may therefore contain stars formed
from gas enriched only by Population III stars, thereby providing direct 
information on the first stellar yields.

However, in the systems relevant to our work (ultra-faint dwarfs, very metal 
poor DLAs), some of the detected elements (e.g. Sr) can only have arisen from supernova
ejecta. Thus, our minimum mass models have the more stringent requirement 
that they can survive one or more supernova events during their lifetime. 
one-shot dwarfs those where the gas never recovers from the first SNe.
In order to experience more than a single star-formation
episode, a very small galaxy must retain enough gas in the face of the
energy outflow from a single star--forming episode for a second one to
occur, perhaps at a much later time. A large burst of 
energy such as a supernova could either
heat the galaxy ISM and halt star formation or completely
evacuate the star-formation region of gas. This will cut off 
star formation until such time as the halo re-accretes the blown-out gas, 
assuming this is allowed by the environment.
Thus, a `first galaxy' is one that must retain sufficient gas to
form stars in the wake of at least a single supernova event.

This is not a contrived circumstance and we show below (\S\ref{s:discuss}) that there may 
already be observations to support such a scenario. It is likely
that there are situations in which only one or two supernovae occur within
a single star-forming ``burst'' \citep{karlsson13,ritter15}. \citet{hurley98} identify several
starburst epochs in the Carina dwarf with the first of these taking place
of order $\sim$10 Gyr ago. At that time, the 
surface density of the star formation rate ($s_\star$) within the core
was roughly 10$^{-3}$ M$_\odot$ yr$^{-1}$ kpc$^{-2}$ over the
period of the burst, $T_0$. This timescale is now well
established in starburst and dwarf galaxies such 
that $T_o \lesssim 10$ Myr \citep{tolstoy09,sharp10}.

\begin{figure}[htb!]
   \includegraphics[scale=0.6]{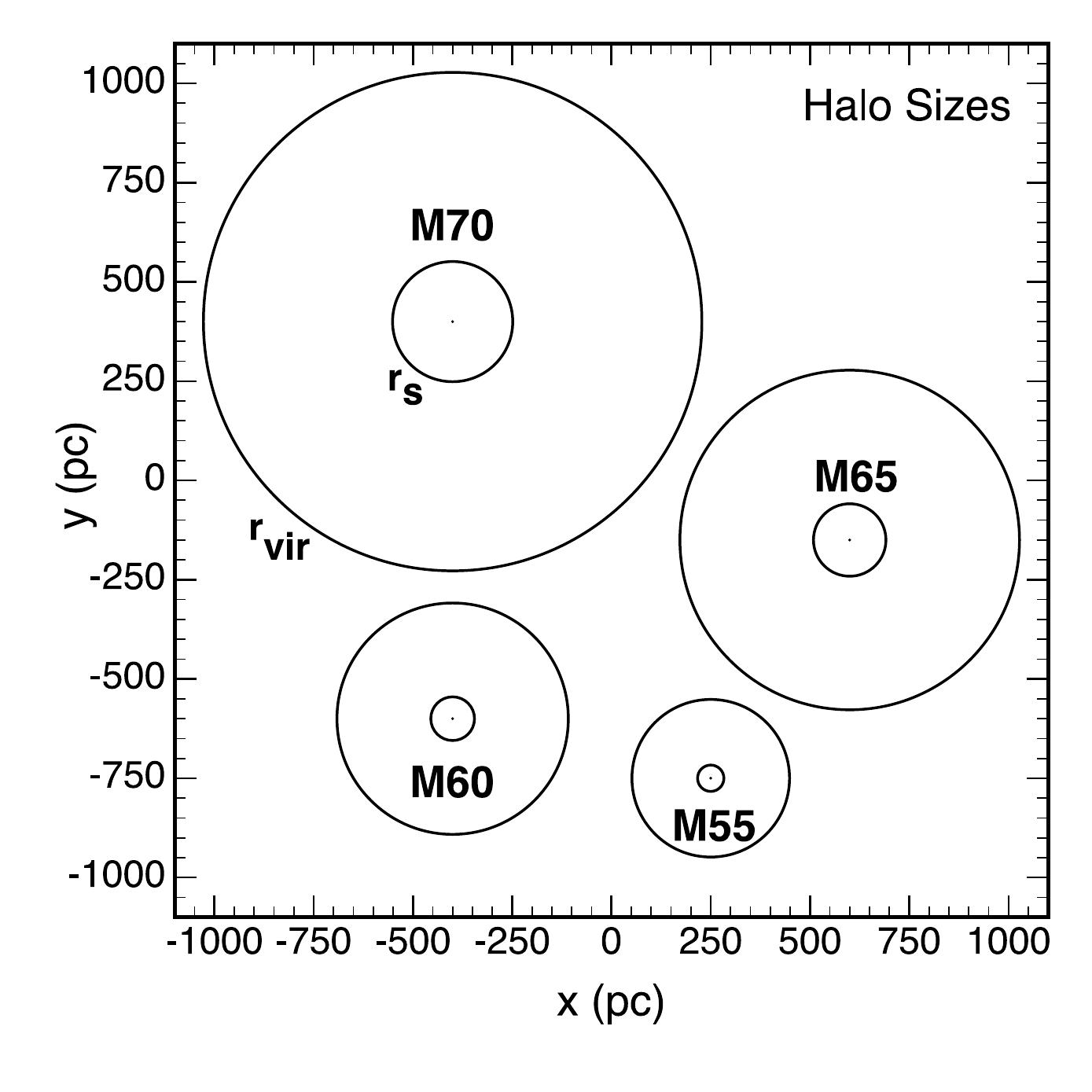} 
   \caption{Einasto dark matter halos considered here showing the relative sizes of the
   virial and scale radii at $z=10$. The virial masses of the four haloes are 
   $\log[M_{\rm vir} ({\rm M_\odot})]$ = 5.5, 6.0, 6.5 and 7.0; 
we refer to these models as M55, M60, M65 and M70 respectively. For all halos, the virial
mass and radius continue to grow to $z=0$ (see Fig.~\ref{f:mass_radius}).
   }
   \label{f:halos}
\end{figure}

\begin{figure}[htb!]
   \includegraphics[scale=0.45]{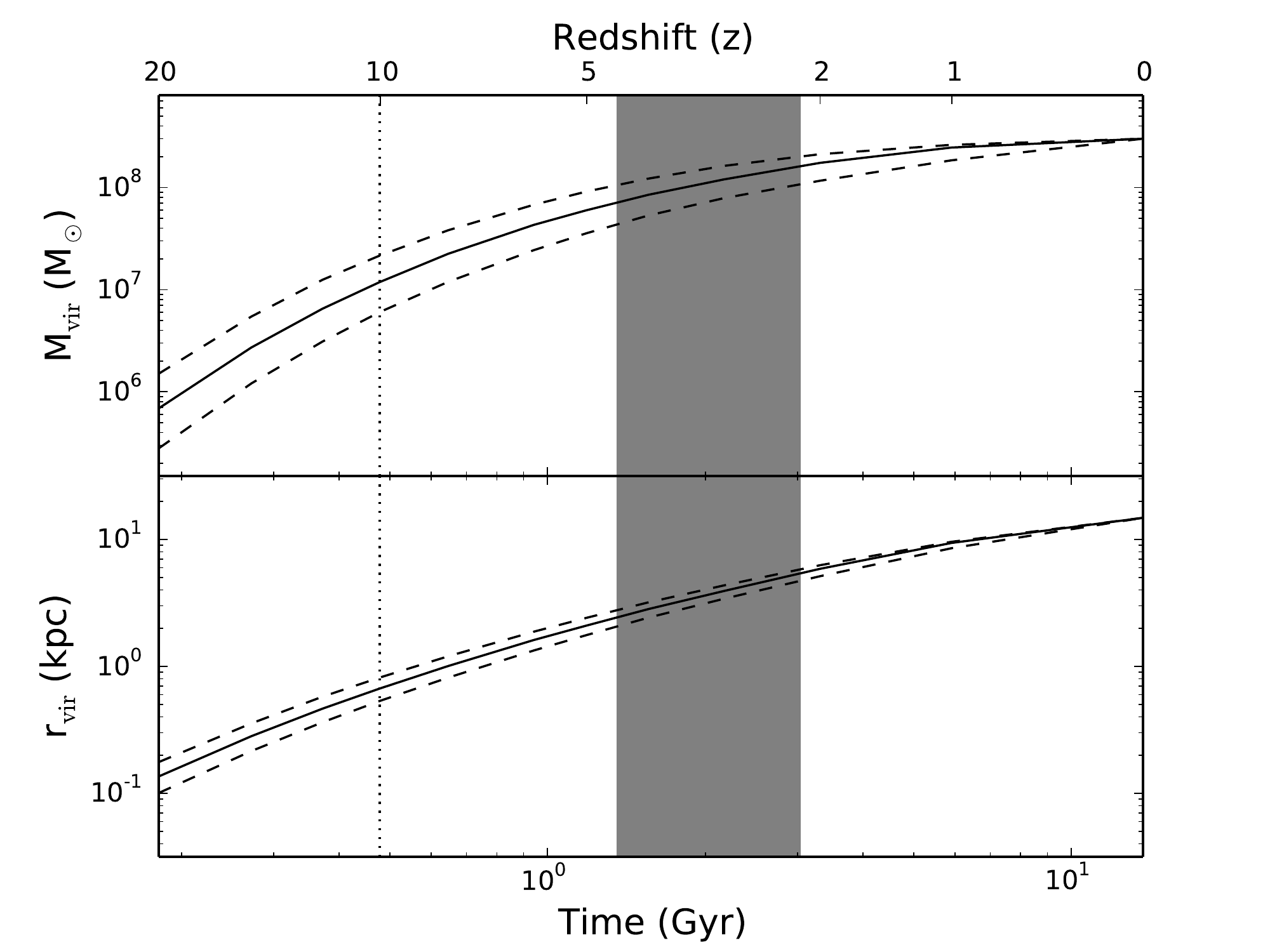} 
   \caption{The predicted growth in the virial mass (top) and radius (bottom) of the M70
   halo at $z=10$ (dotted line) to the present day (see Fig.~\ref{f:halos}). The shaded 
   band is the redshift range of the damped Ly$\alpha$ systems discussed in later sections.
   This plot was derived from 5000 runs of the tree merger code of \citet{parkinson08}:
   the dashed lines encompass 67\% of the predicted halos at each epoch.
   }
   \label{f:mass_radius}
\end{figure}

\begin{figure*}
\centerline{\includegraphics[scale=0.45]{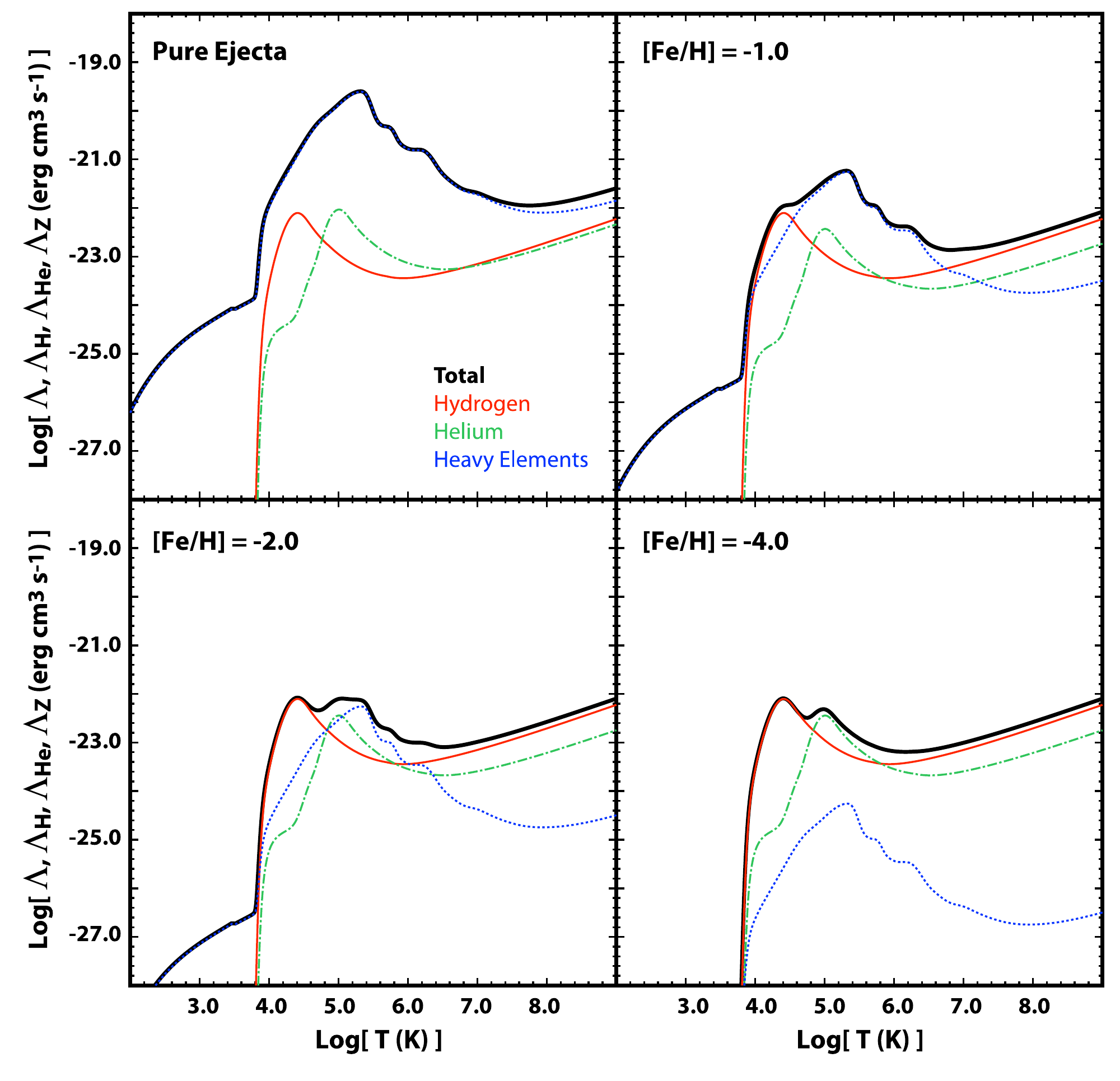}}
\caption{ \label{f:eos} The variable cooling function, using compositions encountered in the simulations.
In each panel, the heavy line is the total cooling, the thin red line is cooling due to hydrogen, the green dash-dot line is due to helium, and the metals combined produce the blue dotted line.  In the single supernova model, the internal metal ratios are fixed by the supernova composition, and are grouped into a single function.}
\end{figure*}
\begin{figure*}
\centerline{\includegraphics[scale=0.6]{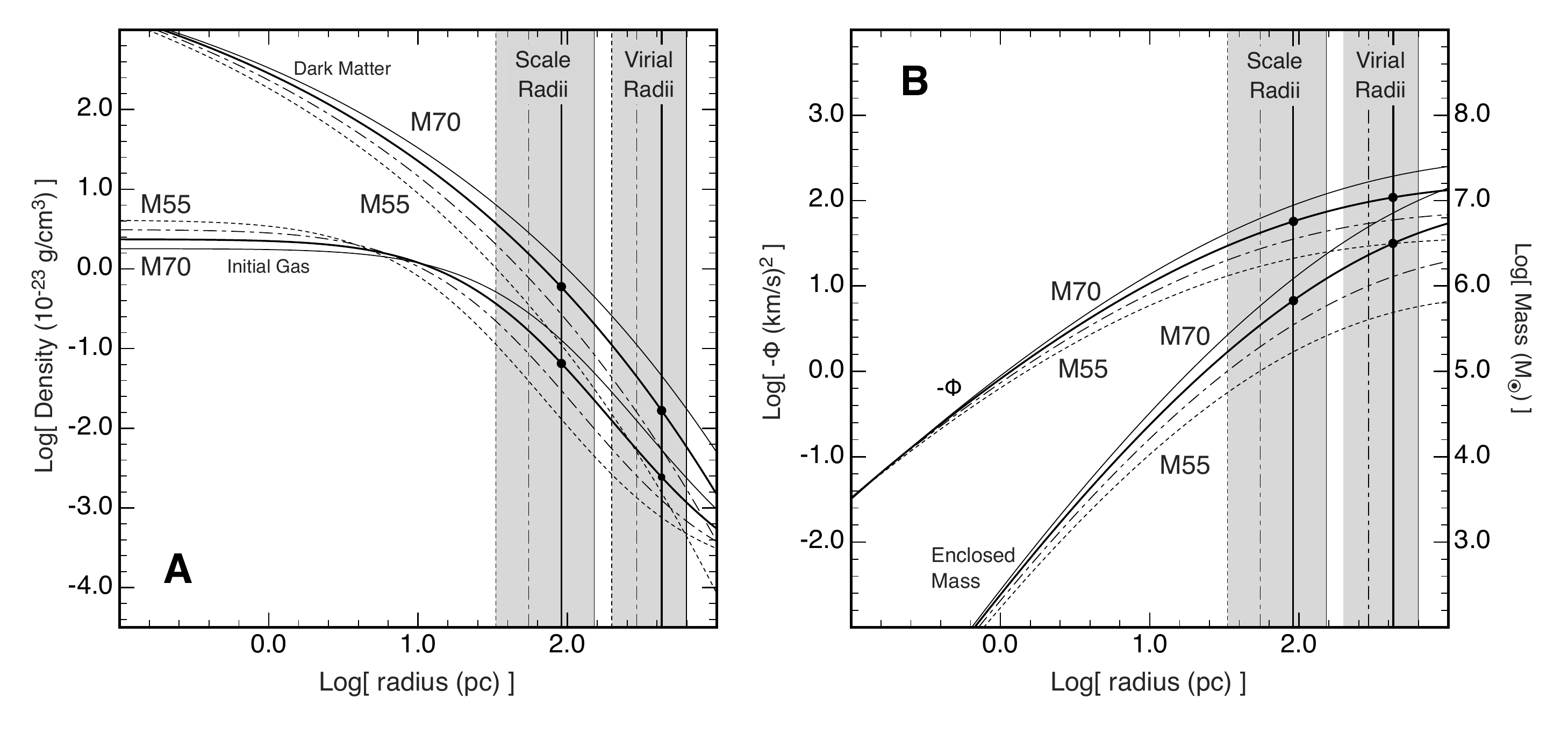}}
\caption{ \label{f:pots}
A: Dark matter and initial gas density (neutral gas, $\mu = 1.21$,  $\beta = 2.0$) profiles for each
of the model halos.
B: The potential functions ($-\Phi$) and enclosed dark matter masses.  
Both panels: The dashed, dash-dot, heavy and thin curves are for the M55, M60, M65 and M70 models respectively. The vertical grey bands cover the radial extent of four scale radii ($r_s$) and four
virial radii ($r_{\rm vir}$), marked with four vertical lines, for the four mass models using the same line
type for each model.
}
\end{figure*}
\begin{figure}
\centerline{\includegraphics[scale=0.45]{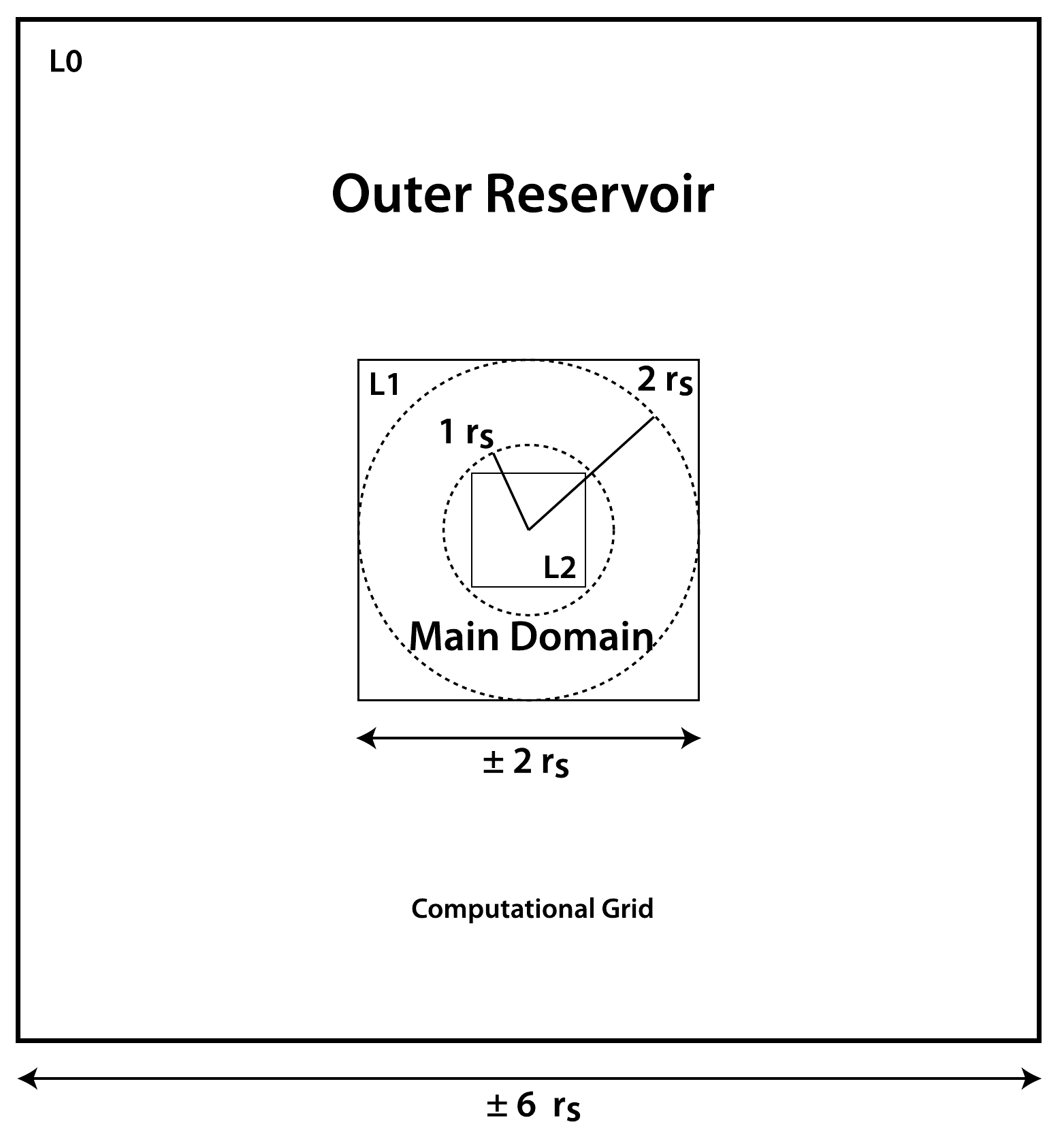}}
\caption{\label{f:grid} A 2D representation of the 3D computational domain. Each square represents a cubic domain enclosing the interior regions.  The main level of interest is L1, where the mass and energy transport during the explosions are measured.  A third inner level, L2, is used to resolve early epoch ($t < 50$~kyr) thin shell phases of the SNR.}
\end{figure}

\subsection{Star formation and supernovae}
\label{s:models}

To test the plausibility of a single supernova event occurring in our models, 
we consider a simple model of star formation. We consider dark-matter
potentials with an Einasto profile over the virial mass range
$\log[M_{\rm vir} ({\rm M_\odot})]$ = 5.5, 6.0, 6.5 and 7.0; 
we refer to these models as M55, M60, M65 and M70 respectively (see Fig.~\ref{f:halos}). 
If we integrate over the Einasto profile, 
the total masses for these models in dark matter 
are $\log[M_{\rm tot} ({\rm M_\odot})]$ = 5.9, 6.4, 7.0 and 7.5.
(These total masses do not include the small amount of baryons.)
In Fig.~\ref{f:mass_radius}, we show how the low mass halos are expected to grow in
mass and radius with cosmic time. Halo growth is not treated in our simulations because
the time frame of our analysis is short. {\it Our results at a given halo mass at $z=10$ 
correspond to more massive systems at later times.}

Here we show complete results for M65 and M70, and partial results for M55 and M60
because these models fail to retain sufficient baryons.
The {\it initial} baryon fraction is assumed to be roughly 10\% within
the virial radius \citep[e.g.][]{krumholz12}. We assume that star formation occurs 
where the gas is densest, i.e. within a radius enclosing 50\% of the baryons ($r_{0.5g}$)
inside one scale radius $r_s$, the scale radius of the Einasto potential (see Section 3.1).
Over this volume, we find that the gas density is within a factor of 6 of the maximum
central value.
%Here the enclosed baryon mass is 10\% of the enclosed dark matter mass reflecting the same 

For the models (M55, M60, M65, M70), the enclosed gas mass within the half gas-mass
radius ($\lesssim$\rs/2) are
($2.8\times 10^3$, $9.5\times 10^3$, $2.5\times 10^4$, $7.5\times 10^4$)\msun\ and
thus the average gas surface densities are (2.1, 2.6, 3.3, 4.2)\msun\ pc$^{-2}$.
In the clumpy gas models, there are order of magnitude variations in local gas density.
From direct observation, we know that star formation takes place at these moderately 
low column densities over a range of metallicities down to [Fe/H] $\approx$ -1 
\citep[e.g.][]{bigiel08}. As these authors show, the star formation efficiency is observed to vary 
wildly below 9\msun\ pc$^{-2}$ with a median value of only a few percent. Little is
known about how the star formation rate depends on metallicity, but it seems likely that
it declines with [Fe/H] due to the difficulty of forming H$_2$. We stress however that
the inferred value of $s_\star$ in Carina is comparable to what is observed in the
outer metal poor regions of spiral galaxies \citep[e.g.][]{bigiel08}, and this dwarf galaxy 
somehow managed to form long lived stars at much lower metallicities.

For a star formation efficiency of 2\% \citep{kennicutt89}, the expected {\it total} stellar mass 
over the lifetime of the system is (56, 190, 670, 2400) M$_\odot$. If we assume a Kroupa mass 
function, roughly one fifth of the mass in stars will end their lives as supernovae. These 
estimates are susceptible to a lot of scatter in the low mass/surface density limit \citep{bigiel08}.
If we scale the projected star formation rate per unit area to the largest burst in Carina 
($\sim$ 10$^{-3}$\msun\ yr$^{-1}$ kpc$^{-2}$), the timescales required to form these stellar 
masses are (44, 54, 71, 86) Myr, for a half gas-mass radius of (20, 34, 55, 94) pc (see 
\S\ref{s:numeric} for
further details on the halo properties). If a typical burst lasts 10 Myr as in Carina \citep{tolstoy09}, 
we would expect multiple bursts over the star-forming lifetime of the galaxy.

So can we consider an isolated supernova event to be applicable to all of our
models? It is conceivable that more
than one event can occur within the timescale of a single burst $T_o$.
The results 
given in Table~\ref{t:SNrate} show that multiple SN events are rare in the M55 and M60 models,
are relatively unusual in the M65 model although occur half the time in the M70 model. Multiple
supernovae in more gas-rich M70 models happen half the time, but in practice we can consider
most of these eventualities to be ``single supernova events" because one of these is highly likely
to be a low mass event and physically separated from the other event. These events
are likely to enrich their own immediate environs,
although substantial mixing can occur between the events at late times \citep{ritter15}.
While different choices of the size 
of the star-forming region or baryon fraction may alter these results, we would not expect this to 
alter the overall conclusion of a single supernova event.  We note that simulations
of more massive halos will need to consider multiple events within the same burst.

\begin{table}[htdp]
\caption{\label{t:SNrate} Likelihood of supernova events scaled to Carina star formation rate for a typical burst duration lasting 10 Myr; $t_{SN}$ is the median length of time between supernovae in Myr.}
\begin{center}
\begin{tabular}{r r r r r}
 &M55&M60&M65&M70\\
\hline
\hline
0		&93\%	&80\%	&45\%	&19\% \\
1		&6.6\%	&18\%	&36\%	&32\% \\
$>$1		&0.2\%	&2.0\%	&19\%	&50\% \\
\hline
${t_{SN}}$ (Myr)&100&30&8&4\\             
\hline
\end{tabular}
\end{center}
\label{default}
\end{table}

\subsection{The pre-supernova phase}

Massive stars evolve rapidly in the lead-up to the explosion during which time
the stellar wind and radiative luminosity can be substantial. This pre-supernova phase 
can be an important factor in removing gas, particularly for halos with low baryon fractions. 
In the worst case scenario, a star at the halo centre can ionize an amount of gas, 
$M_{\rm ion} \sim
S_\star m_{\rm p} / \alpha n_{\rm H}$ M$_\odot$ where $\alpha$ is the
total recombination coefficient, $m_{\rm p}$ is the mass of the proton,
and $n_{\rm H}$ is the hydrogen density. Using the \citet{schaerer03} 
models at a metallicity of [Fe/H] $=$ -4.0, the mean ionizing rate averaged
over the Kroupa initial mass function is $S_\star \sim 4\times 10^{48}$
phot s$^{-1}$ for the M55 model, $1\times 10^{49}$ phot s$^{-1}$ for M60, 
and $3\times 10^{49}$ phot s$^{-1}$ for M65. Taking the worst case scenario 
where the most massive star falls near the centre, the total ionized masses 
are $1.7\times10^3$\msun\ for M55, $6.1\times10^3$\msun\ for M60,
$2.2\times10^4$\msun\ for M65 and $5.5\times10^4$\msun\ for M70.
In each case, a large fraction of the neutral 
gas can be ionised. This is likely to have a significant impact in reducing 
subsequent star formation. But the more likely situation is not expected to
be so severe. Because the total amount of star formation is 
small in our models, stochastic effects can dominate. The median radiation
levels expected are more biassed to B stars rather than rare O stars, and
the ionized masses can be much less than 10\%. We include the effects of
pre-ionization in our models below.

\section{Model parameters} \label{s:modpar}

\subsection{Numerical models}
\label{s:numeric}

In our experiments, we look at a series of spherically
symmetric gas distributions in hydrostatic equilibrium, detonate
a single supernova, and track the ability of the potential to retain
significant amounts of gas over 25~Myr or more. For all of our models,
we consider the period prior to the onset of cosmological reionization
when stars were forming for the first time. 
At such early times, substantial amounts of gas have already begun 
to settle into dark matter halos up to and exceeding \mcrit\ \citep{bullock01,ricotti05}.
This is consistent with the
latest results from deep photometry for ultra faint dwarfs where star formation 
appears to have been quenched by reionization in almost all cases 
\citep{brown12,brown14,webster15a}.

We consider both clumpy and smooth gas distributions within the low-mass halos. 
For the clumpy models, we make the important assumption that the confined 
gas has a fractal distribution {\em before} the supernova explodes. There
are two reasons for this assumption. First, the infalling gas becomes 
shocked and highly inhomogeneous close to the dark halo core 
\citep{abel00}. Secondly, a high mass star that is
destined to become a supernova injects mechanical and ionizing energy
into the surrounding gas prior to the explosion. This energy input
drives turbulence in the surrounding gas \citep{elmegreen04}.

In looking for possible key factors in the retention, we consider the
following scenarios: 
\begin{itemize}
 \item {\em Abundances \& cooling:}
We use chemical abundances that approximate primordial conditions
starting at $[{\rm Fe/H}] = -4.0$. This starting metallicity is 
conservative:  \citet{wise08a} show that the initial metallicities of the first 
galaxies could have been as elevated as [Fe/H] $\sim$ -3.0 or higher.
%\citet[Fig. 7]{bromm11}
We compute models with cooling, and models ($\gamma=\frac{5}{3}$) 
where we have artificially turned off microphysics affecting
energy loss (e.g. cooling through metal and recombination lines).
The cooling function
%and the mean molecular weight, $\mu$, as a function of gas temperature
is presented in Fig.~\ref{f:eos}; the abundances are given in
Appendix A.
 \item {\em Clumpiness:} We consider models with
quasi--fractal cool star--forming gas, and comparison models where the
gas is homogeneous and smooth. The key point here is that the degree of
coupling between the supernova energy and the surrounding medium will
depend on whether the medium is smooth or structured.
 \item{\em Central and off--centre explosions:} We
consider a single $10^{51}$~erg supernova explosion (SNe), located
either at the geometrical centre of the halo potential, or off-centre at
a radius that encloses half the mass contained inside one scale radius
of the potential. Supernovae that occur away from the centre have an
asymmetric environment. 
 \end{itemize} 
This gives, in combination, 8 scenarios for each of
the halo masses considered.  In each case, the mass and energy contained
in the inner scale radius is integrated at 25~kyr intervals and plotted
over the course of the simulation to obtain estimates for the masses
retained at late times.

In the {\em Fyris Alpha} models presented here, we use a fixed dark matter
potential that dominates everywhere. Future models will include
self-gravity in the gas. Therefore we limit the {\it initial} baryonic mass inside a
scale radius ($r_s$) to be 10\% of the dark matter mass
inside that radius.  This results in an overall baryonic fraction out to 
the virial radius ($r_{\rm vir}$) of \fbary\ $\approx$ 12\%, less than the
canonical \fcrit\ $\approx 17$\% global baryonic mass fraction \citep{ade14}.  
Our choice of \fbary\ is a conservative assumption because a lower gas
budget generally assists the supernova in accelerating and heating the gas.  If
the models retain gas with these lower mass fractions, they can be
reasonably expected to do so with more gas in place.

 %[Gas with $\beta$, $T_*/T = 2.0$ (wrt $v_c$ at $r_s$), assuming equipartition of thermal and turbulent 

\begin{table*}[htdp]
\footnotesize{
\caption[Parameters for models with preionisation]{Model Parameters: the cosmology is fixed by the halo formation redshift, $z$, the normalized Hubble constant 
$h=H_o/100$ km s$^{-1}$ Mpc$^{-1}$, the dark energy density $\Omega_\Lambda$ and the dark matter
density $\Omega_M$. The Einasto halo properties (see text) are fixed by the concentration parameter, $\alpha$,
and the halo dark matter density contrast with the ambient medium, $\Delta$, defined at redshift $z$. This density
contrast can be expressed as $\Delta_c$ averaged over the virial radius and normalized to the mean critical density $\rho_c$, or as $\Delta_u$ normalized to the universal mass density $\langle\rho_u\rangle$. The halo potential is
defined by its scale radius $r_s$, the mean density and mass within this radius $\rho_s$ and $M_s$. The halo virial 
mass is $M_{\rm vir}$ within the virial radius $r_{\rm vir} = x_{\rm vir} r_s$; M$_{300}$ is the mass within 300 pc and 
M$_{\rm tot}$ is the mass integrated to infinity. For the baryons, $c_s$ and $c_{s,\rm iso}$ are the adiabatic (physical) and 
isothermal sound speeds in the gas with mean temperature $T_{\rm gas}$. This gas has a central density
(number density) of $\rho_{\rm gas,0}$ ($n_{H,0}$) with a virial (total) mass of $M_{\rm gas,vir}$ ($M_{\rm gas,tot}$).
Note that the initial baryon fraction $f_b \approx M_{\rm gas,\it s}/M_s \approx M_{\rm gas,vir}/M_{\rm vir}$ $\approx$ 10\% 
and the radius containing half the initial gas within $r_s$ is $r_{0.5g}$. 
}
\begin{center}
\begin{tabular}{c c c c c}

%----------------------------------------------------------
&&{\bf Cosmology}&&\\
\\
 &$z$&$h$&$\Omega_\Lambda$&$\Omega_M$\\
  & 10.0 &0.7&0.7&0.3\\
  \\
$\alpha$&$\Delta_c$&$\rho_c$(g/cm$^3$)&$\Delta_u$&$\langle \rho_u \rangle $(g/cm$^3$)\\
0.18&177.5&3.68E-027&591.7&1.11E-027\\
\\
%----------------------------------------------------------
&&{\bf Dark Matter}&&\\
\\
     &$r_s$&$\rho_s$&$x_{\rm vir}$&$r_{\rm vir}$\\
Model&(pc)&(g/cm$^3$)&&(pc)\\
\hline
\hline
M55&33.2&1.02E-023&5.98&198.5\\
M60&55.0&7.73E-024&5.29&291.3\\
M65&91.3&5.87E-024&4.69&427.6\\
M70&151.3&4.48E-024&4.15&627.7\\
\\
   &$M_s$&$M_{\rm vir}$&$M_{300}$&$M_{\rm tot}$\\
Model &$({\rm M}_\odot)$&$({\rm M}_\odot)$&$({\rm M}_\odot)$&$({\rm M}_\odot)$\\
\hline
\hline
M55&5.64E+04&3.16E+05&4.10E+05&7.90E+05\\
M60&1.94E+05&1.00E+06&1.02E+06&2.73E+06\\
M65&6.73E+05&3.16E+06&2.36E+06&9.44E+06\\
M70&2.34E+06&1.00E+07&5.08E+06&3.29E+07\\
\\
%----------------------------------------------------------
&&{\bf Baryons}&&\\
\\
   &$c_s$&$c_{s,{\rm iso}}$&$T_{\rm gas}$&$M_{\rm gas,vir}$\\
Model &(km/s)& (km/s)&(K)&$({\rm M}_\odot)$\\
\hline
\hline
M55&2.70&1.91&5.33E+02&3.87E+04\\
M60&3.90&2.76&1.11E+03&1.19E+05\\
M65&5.63&3.98&2.32E+03&3.66E+05\\
M70&8.16&5.77&4.86E+03&1.15E+06\\
\\
    &$r_{\rm 0.5g}$&$\rho_{\rm gas,0}$&$n_{\rm H,0}$&$M_{\rm gas,\it s}$\\
Model &$({\rm pc})$ &(g/cm$^3$)&(cm$^{-3}$)&$({\rm M}_\odot)$\\
\hline
\hline
M55&20.8&4.09E-23&17.2&5.63E+03\\
M60&34.5&3.11E-23&13.1&1.94E+04\\
M65&57.1&2.36E-23&9.91&6.73E+04\\
M70&94.8&1.79E-23&7.55&2.34E+05\\
\hline
\end{tabular}
\end{center}
\label{t:halos}
}
\end{table*}%
 %small

\subsection{Dark matter halo} \label{s:einasto}

In recent years, high resolution CDM models have identified the need for
a halo density profile that has an additional parameter beyond those
used in the NFW profile \citep{springel2008, navarro2004}.  The profiles
seen in the latest simulations display local density
derivatives that are power laws of radius, unlike the NFW profile that
has asymptotic behaviour at small radius. They invoke a generalised
exponential function that has this property, and note that it has been
used in the past by \citet{einasto65} in the context of galactic halo
spatial denities\footnote{This has the same functional form as the
projected Sersic profile but the 2D function does not transform to the
same underlying radial density distribution.}.

We use the Einasto function, a generalised exponential, to form a dimensionless potential
defined by the parameters $\alpha$, a halo concentration factor $x_{\rm vir}$ (often
referred to as $c$) and the density
contrast of the halo $\Delta$ ($\approx$178) with respect to the universal background
\citep{nichols09}. We stress that these parameters must be specified {\it at a fixed redshift
$z$} because of their strong cosmic evolution \citep{power03,duffy08}.

We consider four halo mass models
(M55, M60, M65, M70) defined in Section~\ref{s:models}.
The total halo masses are (3.0, 2.7, 2.5, 2.3) times the virial mass respectively
(e.g. Fig.~\ref{f:pots}). The halo and gas properties are
summarised in Table \ref{t:halos}. Our normalization uses a density
function value of $1/e$ at the scale radius, $r_s$. This gives a slightly
different set of scaling constants, but recovers exactly the same
physical quantities as the $\frac{1}{4}$--scaling used in
\citet{nichols09}. This change was motivated by wanting to compare with
dimensionless isothermal potentials used by \citet{sutherland2007}, which
also obtained a scaled density value of $1/e$ at the scale radius. (We
refrain from using the popular `core' radius because, unlike the
isothermal potentials, the generalized Einasto functions do not have
a distinct core region; here we use $r_s$ throughout.)

Our adopted halo parameters are based on those of \citet{duffy08} and 
\citet{springel2008}. As their fits to the simulated halos are measured over 
a relatively low redshift range, we extrapolate these properties back to 
a redshift of $z = 10$. At the time of writing, we became aware of new
work \citep{correa15} where the parameters are extended to back to
$z=20$ but these appear to be consistent with our estimated values,
particularly for the important M65 and M70 models.

\subsection{The cold interstellar medium} \label{s:fractal_ISM}

In our simulations, we introduce a non--uniform medium to study the
influence of inhomogeneity on the dynamical interaction between the supernova
ejecta and the surrounding gas. To assess the importance of
inhomogeneity, we run both homogeneous (smooth) and inhomogeneous
(clumpy) models and compare the outcomes. Following
\citet{sutherland2007}, to establish a non--uniform medium, we make use of
an analogy with a turbulent medium. Rather than attempt to review this
huge topic, we refer the reader to astrophysically oriented
reviews as a starting point for background material \citep{elmegreen04,
scalo04}.

We represent the non--uniform properties of the turbulent medium using
three statistical characteristics: the variance, $\sigma_F^2$, the
intermittency (by using log--normal distributions), and a self-similar
power--law structure.  We achieve both the property of intermittency and
power-law structure simultaneously with an iterative scheme first
developed by \citet{lewis02}; see Appendix B for more details. Hence the
initial distribution of the ISM that we employ should be regarded as a
physically motivated generalization of a homogeneous model.

We further simplify our turbulent medium by neglecting the velocity
structure, and focus on choosing statistical parameters to describe the
density alone.  This is justified numerically by the relatively small
turbulent velocities expected when compared to the very large velocities
found in our global SNe--ISM interaction.  The velocities observed in
typical warm ISM conditions fall in a range of transonic to mildly
supersonic values, Mach 1-5 \citep[e.g.][]{heiles04}.  At
temperatures at or below $10^3$~K, this corresponds to velocities $\ll
4$\kms. In our 0.1-1~kpc simulations over $10^7$~yr timescales, the
resulting displacements amount to only a few cells at most, and are
insignificant compared to those of the energy bubble generated by the
supernova outburst, where  velocities of many hundreds of \kms\ occur. 
Consequently, we do not impose a turbulent velocity field on the cold
ISM medium.

\subsection{ Hydrodynamic calculations} \label{s:hydro}

The modelling was carried out using the {\em Fyris Alpha} code
\citep{sutherland2010, hawthorn2007}.  It solves ideal Euler hydrodynamic
flows of gas with cooling and an equation of state appropriate to
astrophysical gasses, in the presence of a fixed gravitational field. 
The code uses a third order semi--Lagrangian shock--capturing method
that is robust and well suited to the extreme temperatures and Mach
numbers encountered here.

The computational domain consists of a three dimensional nested series of 
cubic meshes, as illustrated by the schematic in Fig.~\ref{f:grid}. Each level is 
a factor of 3 smaller in physical domain and higher resolution than its
outer containing level, covering the central third of the 
enclosing level. Each level is $216^3$ cells with an overall effective 
resolution of $1944^3$ cells.  

A relatively low resolution outer level, L0, covers the halo out to $\pm 
6$ scale radii, and provides an outer reservoir and boundary 
condition for the main level of interest, L1.  The outer reservoir reduces 
or avoids completely any boundary artefacts on L1 due to 
gravity bringing new gas onto the main level when infall conditions apply, 
keeping level L1 as physically consistent as possible at all 
times.  L0 contains the virial radius. Velocities at the outer boundary 
are extremely small at all times considered here and the 
outer L0 boundary is essentially fixed.

The main level, L1, covers a region of just over $\pm 2$ scale radii, and 
is the primary region where the movement and evolution of mass 
and energy are measured.   Early epochs require higher resolution to resolve thin spherical shell 
radiative blast--waves, so a third, inner injection level, L2, is 
added for to resolve the early phases of the SNe.  After 25 kyr, L2 is disabled and the 
remainder of the simulations are carried out on L0 and L1.

For each halo mass, the scale radius varies and so the resolution is 
variable.  The highest resolution, on level L2, is $\Delta x \sim 
0.25$, 0.37, and 0.62~pc per cell for the M55, M60 and M65 models.  
Thus the effective mass resolution is $\sim 10^4$ M$_\odot$ / 216$^3$
or of order 10$^{-3}$ M$_\odot$; in practice, cell densities can vary by more
than 4 orders of magnitude across the grid.

The remaining key features of the calculations can be summarised:
\begin{itemize}
\item A fixed gravity potential,
interpolated locally by 3rd order PPM interpolation to match the
interpolation of the hydrodynamical algorithm, taking conservative 
potential differences to compute mean accelerations. 

\item SNe simulated by depositing $10^{51}$ erg of
energy as internal gas energy in a smoothed spherical region, as small
as possible while retaining spherical geometry well enough to prevent
gross non-spherical `pixel' errors, typically 6 cell radius on L2, corresponding to 
$1.5 - 3.7$~pc.  This then takes the form of a pulse of low density hot over-pressure gas
which then expands, converting internal energy to kinetic energy and
doing work against the potential.  Energy is also lost via cooling.

\item To allow for significant changes in the local metallicity in the models as the metal 
rich-ejecta propagate through the models, a generalised cooling model has been adopted 
that allows the local composition of each cell to be determined during the simulation.  
Instead of a single cooling function and uniform molecular weight, the cooling was 
separated into three components, representing hydrogen, helium and heavier metals, and 
ionization was taken into account.

\end{itemize}

The metal-rich supernova ejecta were followed using a scalar tracer variable, advected passively, 
representing the supernova material, and hence the local helium fraction and metal fraction.
The local enrichment of the gas was thus computed at each time step.
Furthermore, the ionization fraction of the components, precomputed with the MAPPINGS~IV code 
\citep{allen08,sutherland93} as a function of temperature, allowed the equation of state to vary with the mean molecular weight, ensuring more accurate shock temperatures, and allowing for a wider range of cooling timescales caused by strongly varying composition. Fig.~\ref{f:eos} shows the resulting computed cooling functions for a range of 
encountered compositions on the grids.

We acknowledge that there are large uncertainties in the yields from essentially zero metallicity 
stellar supernova models. The yield of metals, particularly oxygen, could have an impact on the outcome, lower mass supernovae giving less enrichment and cooling, while higher mass stars could give more cooling. More detailed future work incorporating a range of supernova progenitor masses will investigate the impact of different supernova assumptions.
By allowing for the ionization and composition to vary from cell to cell, we obtain more accurate species columns, for example, for neutral and ionised hydrogen.

Our focus is on the hotter atomic and ionised phases of high pressure gas, and the cooling therein,
as that gas is volume filling and will have the primary effect on the global dynamic evolution if any. 
Gas in the molecular phase is likely to
be confined to the centres of the densest cores, if present at all.  As
the outflows are all warm ($> 10^4$K), they are expected to be atomic or
ionised in any case.  In the face of uncertainty in the formation of H$_2$
under primordial conditions and little or no
information on CO cooling, we leave molecular cooling out. It would be essential 
to include it if we were tracking the (much longer) time required for the hot gas to re--cool back to star--forming
temperatures $\ll 100$~K, but here we are focussed on the initial halo gas retention
problem. We defer the star formation problem to more substantial investigations.

That we neglect molecular cooling in these simulations can be considered a
conservative approximation, but we expect it to have little effect on
the global dynamics of gas retention. More than 99.9\% of the
internal energy is lost  by the time the gas cools back below $10^4$~K
from early temperatures of more than $10^{7}$~K, so only a tiny fraction of
the SNe energy is available to the molecular cooling in the present
simulations.

In Table \ref{t:effic}, we calculated the thermalization efficiency for a subset of the models
as a check that our simulations correctly handle cooling. 
The thermalization efficiency was estimated by finding the ratio of the total energy at the time 
when gas is first leaving the grid to the total initial energy.
Our efficiencies should be considered upper bounds as it is likely that some cooling will 
occur at later times. From \citet{thornton98}, we expect the thermalization efficiency to be 
$\sim 10\%$, which is consistent with the results of our models.

\subsection{Pre-ionization from the SN progenitor}
\label{s:preion}

The pre-ionization of the gas by the supernova precursor plays a key role in the loss 
or survival of gas after the supernova explosion, particularly in low-mass dark matter
halos. We consider in the detail the impact of an ionised HII region that
forms around the progenitor star prior to the explosion.  The key parameters
of this model are the stellar properties ($T_{\rm eff}$, $L$) and the associated stellar atmospheres. 
These give the ionising photon flux as a function of time up to the SN
event.  Transforming the grid coordinates to spherical coordinates around the star, radial summations to 
approximate the optical depth integrals are performed, resulting in thermal and ionization structures 
calculated from the {\em MAPPINGS~IV} ionization code.
These are then mapped back to cartesian coordinates and used in the {\em Fyris} fluid computations.  
It was necessary to perform this remapping for every timestep in the fluid dynamics simulation in order 
to track the evolution in opacity. This slowed the code by a factor of 5--10. We considered both
clumpy and smooth external media with and without gas cooling. 

\begin{figure*}[htb!]
\centerline{
   \includegraphics[scale=0.55]{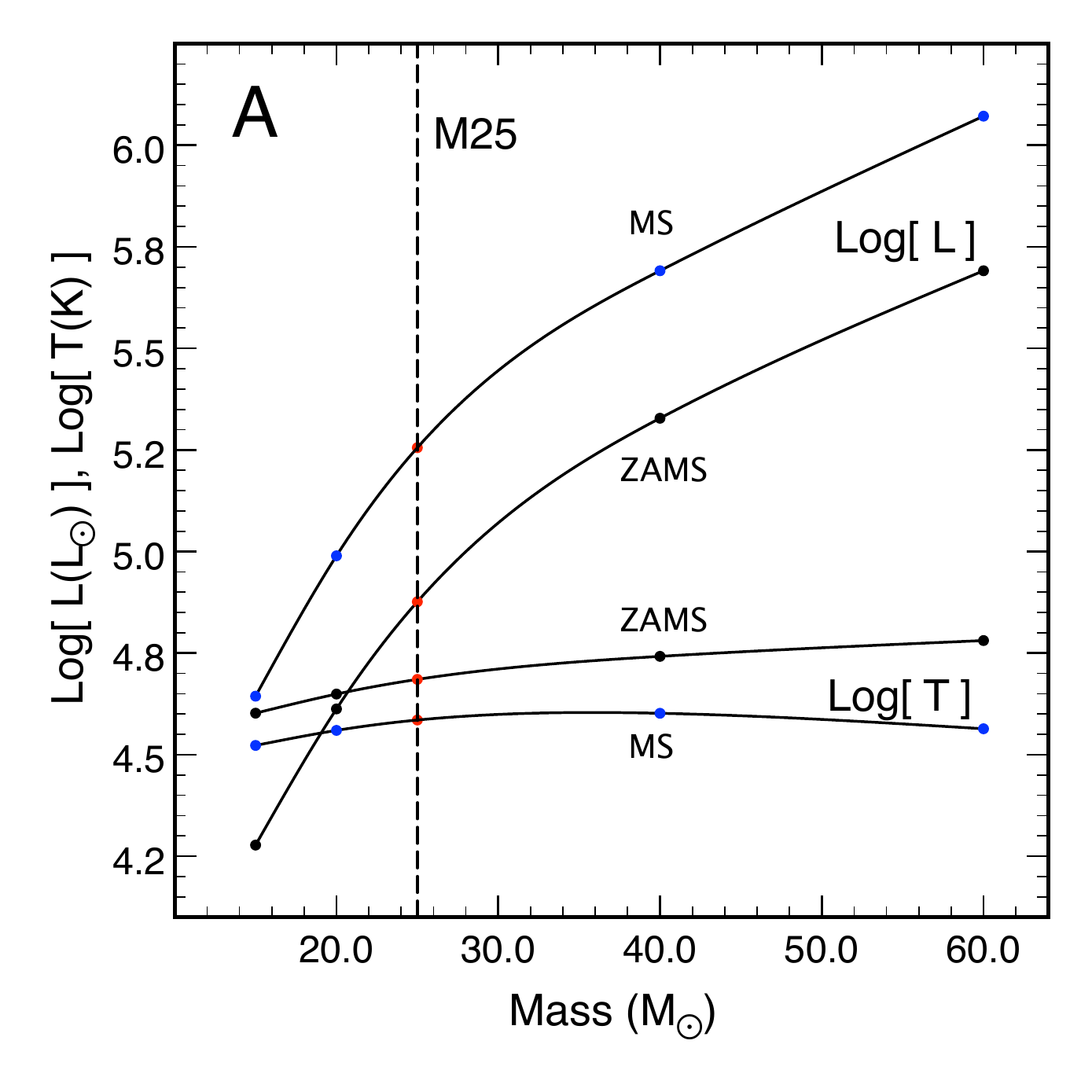} 
   \includegraphics[scale=0.55]{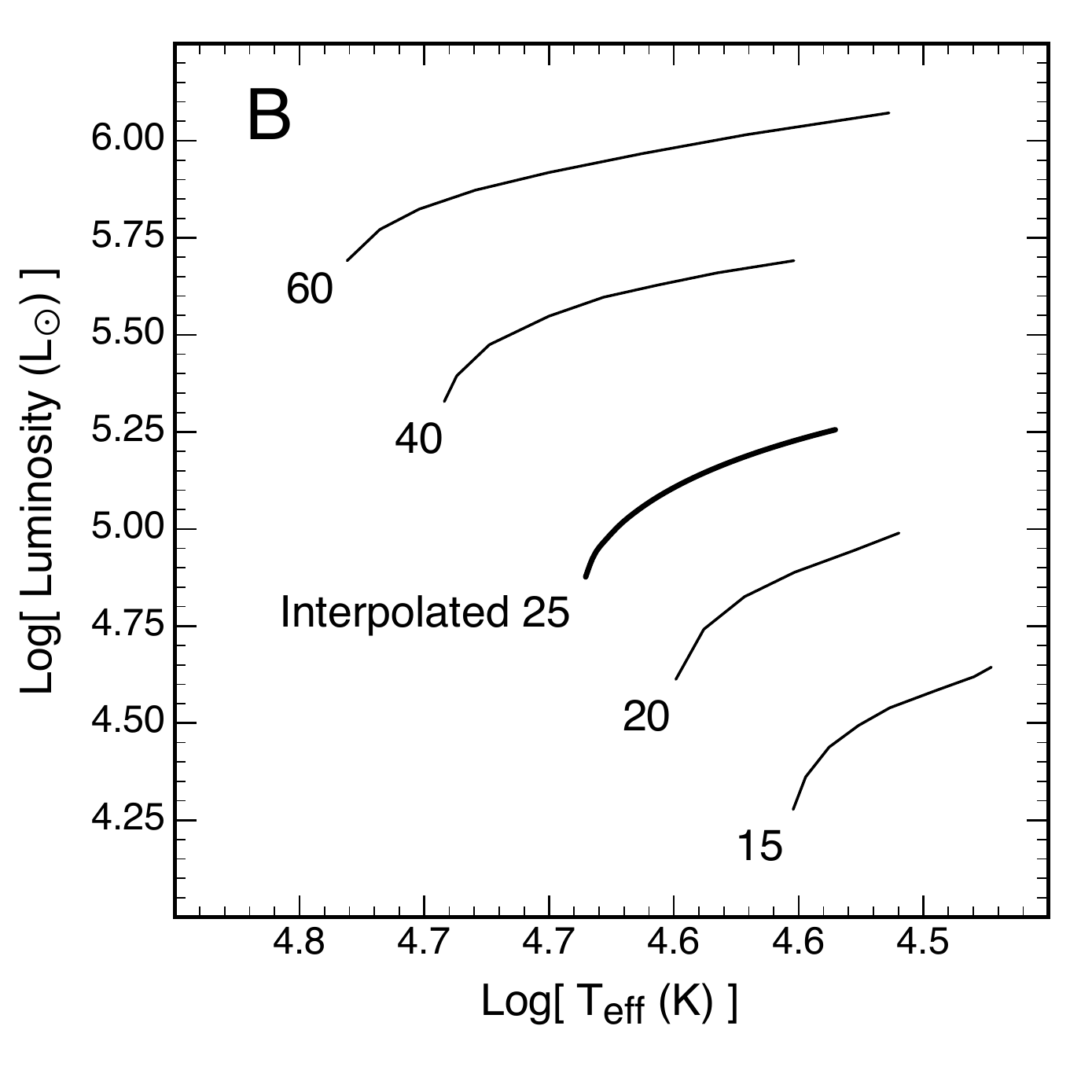} 
   }
   \caption{A: Interpolated luminosity and temperature for the 25\msun\ model in the MM02 grid.  The zero age main sequence is labelled ZAMS and the end of the main sequence as MS. 
B: Evolutionary tracks for the four most massive rotating (300 km s$^{-1}$) star models in MM02 at [Fe/H]~$\approx$~-4. We adopt a 25 M$_{\odot}$ star model interpolated from these tracks with cubic splines using the endpoints in A. For this star, the time span from the zero age main sequence (ZAMS) to the explosion is 6.1 Myr.}
   \label{f:m25spline}
\end{figure*}

We model the supernova progenitor with a low metallicity star of 25\msun.
Higher mass stars produce stronger ionising intensity
and therefore assist gas loss by overheating the gas prior to the supernova 
explosion. But since 80\% of all supernova events in a conventional initial
mass function have lower mass and therefore weaker ionising radiation fields,
we consider our choice of the upper mass limit to be conservative.
We use the evolutionary tracks of \citet[hereafter MM02]{meynet2002} for a rapidly 
rotating, low-metallicity massive star. 
The 25\msun\ track was missing from that work, and so we have 
interpolated the main sequence phase of this model from the other stellar models. 
In Fig.~\ref{f:m25spline}A,
we show the interpolated values of the luminosity and temperature for
the 25\msun\ model at [Fe/H] $\sim$ -3.7 (MM02). 
These are the closest published tracks for our required starting
metallicity of [Fe/H] = -4.0. We used the tabulated 
MS lifetime of 6.1 Myr; the interpolated temperature-luminosity
track is shown in Fig.~\ref{f:m25spline}B.

\begin{figure*}
\plotone{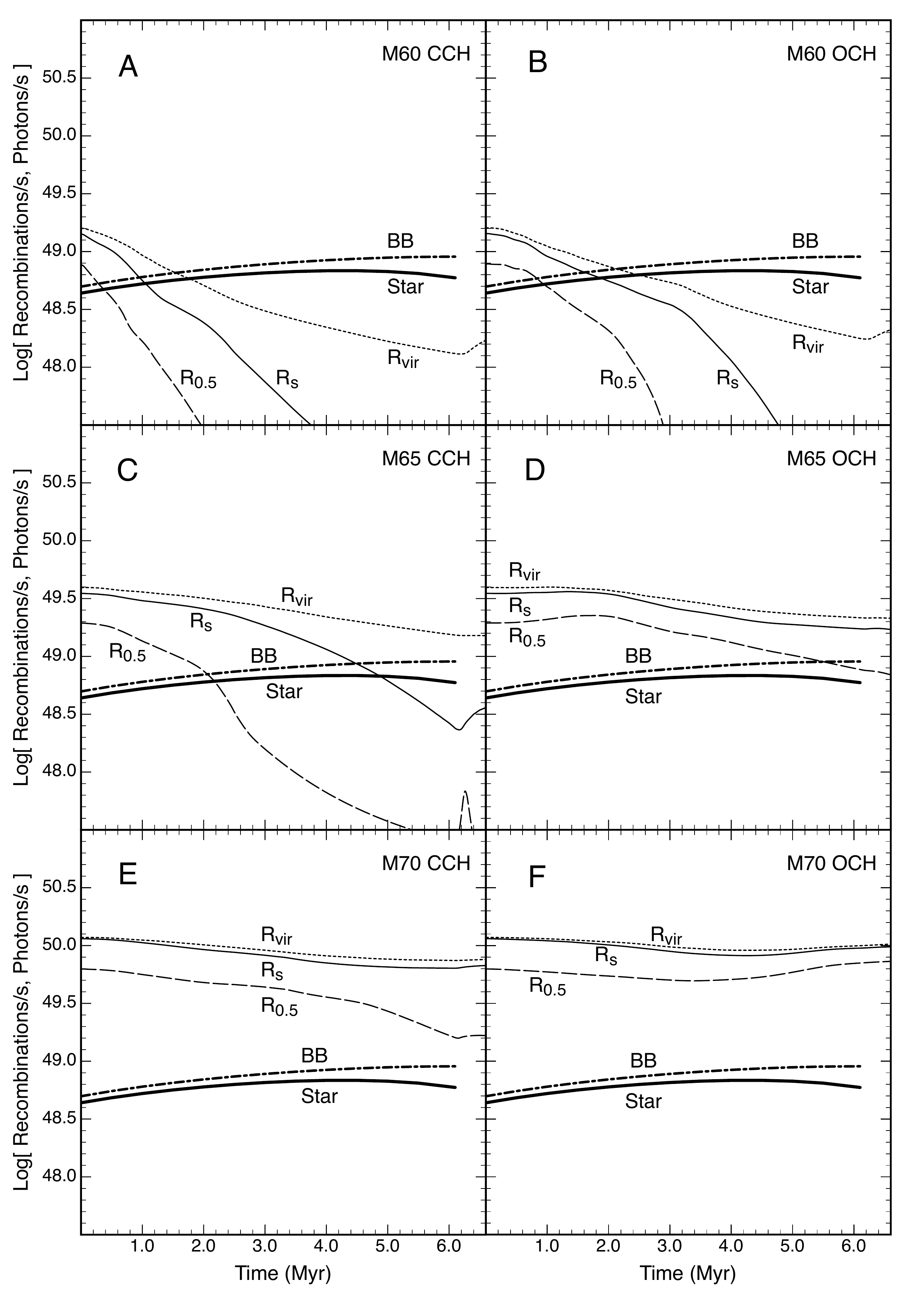}
\caption{ \label{f:recphot}
Comparison of the instantaneous photon rate (heavy lines) compared to the
total available recombinations as the models progress in the pre-supernova
phase. The lefthand figures (A, C, E) correspond to the centred models;
the righthand figures (B, D, F) refer to the off-centred models.
The two heavy lines represent a pure black body model (BB) and an extreme
low metallicity stellar atmosphere (Star), and the difference between them gives
an indication of the magnitude of any uncertainties in the atmosphere
models. Available recombinations are computed by integrating $n_e n_H^{+} \alpha(T_4)$
in cm$^{-3}$ s$^{-2}$ over the respective volumes where $\alpha$ is the recombination
coefficient summed over all possible transitions and $T_4$ is the gas temperature in units
of 10$^4$K. This yields the total number of recombinations
per second if all of the gas can be ionised, ignoring optical depth and geometry of the
ionised volume.}
\end{figure*}

In Fig.~\ref{f:recphot}, we show the impact of the total ionising budget during the
pre-supernova phase for our adopted stellar model and a perfect black body of the same
surface temperature. The panels A, C and E refer to the models with the progenitor
at the centre; 
panels B, D, F are for the off-centred models.
This gives some indication of the sensitivity to the stellar models.
If the thin lines exceed the thick lines, then a Str\"{o}mgren sphere is possible within the
radius, and the stellar radiation would be incapable of extending beyond that, which is the initial case
for all halos.  However as the gas responds to heating and evolves, the density is smoothed 
and falls on average due to outflows.  The quadratic dependence of recombinations means
that the available recombinations falls rapidly, and once the photon curves
exceed the recombination curves, the photoionised zones can grow rapidly and
ionise more of the halo.

In the upper panels, the M60 halo will be easily fully ionised after the first
2~Myr.  The middle panels show the M65 models are marginal if the photons are uniformly 
spread throughout the halo, and capable of ionising the inner core region
and possibly beyond before the supernova occurs.  The actual result will depend
on the details of the radiative transfer, whereas this is an indicative integral
view of the processes which does not take the geometry of the ionised volume 
into account. The lower panels for M70 indicate that the stars are insufficient 
to photoionise a significant volume of the halo.

These indications are broadly supported by the simulations: M60 is rapidly 
ionised, M65 is able to break out of the core radius before the SNe, and
the ionised region is contained in the M70 models.

\begin{figure}[htb!]
 \includegraphics[scale=0.55]{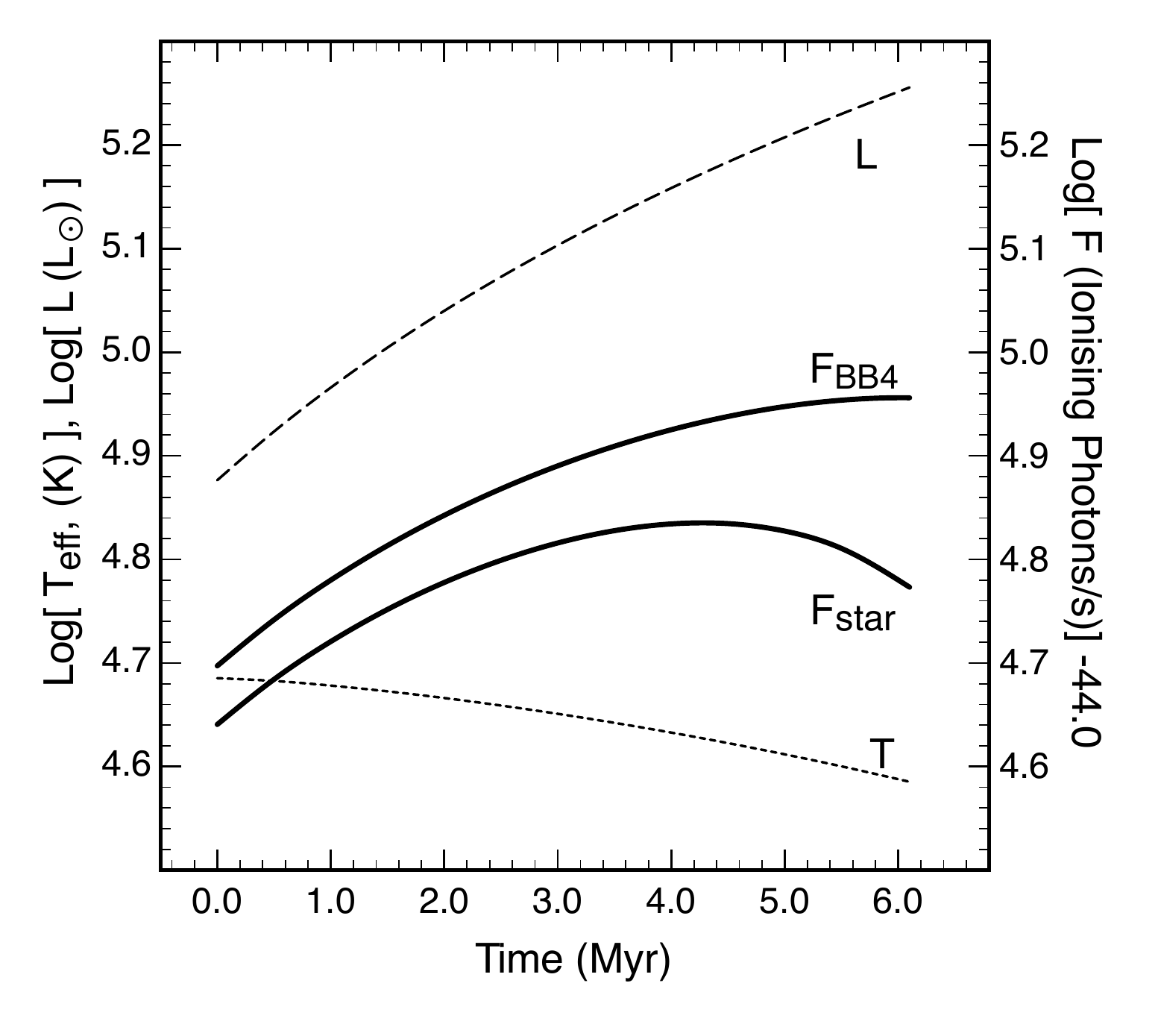}
   \caption{ Interpolated $L(t)$ [long dash], $T_{\rm eff}(t)$[short dash] and ionising photon fluxes, [heavy lines] for M25.  $F_{\rm star}$ assumes stellar atmosphere models, and a comparison is shown, $F_{\rm BB}$, which is the flux assuming a simple black body approximation to the stellar spectrum.  The star emits between $5\times10^{48}$ and  $7\times10^{48}$ ionizing phot s$^{-1}$ over its lifetime. }
   \label{f:m25lum}
\end{figure}
\begin{figure}[htb!]
\includegraphics[scale=0.47]{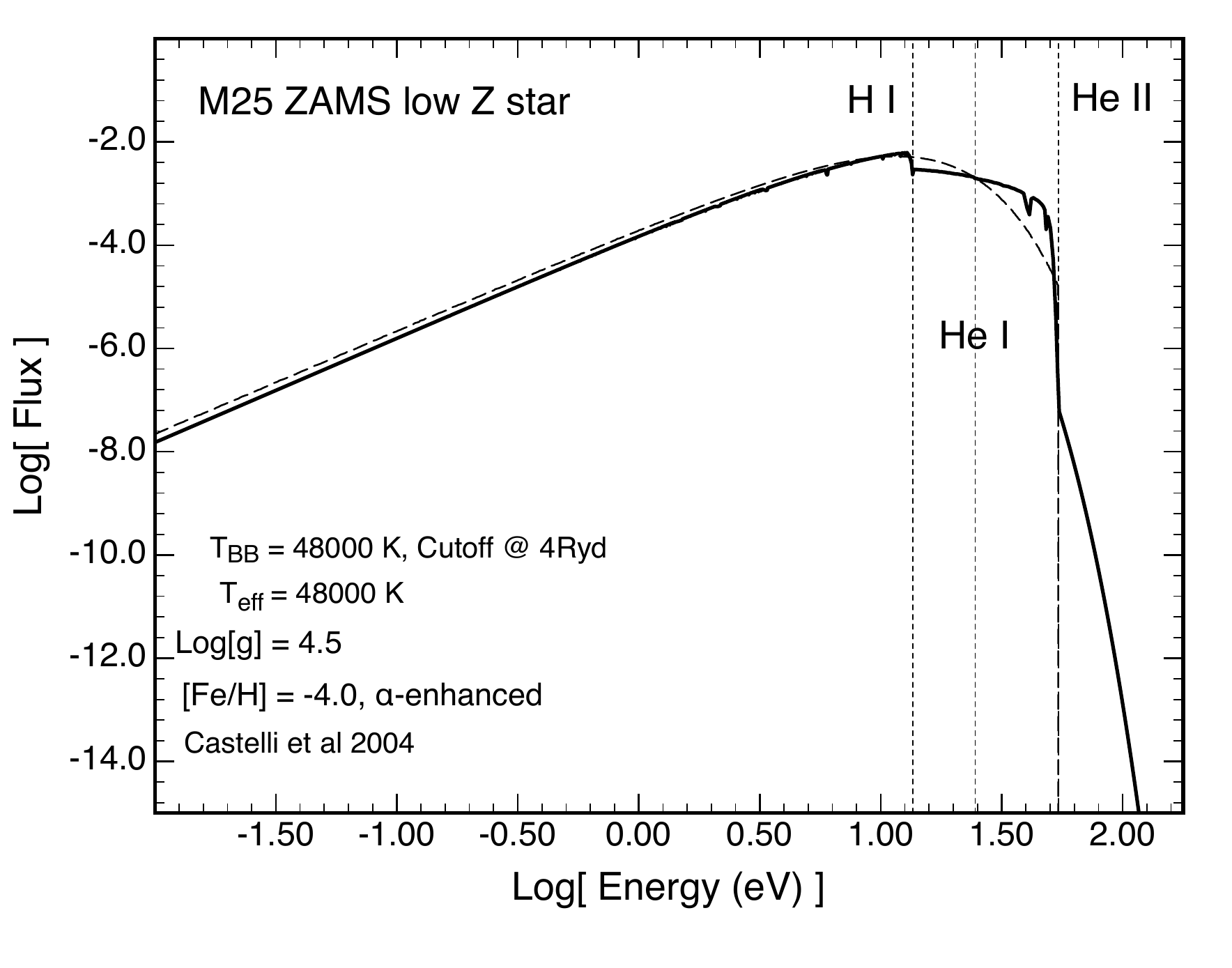}
\caption{Stellar spectrum (solid line) of our adopted 25 
M$_{\odot}$ star model using an ATLAS9 ZAMS stellar atmosphere at 
[Fe/H]~$\approx$~-4 \citep{castelli04}. The dashed line is the blackbody curve for the same surface temperature 
T$_{\rm{eff}}$ = 48 000~K.
   \label{f:m25spec}
 }
 \end{figure}
 
\begin{figure*}
\centerline{\includegraphics[scale=0.55]{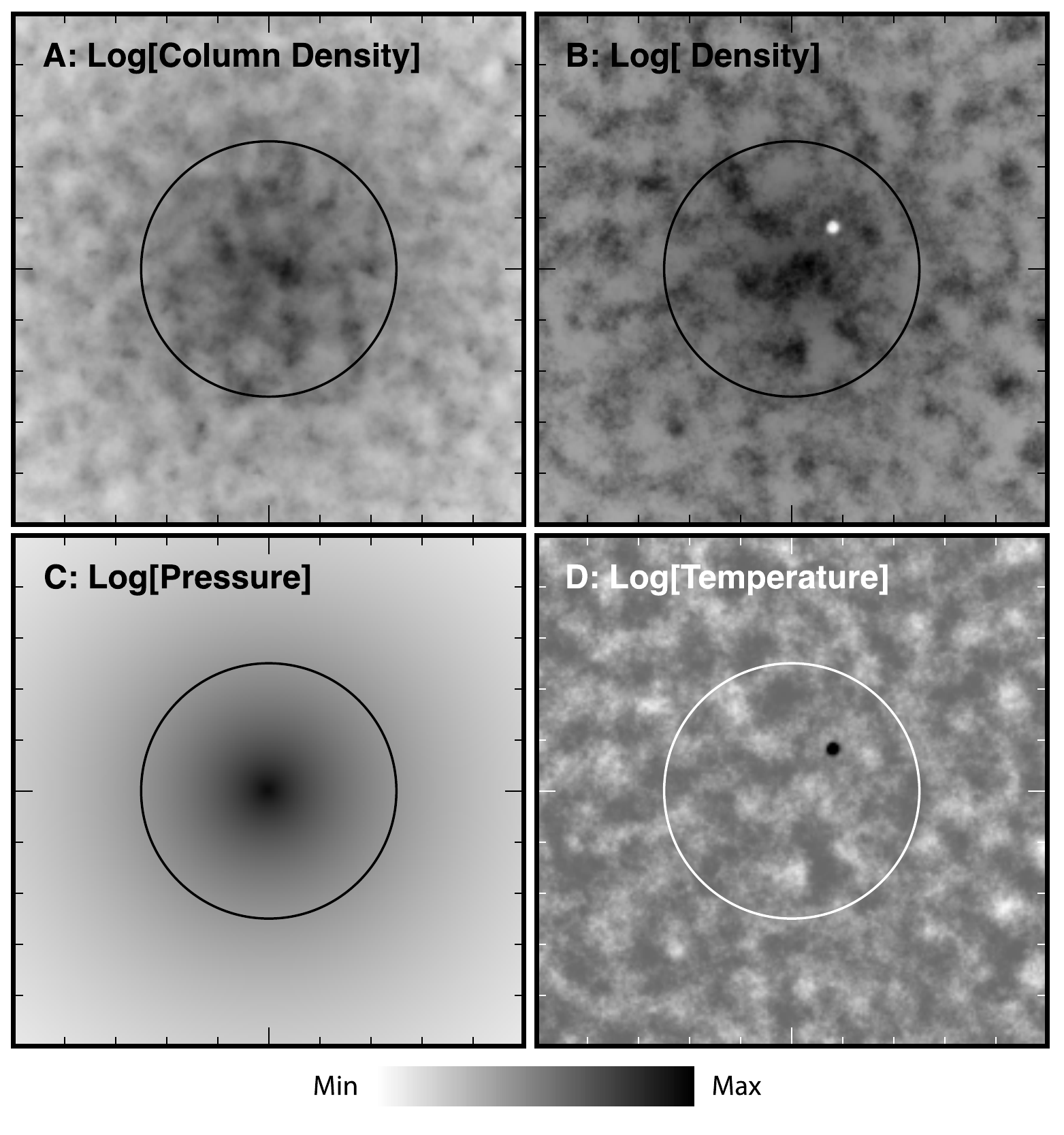}}
\caption{\label{f:setup}
The initial distributions of density, pressure and temperature (level L01) for the M60 
off-centred supernova model; the distributions are similar for M55, M65 and M70. While
the density and temperature structure are fractal, the overall pressure is constant
at the start. Each panel shows a slice through the middle of the halo and the quantity shown
is scaled logarithmically; the circle has a radius of $r_s$, the scale radius.
{\bf A:}  relative column density over the range -4.0 (white) to 0.0 (black). {\bf B:} relative density over the range -5.0 (white) to 0.0 (black). The off--centred SN is visible as a white dot.  {\bf C:}  relative pressure ranging from -5.0 (white) to 0.0 (black). {\bf D:} temperature ranging from 1.0 (white) to $> 4.0$ (black).}
\end{figure*}

\begin{figure}
\centerline{\includegraphics[scale=0.55]{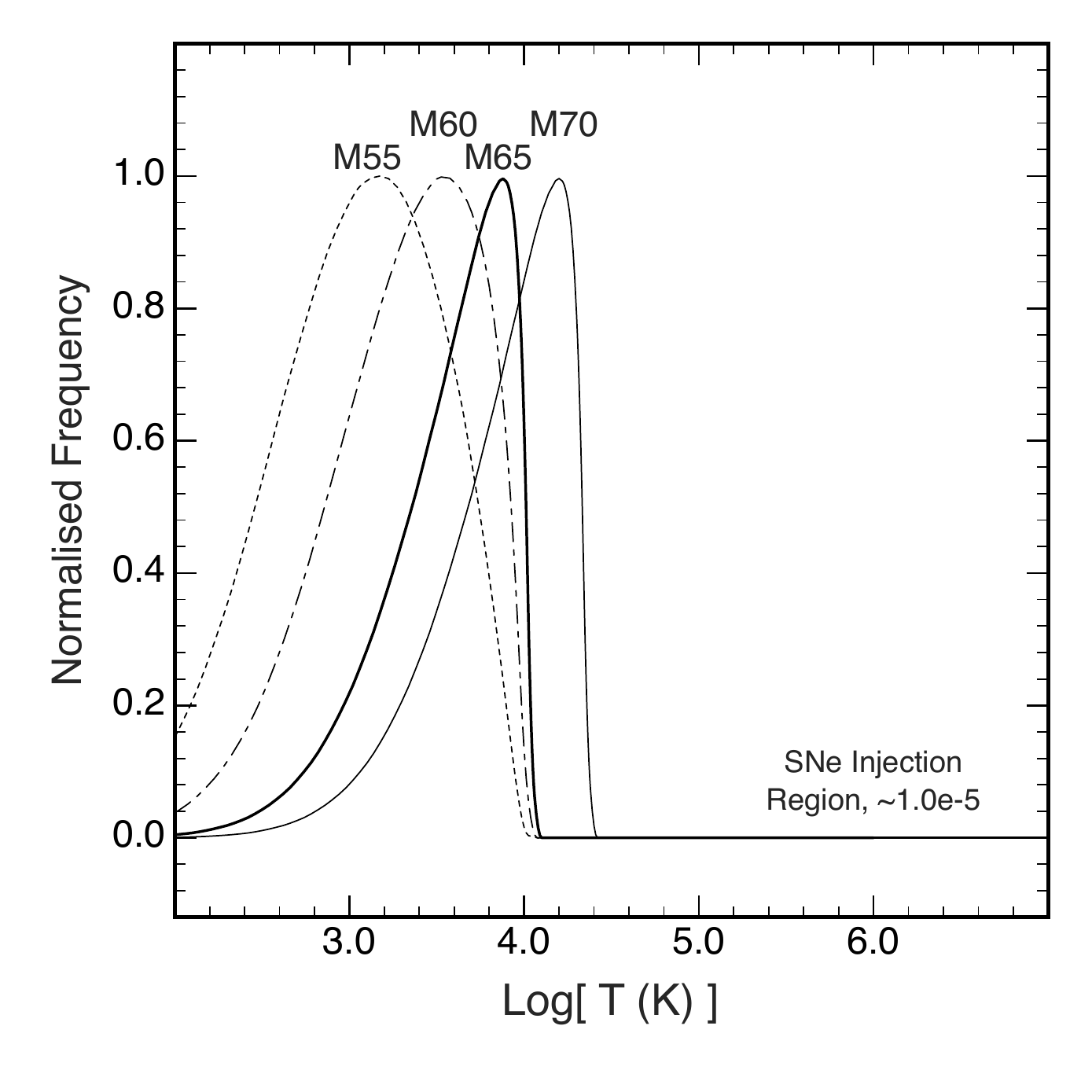}}
\caption{\label{f:tdist} 
The initial gas temperature distribution for the material inside one scale radius, $r_s$, for the models with fractal clumpy media.  Voids have higher temperatures and occupy a larger volume than the cooler, denser clouds.   
The material between $10^{4.4}$~K and $10^6$~K arises in the small spherical region that is evacuated prior to the supernova injection onto the grid.  There the local density is reduced to 0.01 of its original values.  The fractional volume this occupies remains small compared to the entire region inside the scale radius.  See text for details.
}
\end{figure}

\smallskip
\noindent{\sl Stellar winds.}
To avoid the complexities of rapid post-main sequence evolution, we assume that the star simply explodes as a supernova at the end of the main sequence.  \citet{ekstrom2006} showed that mass loss during the main sequence in similar stars at very low metallicity may only amount to 1\% of the initial mass, and \cite{kudritzki2002, kudritzki2005} showed that the stellar wind luminosities of low metallicity ([Fe/H]$\le -4.0$) O stars are typically $10^{34}$ erg s$^{-1}$ or less, several magnitudes below an equivalent solar metallicity star.  We combine these values to get a mean wind velocity, for a main sequence mass loss of 1\%:
\begin{equation}
v_w \sim  880 \left[\frac{L_w}{10^{34} {\rm erg\: s^{-1}}}\right]^{1/2}
\left[\frac{M}{25 {\rm M_\odot}}\right]^{-1/2}
\left[\frac{t_{\rm MS}}{6.1 {\rm Myr}}\right]^{-1/2} \; {\rm km\: s^{-1}}.
\end{equation}
This wind is included in the pre--supernova evolution, but the small mass flux and low ram pressure meant that it had negligible effect on the simulations.  Once the HII region pressurised the surrounding region, the stellar wind was unable to blow a wind bubble of any significant size.

The key parameter for these near zero metallicity O stars is their unusually high effective temperatures for a given mass.  At 25~$M_\odot$, we estimate the initial temperature to be 48,000~K , compared to a similar mass star at solar metallicity of only 40,000~K (see for example \cite{meynet2000} and more recently \citep{georgy2012}).  This means the star will be more effective as a source of ionising photons than the equivalent star in the solar neighbourhood. To convert the evolutionary tracks, $T_{\rm eff}(t)$ and $L(t)$, to an ionising photon luminosity (see Fig.~\ref{f:m25lum}), we used the ATLAS9 atmospheric grid \citep{castelli04} which uniquely contains a set of atmospheres for [Fe/H]$ = -4.0$  
with alpha element abundance enhancements, perfectly matching the initial gas composition in our
simulations.  The interpolated stellar temperature, luminosity and ionising fluxes are shown in 
Fig.~\ref{f:m25lum}.  The derived ionizing spectrum is shown in Fig.~\ref{f:m25spec};
for comparison, a photon flux assuming a blackbody instead of the stellar atmosphere is included.

\section{Simulations}

In the absence of considering the full initial value problem (i.e. infall, self gravity), we assume that 
sufficient gas has already settled onto a spherical dark-matter halo, and that it has dissipated 
energy and come to rough dynamical equilibrium.
Fundamentally, an isothermal hydrostatic equilibrium in a gravitational potential, $\Phi$, is
only possible with a uniform medium, with the requirement that the local
pressure gradients oppose the local potential gradient, that is:
\begin{equation} \frac{1}{\rho} \frac{dP}{dr} = -\frac{d\Phi}{dr}
\end{equation} 
where $P$ and $\rho$ are the gas pressure and density at a radius $r$.
As the potential is smooth, the pressure gradients must
be well behaved. When the density is smooth also, a solution for a given
temperature is possible, with the temperature defining the pressure
scale height.  However, when the density is no longer locally smooth,
corresponding local temperature variations are required to give the
well-behaved pressure gradients needed for hydrostatic
equilibrium. Without a single global temperature, a range of
scale heights exist over the whole domain, and a single structure scale
-- and hence equilibrium -- is not strictly possible. However, in practice it is close
to equilibrium, and the timescales for change are much longer than those
that describe the star formation in our models.

When gas falls into a gravitating potential well, it undergoes complex motions as it settles towards 
dynamical equilibrium. Some of the kinetic and potential energy is likely to be converted to thermal energy 
which is then lost through radiative cooling. If there were no losses, after adiabatically virialising in the halo 
potential, the gaseous medium would have an initial temperature, $T_{\rm init}$, related to the virial 
temperature of the halo such that $T_{\rm init} \sim T_{\rm vir}$.  For the adiabatic case,
$T_{\rm vir}$ $= \mu m_{\rm H} G M_{\rm vir}  / k r_{\rm vir}$ for which $k$ is Boltzmann's constant, 
$\mu$ is the mean molecular weight of the gas and $m_{\rm H}$ is the mass of the H atom.
This is before any star formation takes place.

We now introduce the common $\beta$ notation used in the presence of complex 
multiphase gases \citep{cavaliere76,arnaud09}.
An initial (quasi) hydrostatic gas distribution depends on an effective temperature 
$ T_{\rm eff} = T_{\rm gas} + T_{\rm turb} \sim T_{\rm init} $ arising from a combination of thermal and 
turbulent support (plus other terms due to magnetic fields which we neglect at present). With radiative losses, 
the {\em initial} thermal pressure is lost, and we are left with turbulent support only. Thus, if we assume initial 
equipartitition between thermal and turbulent pressure, $T_{\rm eff} = T_{\rm turb}
\approx 0.5 T_{\rm vir}$ such that $\beta = T_{\rm vir}/T_{\rm eff} = 2.0$.

For all of our models, we adopt a $\beta = 2.0$ gas distribution to recognize that the baryons must undergo some cooling before the first stars have formed. It is conceivable that $\beta$ is larger but such an assumption would concentrate the gas closer to the centre of the potential, thereby assisting gas retention.
Indeed, \cite{wise08b} find that baryonic matter is more concentrated than the dark matter in
their evolving halos (\mtot$=10^8$\msun). These authors use an adaptive mesh code to track the 
development of supersonic turbulence driven by accretion over many orders of magnitude in physical 
scale. In combination with line cooling, the supersonic turbulence leads to central densities that are 
several orders of magnitude higher than what we can achieve with {\it Fyris}. 

While we have the option to modify $\beta$ to increase the central gas densities, this would only assist
our ability to retain gas after the supernova.
Our adoption of a lower constant $\beta$ value has the benefit of reducing any central concentration differences in the models, leaving just the global gravity potential as the key mass model difference.
The resulting temperature distributions for the M55-M70 models are discussed below.

\smallskip
\noindent{\sl Initial state of the gas.}

We adopt a mean equilibrium distribution for the gas density given by $\rho =
\rho(0) \exp(\Phi)^\beta$ for which $\Phi$ describes the dark halo potential (Fig.~\ref{f:pots},
right panel). This is described in detail in Appendix B.
The use of $\beta$ allows for the gas scale length to be different from $r_{\rm vir}$
(Fig.~\ref{f:pots}, left panel).
For completeness, note that the dynamical dispersion
of the dark halo is $\sigma = \beta c_s$. For a given temperature and
composition/ionization state, $P \propto \rho$. So for homogeneous models, there is a
uniform temperature everywhere such that the local density $\rho$ is equal to the mean
density ($\rho/\langle \rho \rangle = 1.0$). For fractal models, $\rho/\langle \rho
\rangle = f(x,y,z)$ for which $f$ is the normalised fractal, giving a distribution $T
\propto P/\rho$ that is the same for all the models using the same fractal modulation. The
distribution $f$ has a mean value of unity and a variance of 5 (see Appendix B).

The projected 2D gas distribution (density, pressure,
temperature) for the inhomogeneous M60 model is shown in Fig.~\ref{f:setup}.
The temperature histogram for Fig.~\ref{f:setup}D, and for all models, is given in 
Fig.~\ref{f:tdist}.  Across the models, the global mean gas temperature spans the range 
500-2000~K, with a small gas fraction extending below a few hundred Kelvin, 
compatible with a star forming region. The bulk of the ISM tops out at about $10^4$~K
and thus covers the range of temperatures 
seen in typical warm and cold neutral ISM media \citep{sternberg02}.

To verify the equilibrium of the initial conditions, we ran adiabatic test models with smooth gas in hydrostatic equilibrium, and the initial distribution remained steady for the $\gtrsim$200 Myr, much longer than the 25 Myr timescale of our supernova model.  An adiabatic clumpy test model could not achieve perfect equilibrium, and evolved towards the smooth solution, smearing out the clumps on timescales of
\begin{equation}
\tau_{\rm cross} \sim r_s / c_s \sim 100\; \mbox{\rm pc} / 1\ \mbox{\rm km\ s$^{-1}$} \sim 100 \; \mbox{\rm Myr}\, ,
\end{equation}
where $r_s$ is the scale radius of the Einasto halo;
$c_s$ is the thermal sound speed for a gas temperature of $\sim 10^3$ K. 
This remains a significantly longer timescale than the supernova
evolution timescale, so we are justified in assuming equilibrium during the 
main evolution models, and a differential experiment where we can compare directly
fractal and smooth models that have the same mean properties averaged over the distribution is valid.

\smallskip
\noindent{\sl Energy injection.}
To model the energy injection from supernovae, we insert a bubble of hot gas with the
equivalent energy of 10$^{51}$ erg, on the assumption that the supernova energy has been
thermalized on a scale much below our parsec-scale resolution. This high pressure bubble
expands and converts internal thermal energy to kinetic energy and radiative losses during
the expansion.

For the smooth models with cooling, it is important to provide additional smoothing to the
boundary of the injection region to avoid the grid cell errors that seed the well known
``carbuncle instability'' (cf. Sutherland 2010). This is especially strong in the case of
smooth radiative models because of the formation of radiative thin shells which are
particularly susceptible to this. We reduce the impact of these effects by using an
injection region with a 3-cell radius smoothed with a gaussian kernel (2 cell FWHM).

We locate our injection region either at the centre of the potential or off-centred at the
half gas-mass radius ($r_{0.5\rm g}$) listed in Table~\ref{t:halos} found from integrating
the mean density curves in Fig.~\ref{f:pots}. In Fig.~\ref{f:setup}, the injection region
is shown for the M60 off-centred clumpy model.

We follow the evolution of the SN for $T_0$ $=$ 25 Myr taking snapshots at 25 kyr
intervals. This timescale is now well established in starburst and Carina-like dwarf galaxies
\citep{sharp10,tolstoy09}. We analyse our results by summing the dynamically cold and total mass,
and internal kinetic energy, within one scale radius of the central potential at each time
step.

%%%%%%%%%%%%%%%%%%%%%%%%%%%%%%%%%%%%%%%%%%%%

\begin{table*}[htdp]
\caption{\label{t:naming} Naming convention for all numerical simulations}
%\footnotesize{
%\caption[Naming convention for all numerical simulations]. 
%}
\begin{center}
\begin{tabular}{r r r r r r r r r}
\multicolumn{9}{c}{{\bf Models without pre-ionisation}}\\ \\
\hline
 &\multicolumn{4}{c}{{\bf Central Explosions}}             
 &\multicolumn{4}{c}{{\bf Off-Centre Explosions}}          \\
\multicolumn{1}{l}{} 
&\multicolumn{2}{c}{Smooth}&\multicolumn{2}{c}{Clumpy}
&\multicolumn{2}{c}{Smooth}&\multicolumn{2}{c}{Clumpy}\\ 
\multicolumn{1}{c}{{\bf Halo Mass}} 
&\multicolumn{1}{c}{Adiabatic} &\multicolumn{1}{c}{Cooling }      
&\multicolumn{1}{c}{Adiabatic} &\multicolumn{1}{c}{Cooling }      
&\multicolumn{1}{c}{Adiabatic} &\multicolumn{1}{c}{Cooling }      
&\multicolumn{1}{c}{Adiabatic} &\multicolumn{1}{c}{Cooling }\\
\hline
\multicolumn{1}{r}{{\bf $10^{5.5}$~M$_\odot$}}&M55CSA&M55CSC&M55CCA&M55CCC&M55OSA&M55OSC&M55OCA&M55OCC\\
\multicolumn{1}{r}{{\bf $10^6$~M$_\odot$}}&M60CSA&M60CSC&M60CCA&M60CCC&M60OSA&M60OSC&M60OCA&M60OCC\\
\multicolumn{1}{r}{{\bf $10^{6.5}$~M$_\odot$}}&M65CSA&M65CSC&M65CCA&M65CCC&M65OSA&M65OSC&M65OCA&M65OCC\\
\multicolumn{1}{r}{{\bf $10^7$~M$_\odot$}}\\
\hline
\hline \\

\multicolumn{9}{c}{{\bf Models with pre-ionisation}}\\ \\
\hline
 &\multicolumn{4}{c}{{\bf Central Explosions}}             
 &\multicolumn{4}{c}{{\bf Off-Centre Explosions}}          \\
\multicolumn{1}{l}{} 
&\multicolumn{2}{c}{Smooth}&\multicolumn{2}{c}{Clumpy}
&\multicolumn{2}{c}{Smooth}&\multicolumn{2}{c}{Clumpy}\\ 
\multicolumn{1}{c}{{\bf Halo Mass}} 
&\multicolumn{1}{c}{Adiabatic} &\multicolumn{1}{c}{Cooling }      
&\multicolumn{1}{c}{Adiabatic} &\multicolumn{1}{c}{Cooling }      
&\multicolumn{1}{c}{Adiabatic} &\multicolumn{1}{c}{Cooling }      
&\multicolumn{1}{c}{Adiabatic} &\multicolumn{1}{c}{Cooling }\\
\hline
\multicolumn{1}{r}{{\bf $10^{5.5}$~M$_\odot$}}&\\
\multicolumn{1}{r}{{\bf $10^6$~M$_\odot$}}&&&&M60CCH&&&&M60OCH\\
\multicolumn{1}{r}{{\bf $10^{6.5}$~M$_\odot$}}&&&&M65CCH&&&&M65OCH\\
\multicolumn{1}{r}{{\bf $10^7$~M$_\odot$}}&&&&M70CCH&&&&M70OCH\\
\hline
\hline

\end{tabular}
\end{center}
\label{t:halos}

\end{table*}%

{\small
\begin{table*}[htdp]
\caption{\label{t:effic} Thermalisation efficiency in the M60 and M65 models}
\begin{center}
\begin{tabular}{r r r r r r r r r}
\hline
 &\multicolumn{4}{c}{{\bf Central Explosions}}             
 &\multicolumn{4}{c}{{\bf Off-Centre Explosions}}          \\
\multicolumn{1}{l}{{\bf Model}} 
&\multicolumn{2}{c}{Smooth}&\multicolumn{2}{c}{Clumpy}
&\multicolumn{2}{c}{Smooth}&\multicolumn{2}{c}{Clumpy}\\ 
\multicolumn{1}{r}{Time} 
&\multicolumn{1}{c}{Adiabatic} &\multicolumn{1}{c}{Cooling }      
&\multicolumn{1}{c}{Adiabatic} &\multicolumn{1}{c}{Cooling }      
&\multicolumn{1}{c}{Adiabatic} &\multicolumn{1}{c}{Cooling }      
&\multicolumn{1}{c}{Adiabatic} &\multicolumn{1}{c}{Cooling }\\
\hline
\hline
M65&100\%&8.1\%&100\%&11\%&100\%&9.8\%&100\%&11\%\\
M60&100\%&11\%&100\%&15\%&100\%&11\%&100\%&16\%\\           
\hline
\end{tabular}
\end{center}
\label{default}
\end{table*}%
}%small

{\small
\begin{table*}[htdp]
\caption{\label{t:massretainhii}Percentage of initial mass retained inside $r_s$ and $r_{\rm vir}$
for cooling models.
The time at 0 Myr is when the SN event takes place
at which point the pre-ionization phase ends. Entries in bold fall below 50\%.}
\begin{center}
\begin{tabular}{r r r r r}
\hline
\multicolumn{1}{l}{{\bf Model}}  
&\multicolumn{2}{c}{{\bf Inside $r_s$}}  &\multicolumn{2}{c}{{\bf Inside $r_{\rm vir}$}}          \\
\multicolumn{1}{r}{Time$^\dag$} 
&\multicolumn{1}{c}{Central}&\multicolumn{1}{c}{Off-Center}
&\multicolumn{1}{c}{Central}&\multicolumn{1}{c}{Off-Center}\\ 
\hline
\hline
\\
\multicolumn{1}{l}{{\bf M60H}}&\multicolumn{2}{c}{$M_{\rm g, s}$: 1.91E+04 M$_\odot$ } &\multicolumn{2}{c}{ $M_{\rm g, vir}$: 1.22E+05 M$_\odot$ } \\
{\bf 0\,Myr }& {\bf 12.7}&{\bf  16.1}&  97.6&  97.1\\
{\bf 5\,Myr }& {\bf 0.03}&  {\bf 0.06}& 89.3& 88.4\\
{\bf 25\,Myr }& {\bf  0.01}& {\bf 0.21}& {\bf    0.98 }& {\bf 6.10} \\
\\
\multicolumn{1}{l}{{\bf M65H}}&\multicolumn{2}{c}{$M_{\rm g, s}$: 6.61E+04\,M$_\odot$ } &\multicolumn{2}{c}{ $M_{\rm g, vir}$: 3.74E+05\,M$_\odot$ } \\
{\bf 0\,Myr }& 50.3& 78.8&  100.0&  99.4\\
{\bf 5\,Myr }& {\bf 2.42}& {\bf 41.7}&  99.7&  97.9\\
{\bf 25\,Myr }&{\bf   13.8}& {\bf 33.9}& 81.5&  73.3\\
\\
\multicolumn{1}{l}{{\bf M70H}}&\multicolumn{2}{c}{$M_{\rm g, s}$: 2.38E+05\,M$_\odot$ } &\multicolumn{2}{c}{ $M_{\rm g, vir}$: 1.16E+06\,M$_\odot$ } \\
{\bf 0\,Myr }& 102.6$^*$&98.2& 100.0& 100.0\\
{\bf 5\,Myr }&  106.2$^*$& 92.8&  100.2$^*$&  100.1$^*$\\
{\bf 25\,Myr }&  121.5$^*$& 107.0$^*$& 102.6$^*$& 101.2$^*$\\
\hline
\multicolumn{5}{l}{\footnotesize {\bf Bold: $ < 50\%$ retained}} \\
\multicolumn{5}{l}{\footnotesize $^\dag$:  0 Myr $ = t_{\rm SNe}$, $= 6.1$ Myr from start of simulation}\\
\multicolumn{5}{l}{\footnotesize $^*$: Increased Mass after some infall }\\
\end{tabular}
\end{center}
\label{t:mass_retention}
\end{table*}
}

{\small
\begin{table*}[htdp]
\caption{\label{t:energy_retention}
Percentage of initial energy retained inside $r_s$ and $r_{\rm vir}$ for cooling models.
The time at 0 Myr is when the SN event takes place
at which point the pre-ionization phase ends. Entries in bold fall below 50\%.}
\begin{center}
\begin{tabular}{r r r r r}
\hline
\multicolumn{1}{l}{{\bf Model}}  
&\multicolumn{2}{c}{{\bf Inside $r_s$}}  &\multicolumn{2}{c}{{\bf Inside $r_{\rm vir}$}}          \\
\multicolumn{1}{r}{Time$^\dag$} 
&\multicolumn{1}{c}{Central}&\multicolumn{1}{c}{Off-Center}
&\multicolumn{1}{c}{Central}&\multicolumn{1}{c}{Off-Center}\\ 
\hline
\hline
\\
{\bf M60H} &\multicolumn{2}{c}{$E_s$: 4.52E+48 ergs}&\multicolumn{2}{c}{ $E_{\rm vir}$:  2.75E+49 ergs}\\
{\bf Pre--SNe}  & 237.7        & 299.1      &  1725     & 1647\\
{\bf Post--SNe} & 1.74E+04 & 1.79E+04 &  4365     & 4374\\
{\bf 5\,Myr}    &{\bf 1.90}   &{\bf 1.00}   & 1242      & 1179\\
{\bf 25\,Myr}  &{\bf 0.06}   &{\bf 0.31}   & {\bf 1.93}&{\bf 9.32}\\
\\
{\bf M65H}&\multicolumn{2}{c}{$E_s$: 3.19E+49 ergs}&\multicolumn{2}{c}{$E_{\rm vir}$: 1.73E+50 ergs}\\
{\bf Pre--SNe}  & 327.4            & 318.1     & 325.3 & 402.4\\
{\bf Post--SNe} & 2613             & 2596      & 722.1 & 797.3\\
{\bf 5\,Myr}    &{\bf 2.71}        & 50.7       & 232.7 & 255.4\\
{\bf 25\,Myr}  &{\bf 20.1}$^*$ &{\bf 29.6}& 101.4 & 91.5 \\
\\
{\bf M70H} &\multicolumn{2}{c}{$E_s$: 3.45E+50 ergs}&\multicolumn{2}{c}{$E_{\rm vir}$: 1.63E+51 ergs}\\
{\bf Pre--SNe}  & 92.2      & 89.8       & 59.4        & 81.8 \\ 
{\bf Post--SNe} & 328.5    & 319.2     &107.2       & 126.6\\
{\bf 5\,Myr}    & 64.6       & 57.6       & 54.1        & 64.3 \\
{\bf 25\,Myr}  & 84.9$^*$& 67.8$^*$& 61.8$^*$& 62.4 \\
\hline
\multicolumn{5}{l}{\footnotesize {\bf Bold: $ < 50\%$ retained}} \\
\multicolumn{5}{l}{\footnotesize $^\dag$:  0 Myr $ = t_{\rm SNe}$, $= 6.1$ Myr from start of simulation}\\
\multicolumn{5}{l}{\footnotesize Pre--SNe 50kyr prior to SNe, Post--SNe, 10kyr after SNe}\\
\multicolumn{5}{l}{\footnotesize $^*$: Increased Mass after some infall }\\
\end{tabular}
\end{center}
\label{default}
\end{table*}
}

\section{Results}

\subsection{Pre-ionization phase}

Our goal is to establish the low-mass limit of a spherical dark-matter halo that can retain 
baryons after a single supernova event. But first we consider the impact of the progenitor 
star on the surrounding gas prior to the supernova explosion.
The mechanical wind energy is completely dwarfed by the star's radiative phase for
any progenitor of a supernova. 

We consider how the UV propagates in each of the four halos for both clumpy and smooth 
media which undergo cooling. This is a complex process and so we include access to 3D
movies that reveal how the ionization front develops. These can be found here:
http://miocene.anu.edu.au/smallgalaxy. We show a time sequence of the full
pre-ionization phase and the subsequent SN explosive phase in two ways: (i) fixed reference
frame; (ii) rotating reference frame. There are two other aids to visualization. First, we include
half-tone figures that show critical time steps for the different models: M60 
(Figs.~\ref{f:M60CCcol}, \ref{f:M60OCcol}), M65 (Figs. \ref{f:M65CCcol}, \ref{f:M65OCcol}), and 
M70 (Figs. \ref{f:M70CCcol}, \ref{f:M70OCcol}). 
Secondly, we include colour composite figures in Fig.~\ref{f:colour}
that directly compare the information embodied in Figs.~\ref{f:M60CCcol} to \ref{f:M70OCcol},
i.e. the neutral and ionised components as the nebula develops before and
after the SN event.

A summary of gas mass and energy retention after the pre-ionization phase 
is given in Table~\ref{t:mass_retention} and Table~\ref{t:energy_retention} respectively. 
The initial gas masses inside two cardinal radii, 
\rs\ and \rvir, are included for each model. 
The M60, M65 and M70 models retain essentially all of the
gas within \rvir\ for the cooling models, while the M55 case (not shown) is evacuated
completely. For the adiabatic cases (not shown), the M60 halo is evacuated completely
and the M65 halo appears to lose more than half the gas after 30 Myr. In this instance,
as the gas reacts to the radiation field, it becomes photoheated, expands, smooths out and 
drops in density as a result. The number of recombinations locally declines and the radiation propagates into more of the surrounding gas.

In Figs.~\ref{f:M60CCcol} to
\ref{f:M70OCcol}, we observe a number of important facts about the pre-ionization phase.
For the centred models, the Str\"{o}mgren radius of the HII region falls outside (M60), onto (M65)
and inside (M70) of the scale radius. From the zoomed figures, we observe how the
photoheating smooths out the substructure. The effect is so strong that we observe a
cellular structure as the ionization fronts of many expanding knots overlap, which
is particularly clear in the colour composite figure (Fig.~\ref{f:colour}).  In the same figure,
we see broad channels 
appear where the radiation is able to escape to the outer halo over a substantial solid angle. 
Even before the SN event, the expanding Str\"{o}mgren sphere has produced a diffuse 
{\it neutral} shell of gas around the hot bubble cavity. In the off-centred models, the 
expanding Str\"{o}mgren sphere compresses the core
region gas and the radiation is blocked from propagating over at least $2\pi$ sr. This
becomes important in the post-supernova phase because substantial gas begins to 
re-accrete to the centre along this axis.

\subsection{Supernova phase}

Our complete set of simulations (see Table~\ref{t:naming}) comprises 24 cases 
for supernova events without pre-ionization with the massive progenitor, and 6
cases where the ionization of the supernova progenitor is included. All 30 cases
were run at L00 and L01, with a subset of the cooling models run at L02 to allow 
adequate sampling of the developing supernova shell (see Fig.~\ref{f:grid}). L02
is always found to be adequate to resolve the metal-enriched shell which starts
as a diffuse hot envelope which becomes thinner under expansion. The models do
not suffer the overcooling through inadequate resolution identified by \citet{dallavecchia12}.

We discuss in detail only the clumpy, cooling models for the centred
and off-centred supernova cases where the progenitor ionization
is included. We present these results for three different halo masses 
(M60, M65, M70)\footnote{All simulations can be viewed at http://miocene.anu.edu.au/smallgalaxy$\;.$}.
Tables~\ref{t:mass_retention} and \ref{t:energy_retention}
include the total gas mass and energy retained within $r$ = \rs\ and $r$ = \rvir\ after periods of 
$t = 5$~Myr and $t = 25$~Myr. 
The halos that retain less than 50\% of their original mass or energy have their retention fractions 
emboldened.

\begin{figure*}
 \includegraphics[scale=0.55]{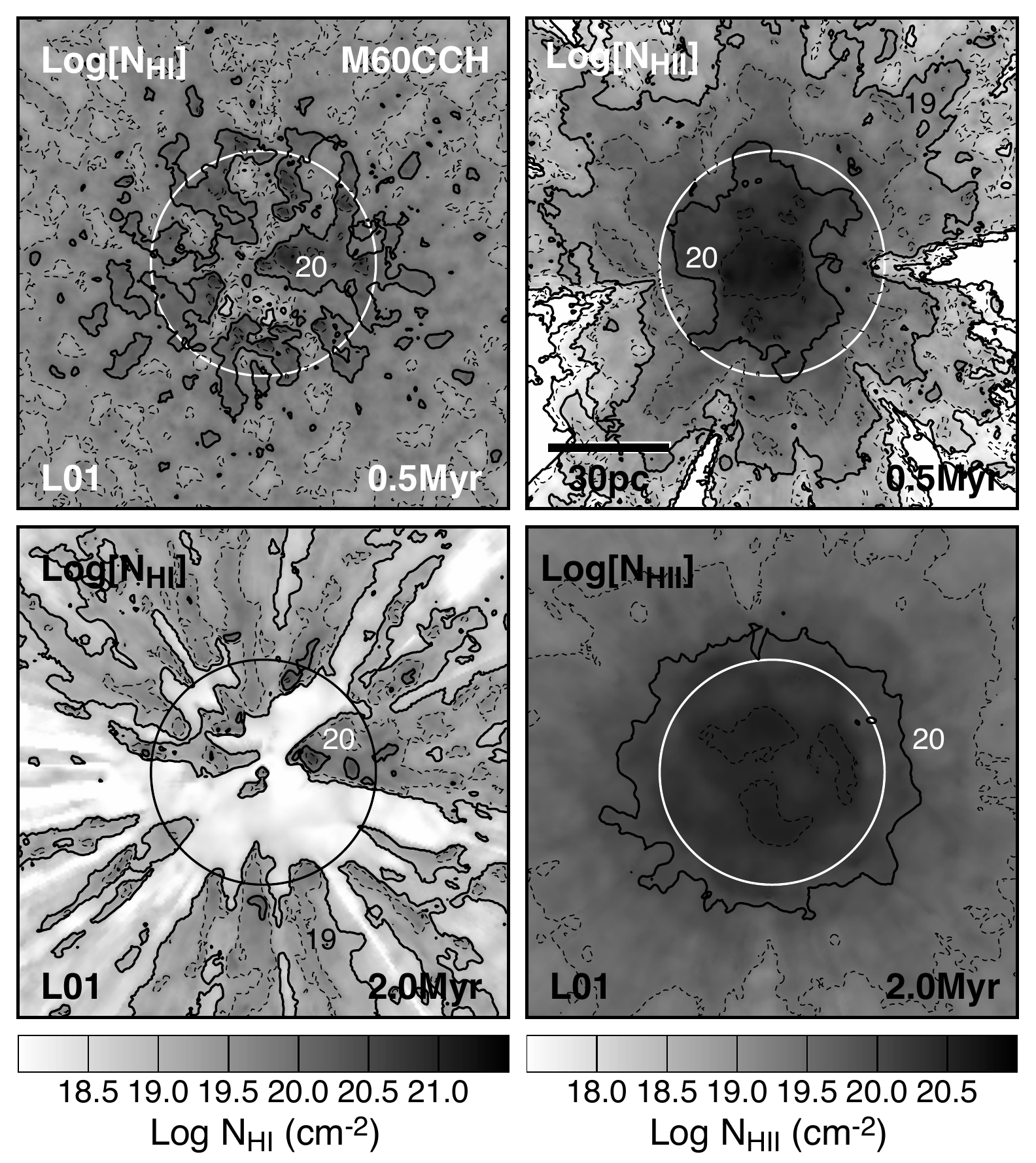}
 \includegraphics[scale=0.55]{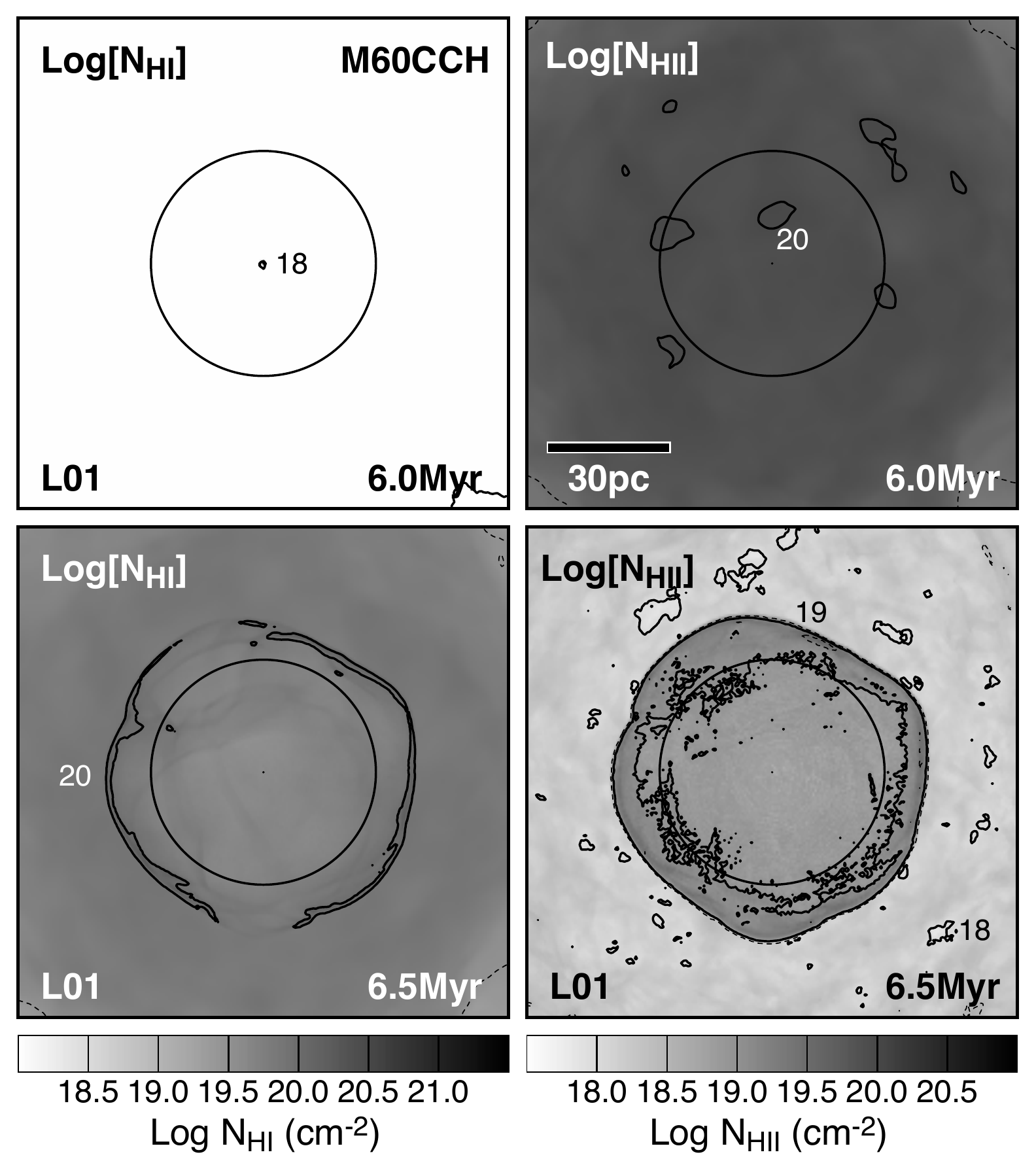}
\caption{ \label{f:M60CCcol}
M60 photoionization (centred source) during the pre-supernova and supernova phases 
at four different epochs (0.5, 2.0, 6.0, 6.5 Myr). The left and right figures in each panel show 
the projected neutral and ionised gas column respectively. The supernova ignites at 6.0 Myr.
The detailed geometry seen here shows how radiation breaks out from the 
inner core by 0.5 Myr along certain directions although the timescale is longer (2.5 Myr) overall.  
The density variations make the ionization fronts complex due to shadowing by the
densest gas. However the dense gas is too sparse to present a significant barrier.
Immediately prior to the SN explosion, the HII region is fully ionised, heated and highly 
smoothed the gas, which improves the SNe - ISM coupling,
further increasing the ability of the SNe to clear the halo.  
The warm gas is less bound than initially as well, and by 6.5 Myr the SNR
has cleared the core region completely.
}
\end{figure*}

\begin{figure*}
 \includegraphics[scale=0.55]{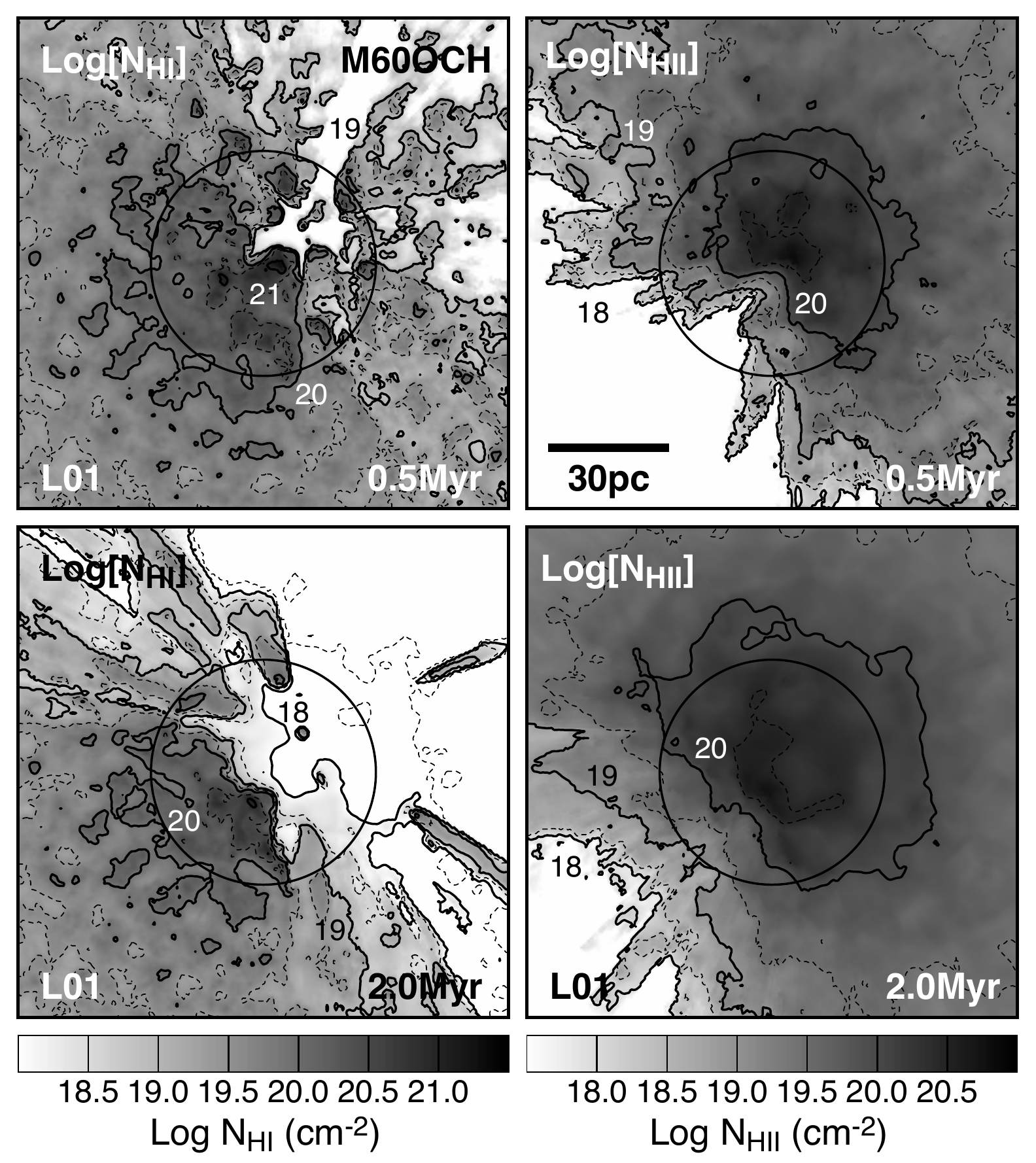}
 \includegraphics[scale=0.55]{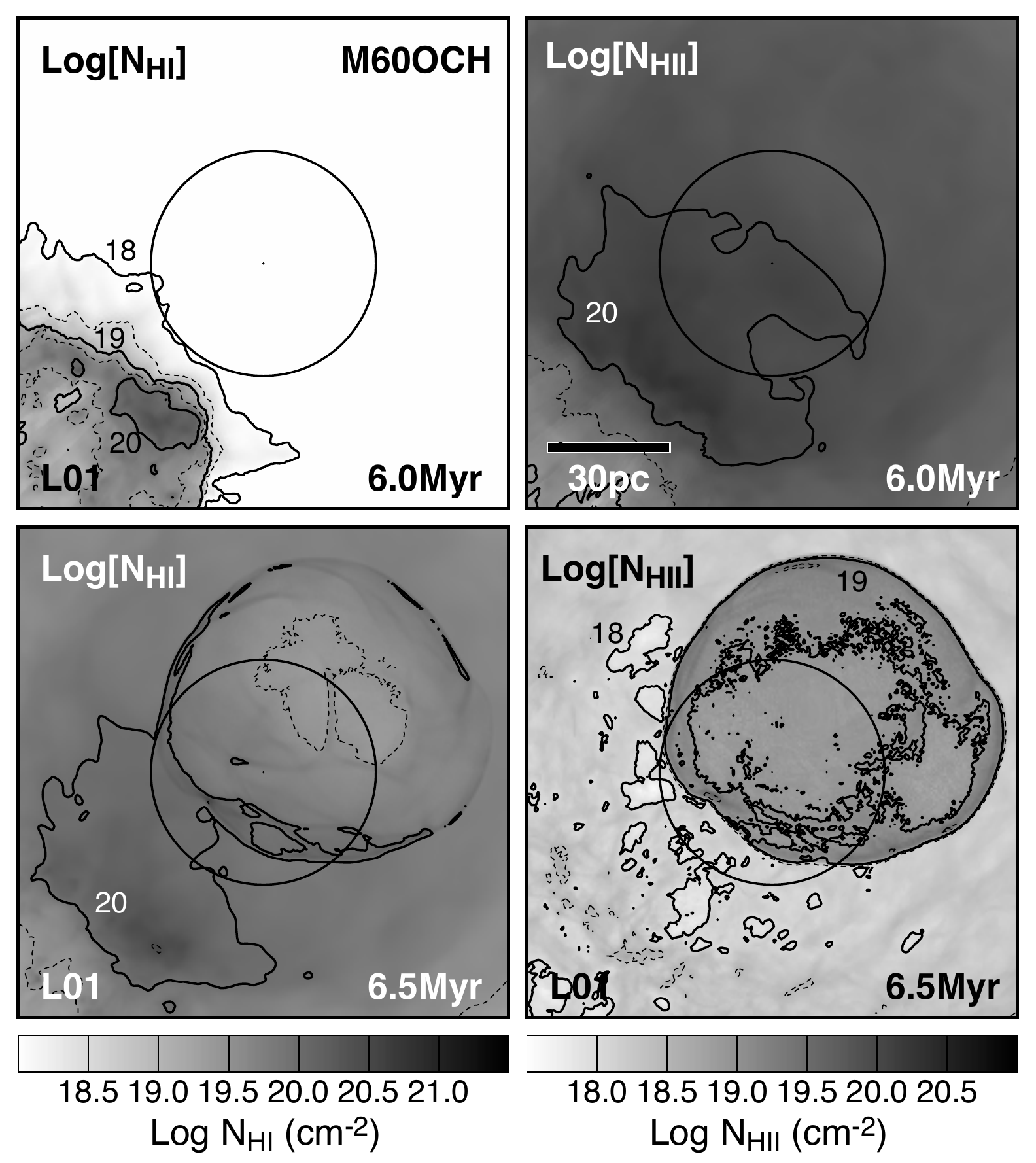}
\caption{ \label{f:M60OCcol} 
M60 photoionization (off-centred source) during the pre-supernova and supernova phases at four different epochs
(see Fig.~\ref{f:M60CCcol}). The supernova ignites at 6.0 Myr. The escape of photons is highly asymmetric and ioization is complete (except on the diametrically opposite side of the core)
by 2 Myr.  By 6 Myr, the core is heated, ionised, and smoothed, so the SN is quite efficient in 
clearing the halo.
}
\end{figure*}

\begin{figure*}
 \includegraphics[scale=0.55]{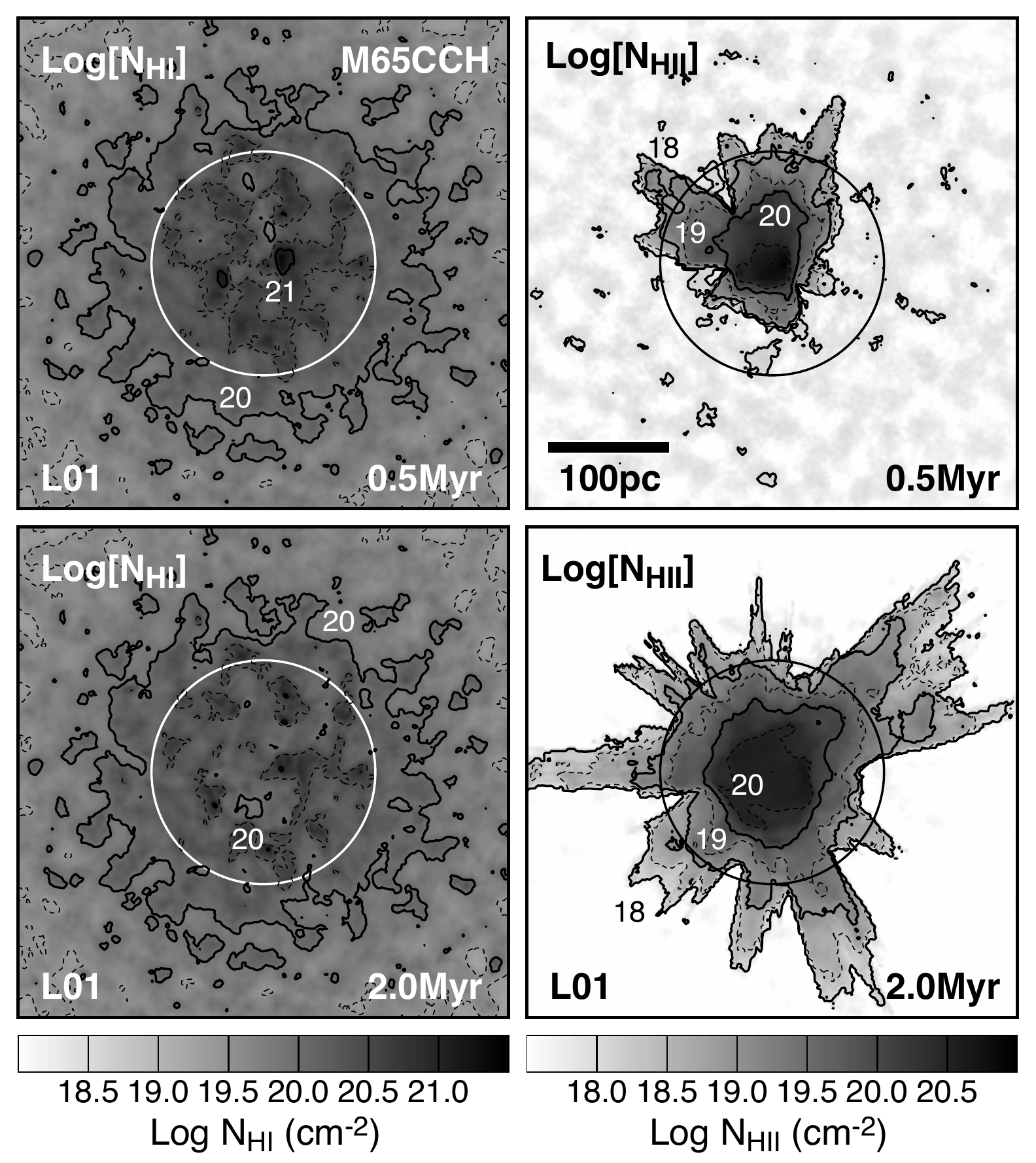}
 \includegraphics[scale=0.55]{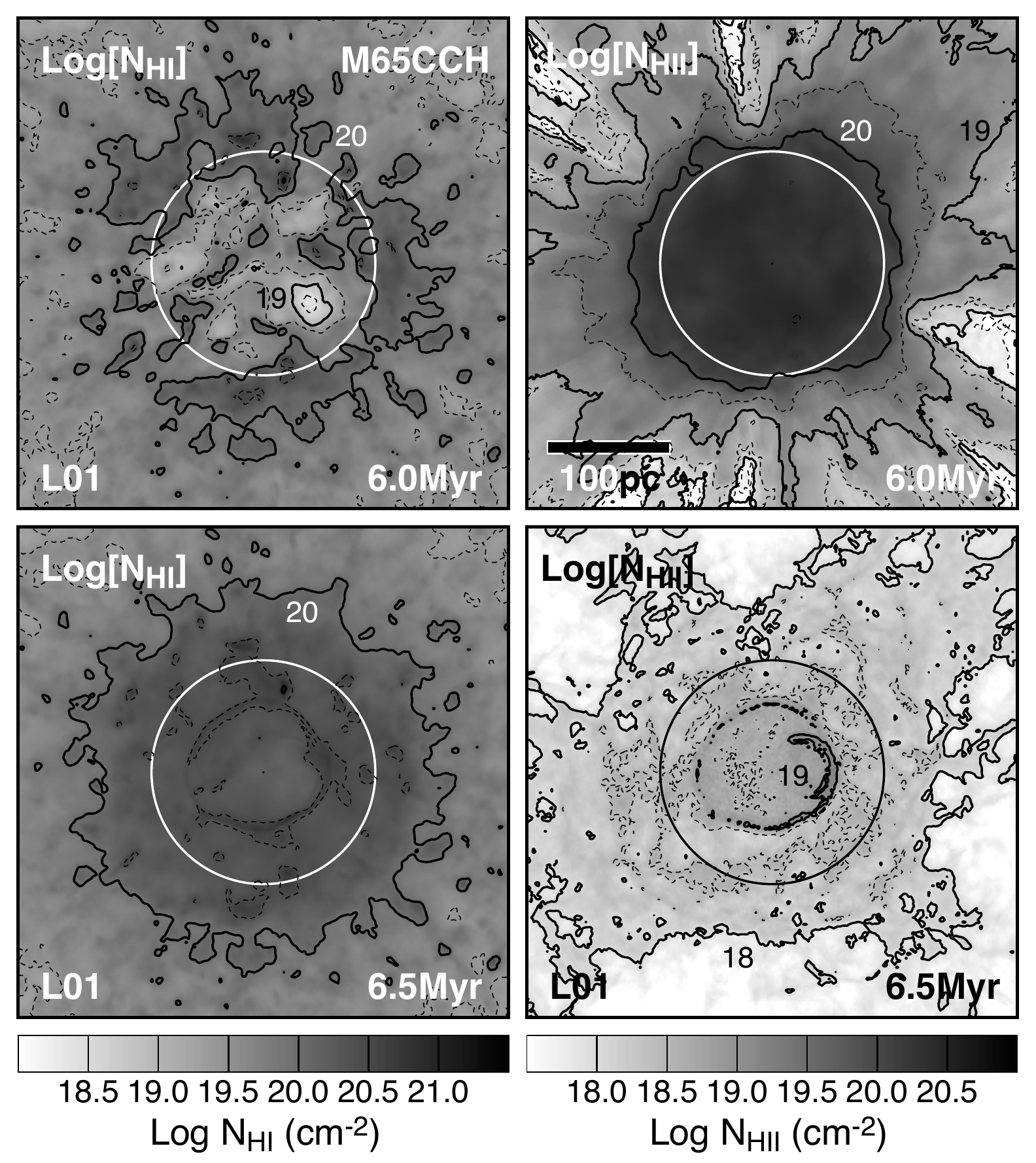}
\caption{ \label{f:M65CCcol}
M65 photoionization (centred source) during the pre-supernova and supernova phases at four different epochs
(see Fig.~\ref{f:M60CCcol}). The supernova ignites at 6.0 Myr.
The non-uniform gas presents low density channels, so that by 2.0 Myr, the spiky ionization fronts have escaped the core in the central model.
By 6.0 Myr, the entire core region is photoionised.
}
\end{figure*}

\begin{figure*}
 \includegraphics[scale=0.55]{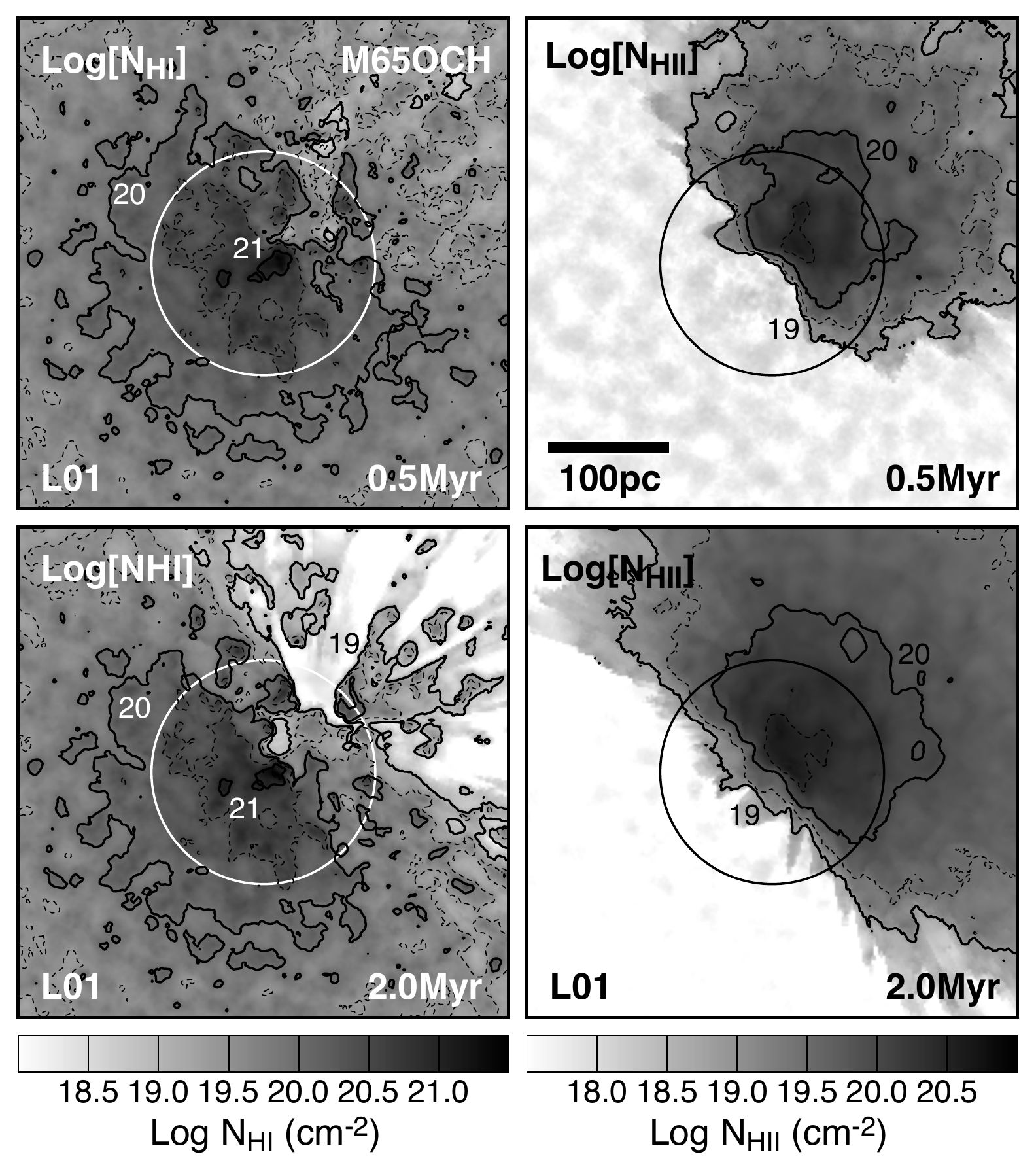}
 \includegraphics[scale=0.55]{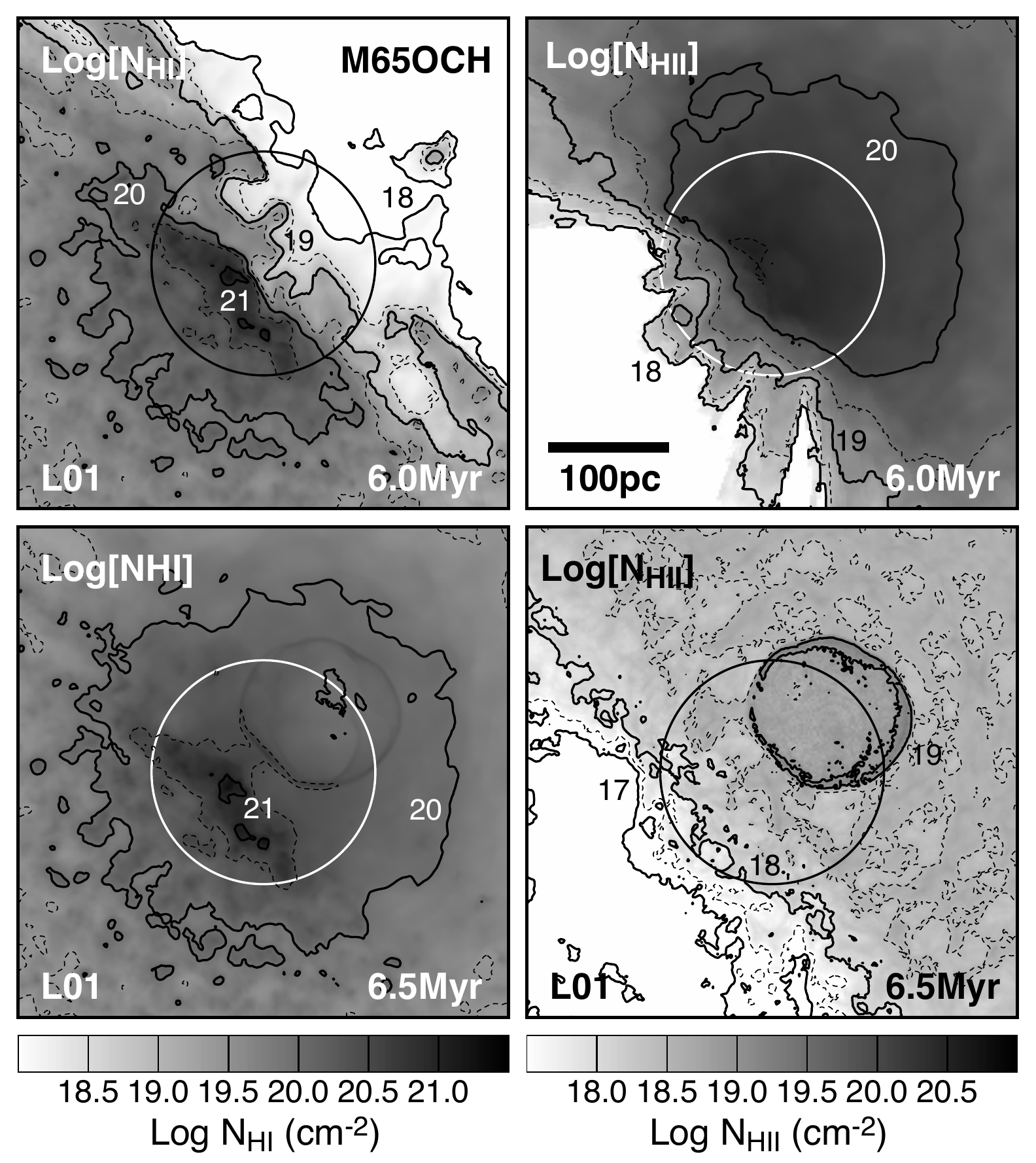}
\caption{ \label{f:M65OCcol}
M65 photoionization (off-centred source) during the pre-supernova phase at four different epochs
(see Fig.~\ref{f:M60CCcol}). The supernova ignites at 6.0 Myr.
The off-centred HII region runs out down the halo density gradient rapidly on the side near the star, but struggles to ionise through the
core to the opposite side.  The geometry allows the ionization to affect a larger volume
in the outer halo than indicated by the integrated analysis in 
Fig.~\ref{f:recphot}.  The SN occurs again in a smoothed region and reaches a significant fraction of
the core radius by 6.5 Myr.
}
\end{figure*}

\begin{figure*}
 \includegraphics[scale=0.55]{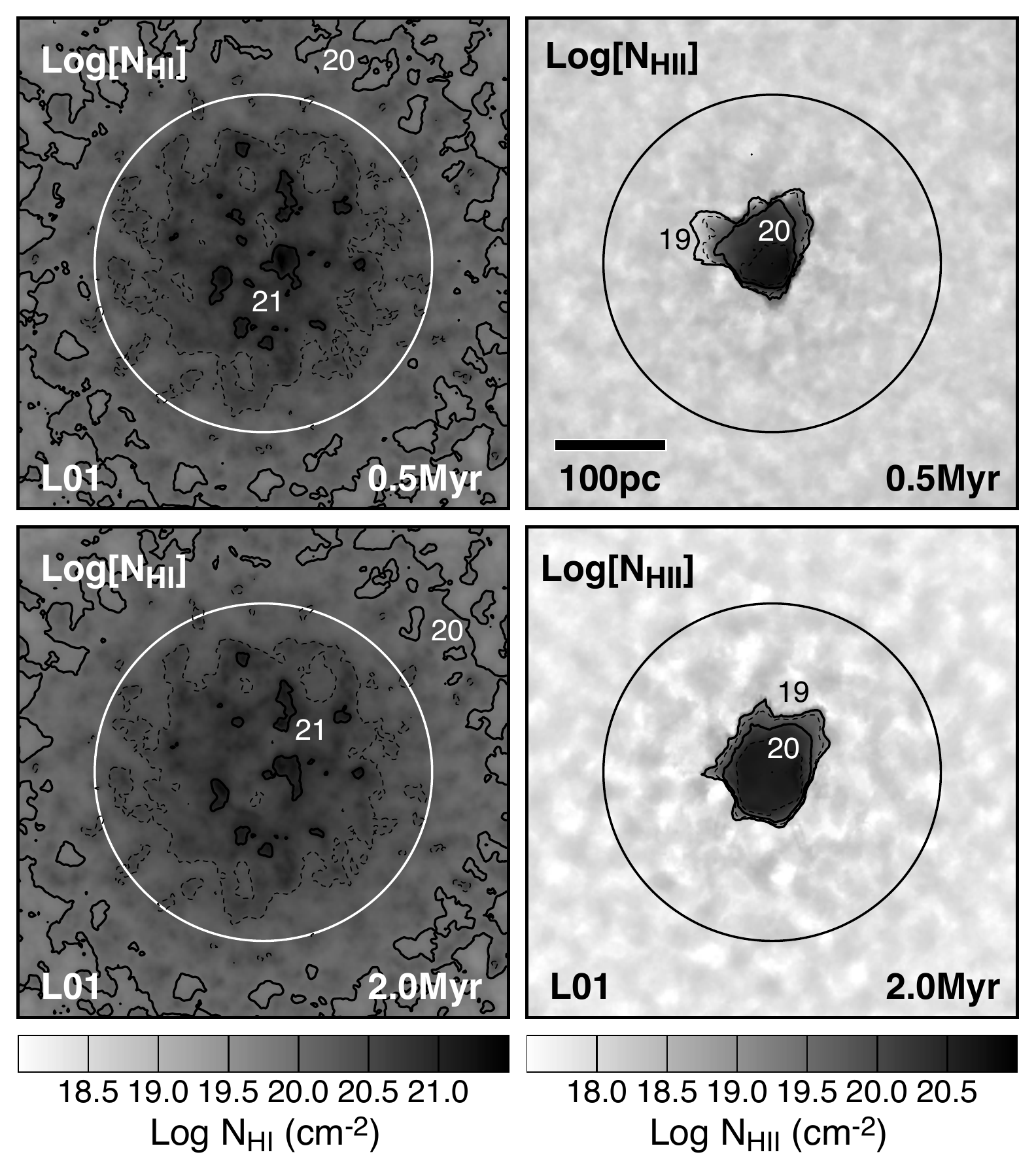}
 \includegraphics[scale=0.55]{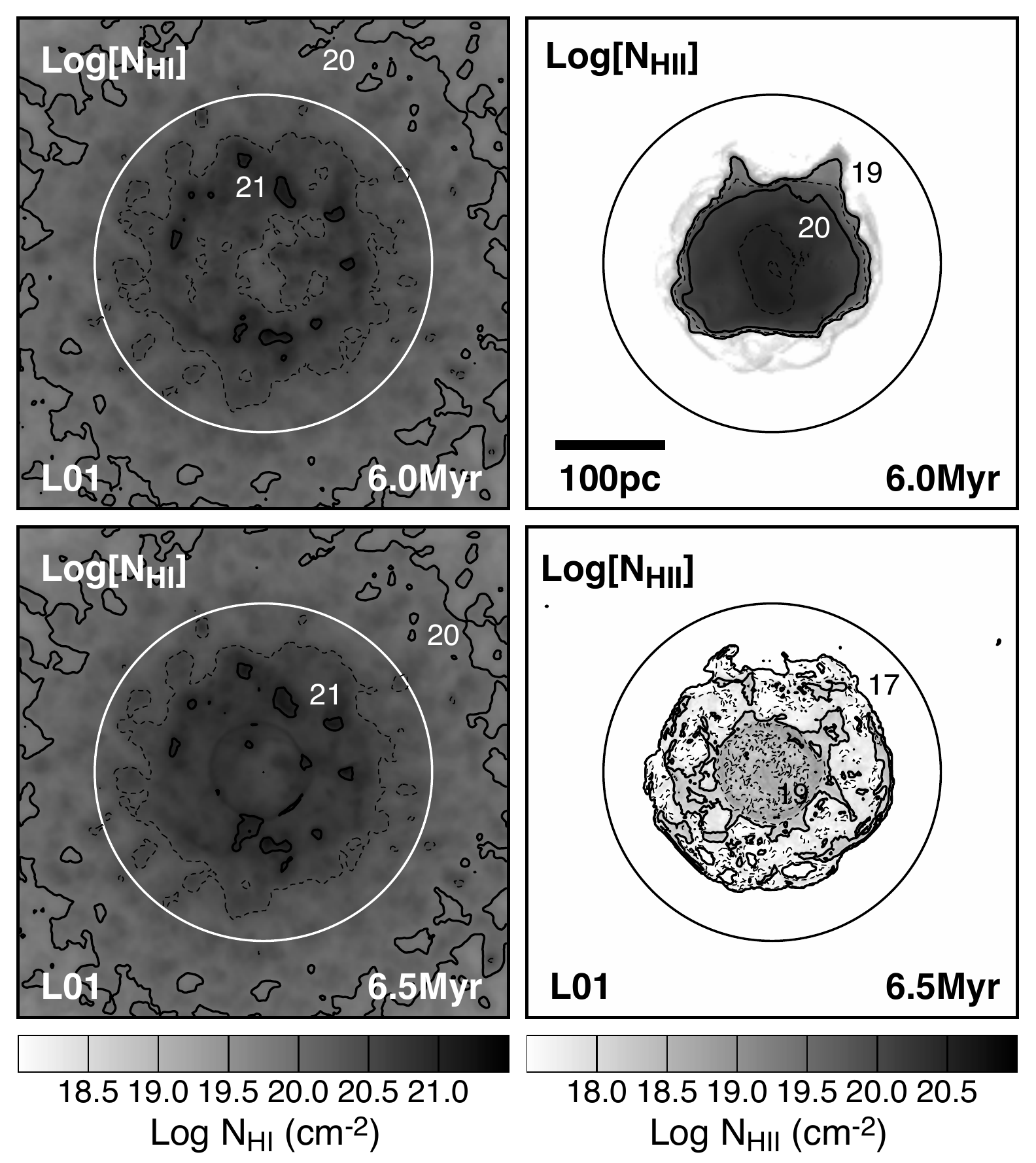}
\caption{ \label{f:M70CCcol}
M70 photoionization (centred source) during the pre-supernova and supernova phases at four different epochs
(see Fig.~\ref{f:M60CCcol}). The supernova ignites at 6.0 Myr.
The photon flux is insufficient to ionise a significant part of the core volume, and the HII
region is stifled.  The SN occurs in a similar environment to the non-preionised models and the outcomes are similar. Essentially all of the gas survives the impact of the SN event.
}
\end{figure*}

\begin{figure*}
 \includegraphics[scale=0.55]{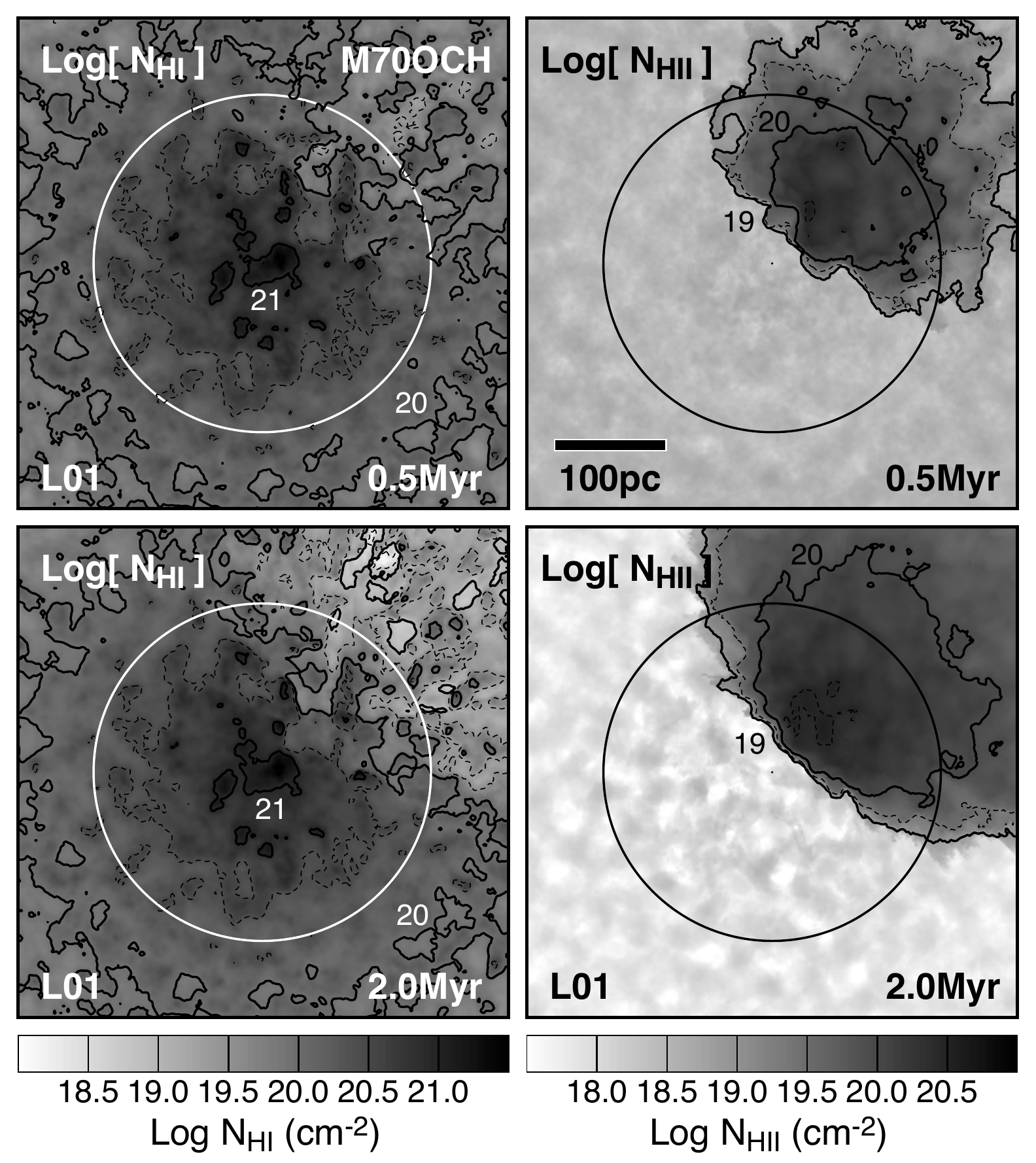}
 \includegraphics[scale=0.55]{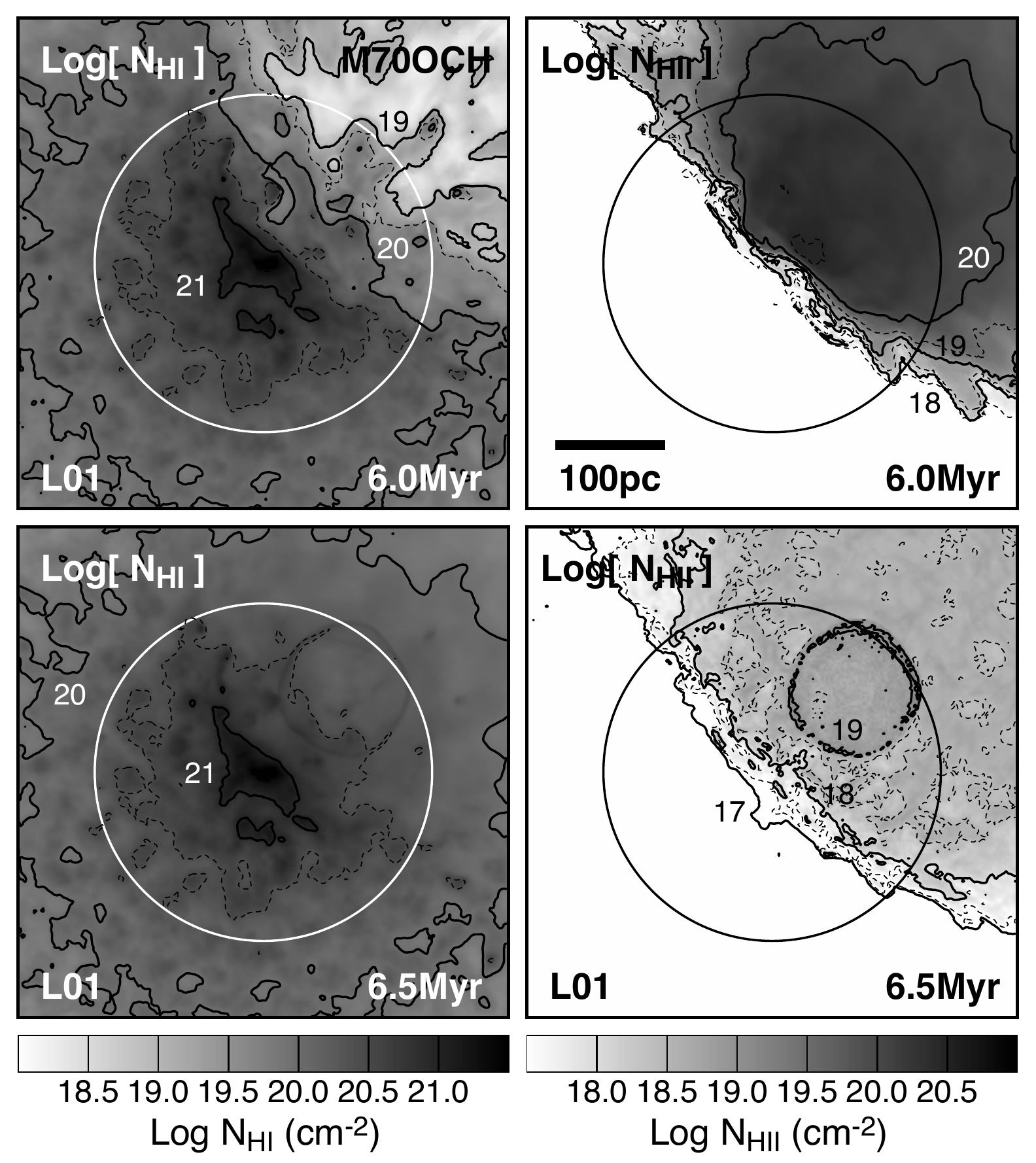}
\caption{ \label{f:M70OCcol}
M70 photoionization (off-centred source) during the pre-supernova and supernova phases at four different epochs
(see Fig.~\ref{f:M60CCcol}). The supernova ignites at 6.0 Myr.
The off-centred HII region is able to reach the edge of the core, and the falling density slope to the outer halo ensures that 
the photoionization fronts accelerate outward on the star side.  However a large volume of the halo remains unaffected.
The SN is able to just reach the core radius by 6.5Myr, but has only a small effect on the bulk off the halo. The SN occurs in a similar environment to the non-preionised models and the outcomes are similar. Almost all of the gas survives the impact of the SN event.
}
\end{figure*}

\begin{figure*}[htb!]
    \includegraphics[scale=0.37]{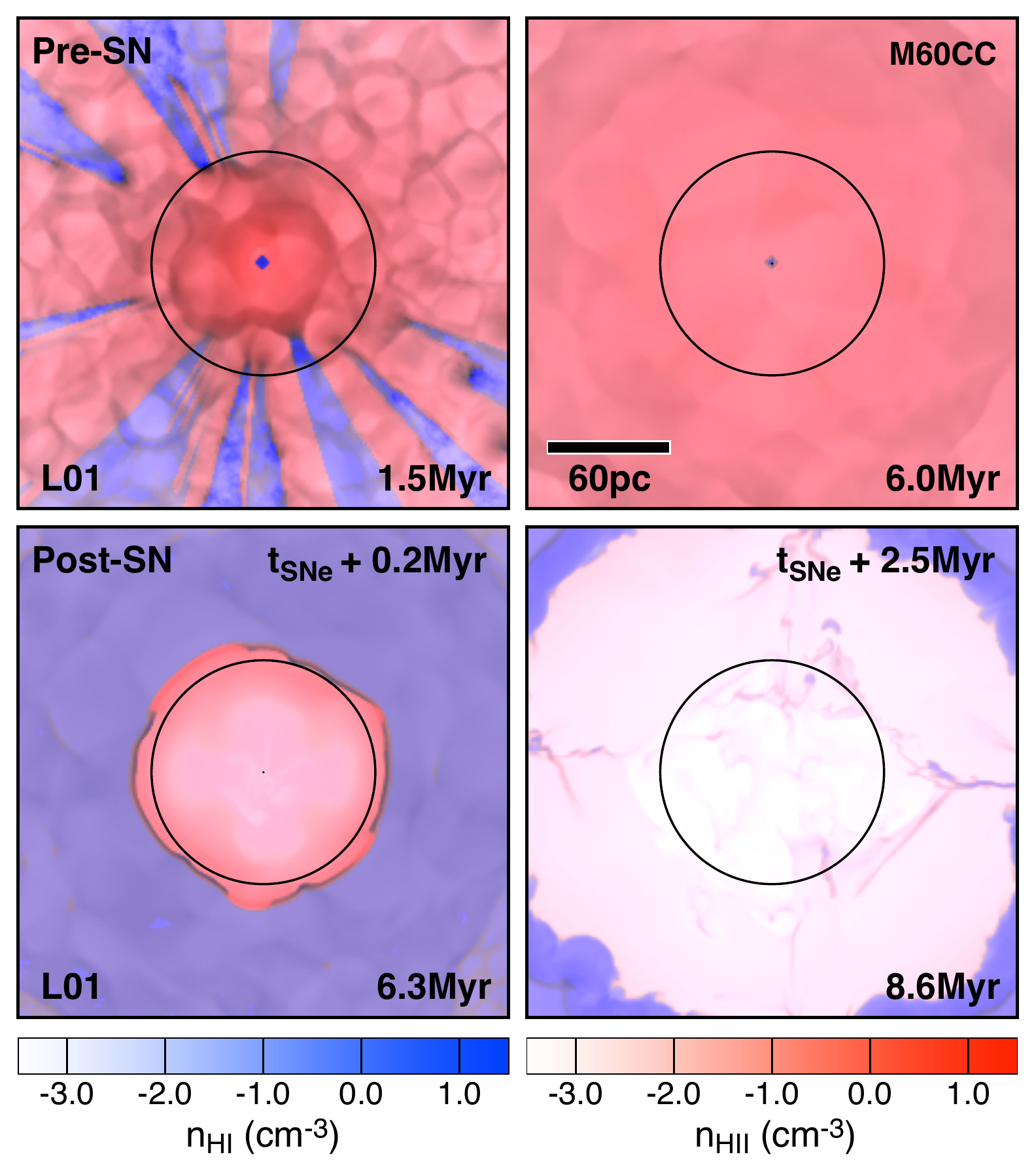}
    \includegraphics[scale=0.37]{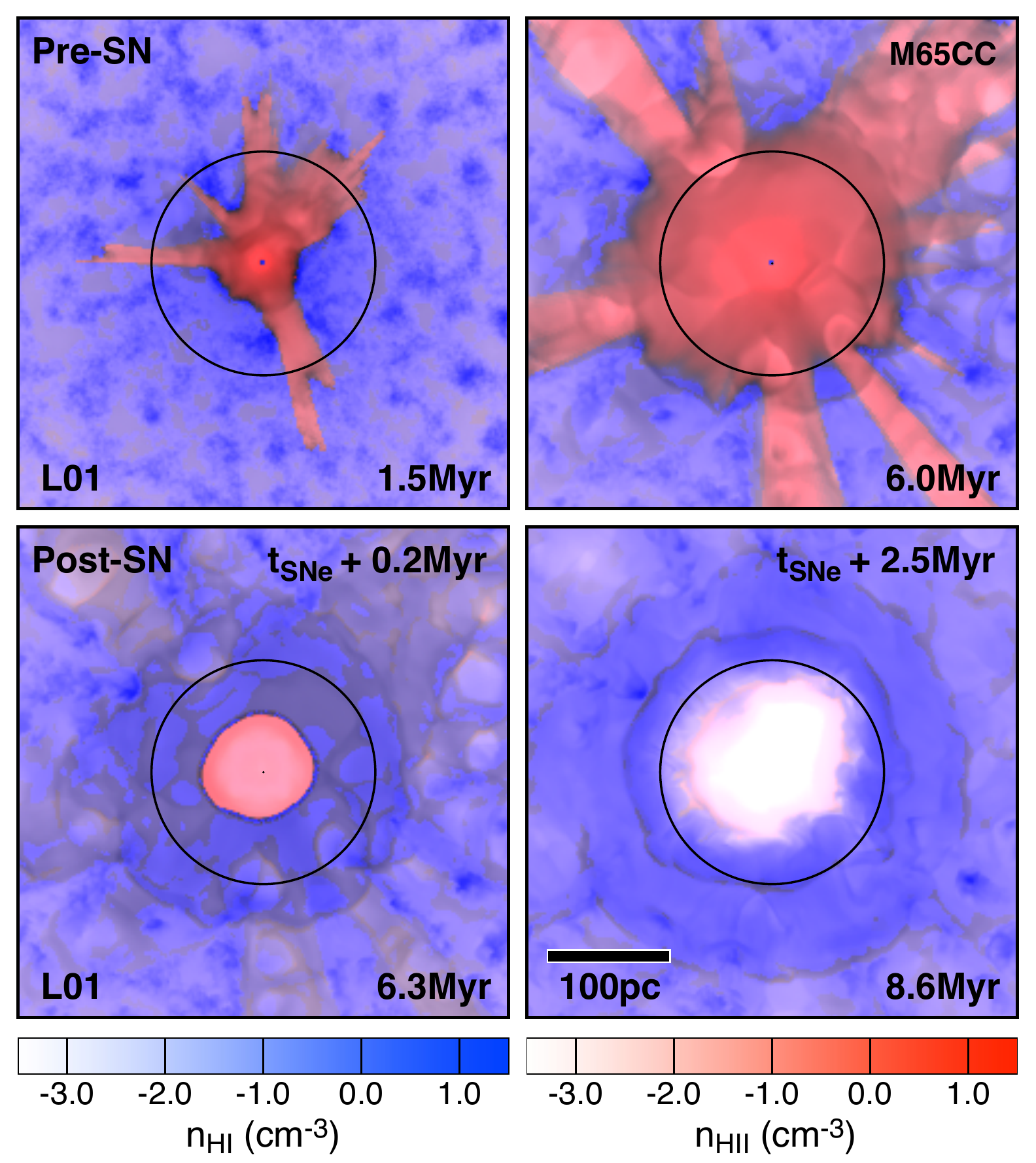}
    \includegraphics[scale=0.37]{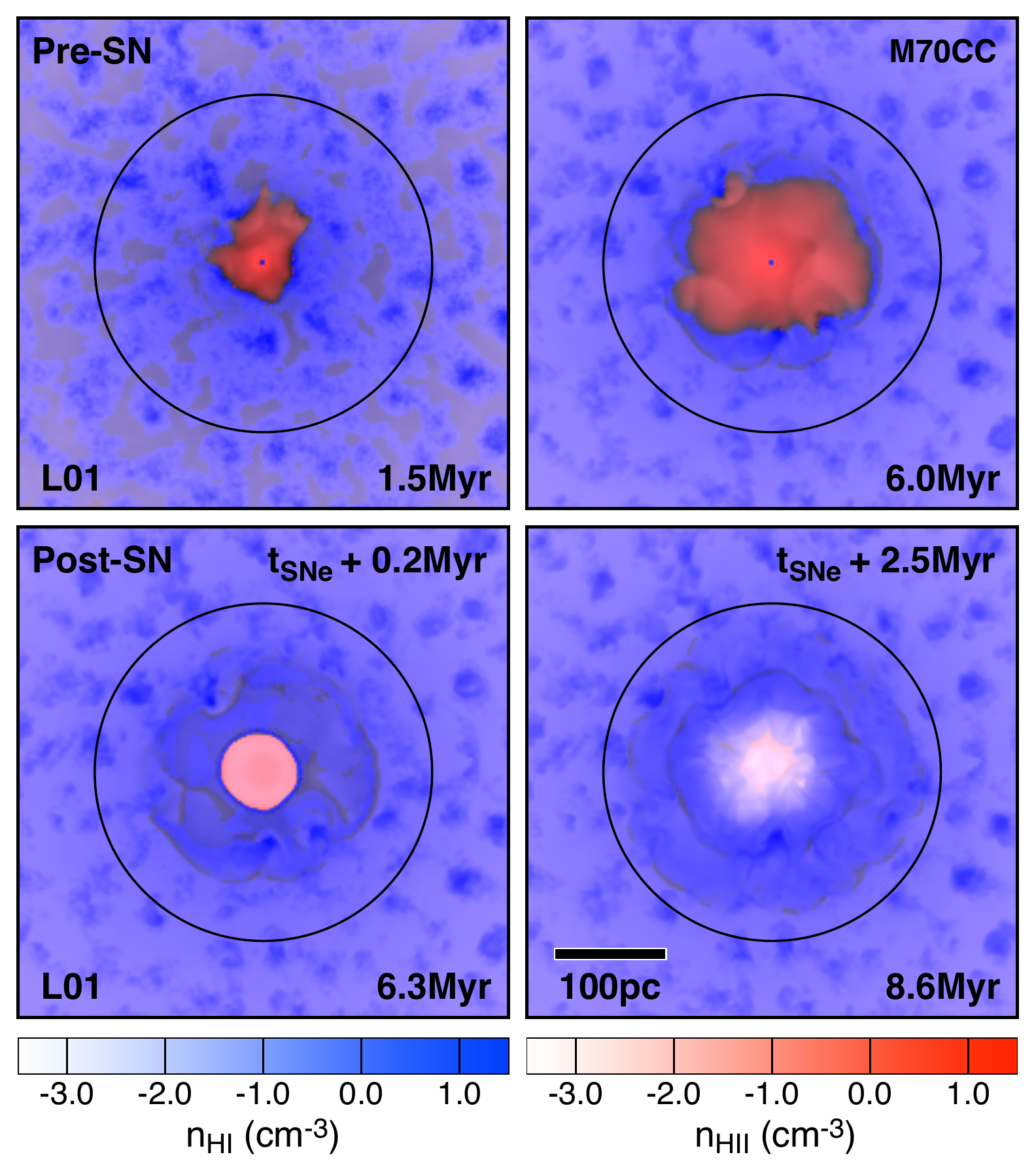}
    \includegraphics[scale=0.37]{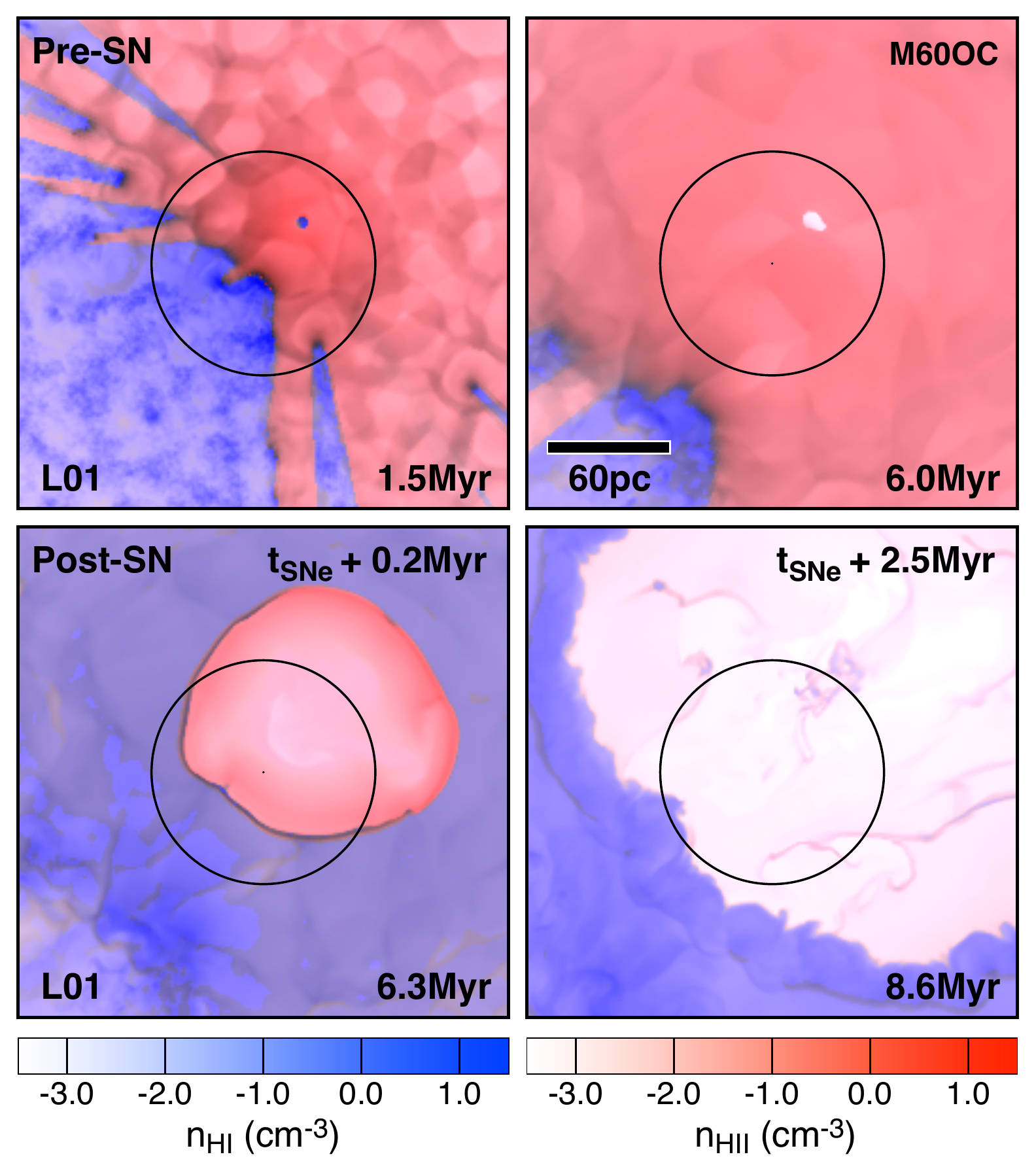}
    \includegraphics[scale=0.37]{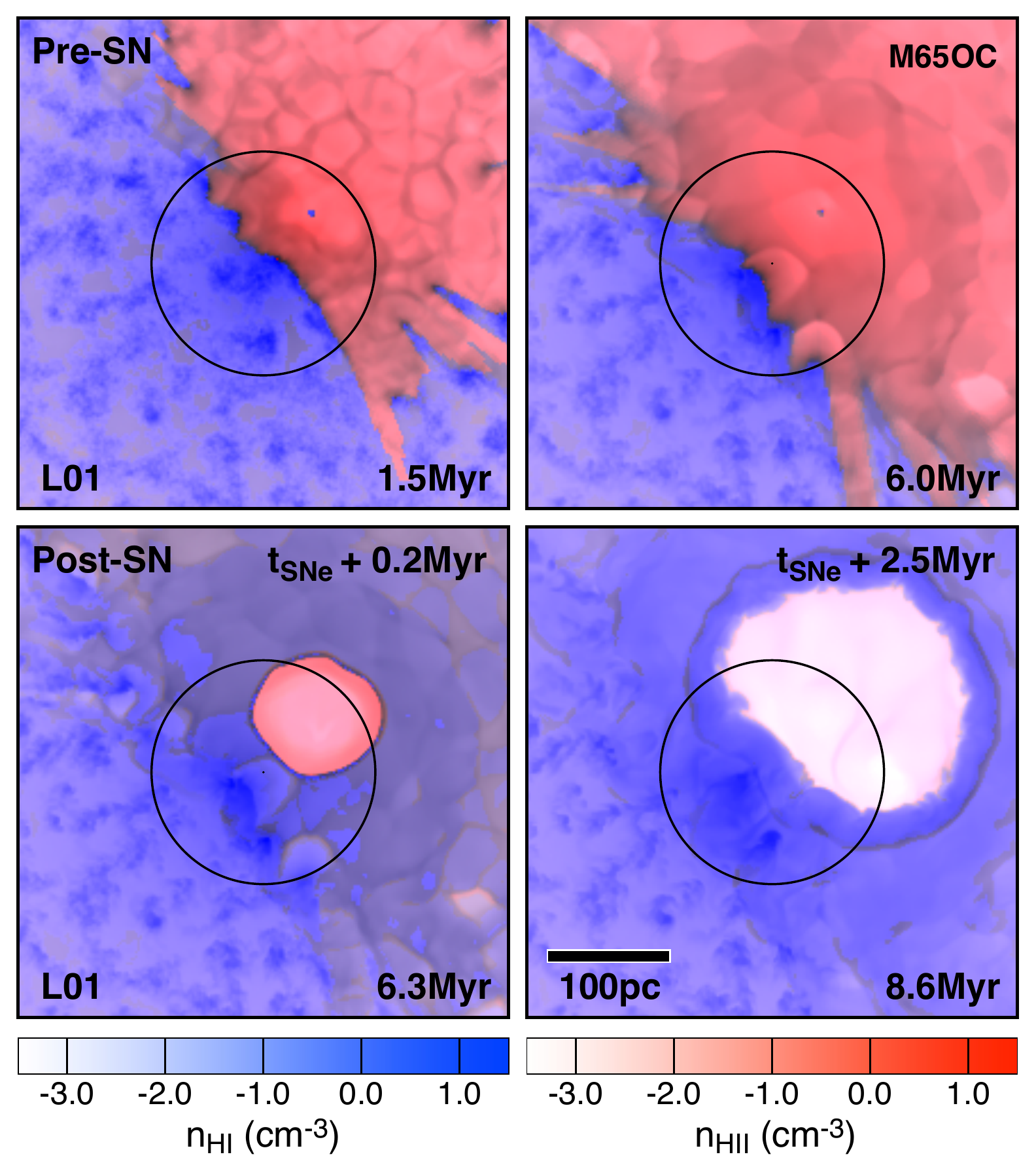}
    \includegraphics[scale=0.37]{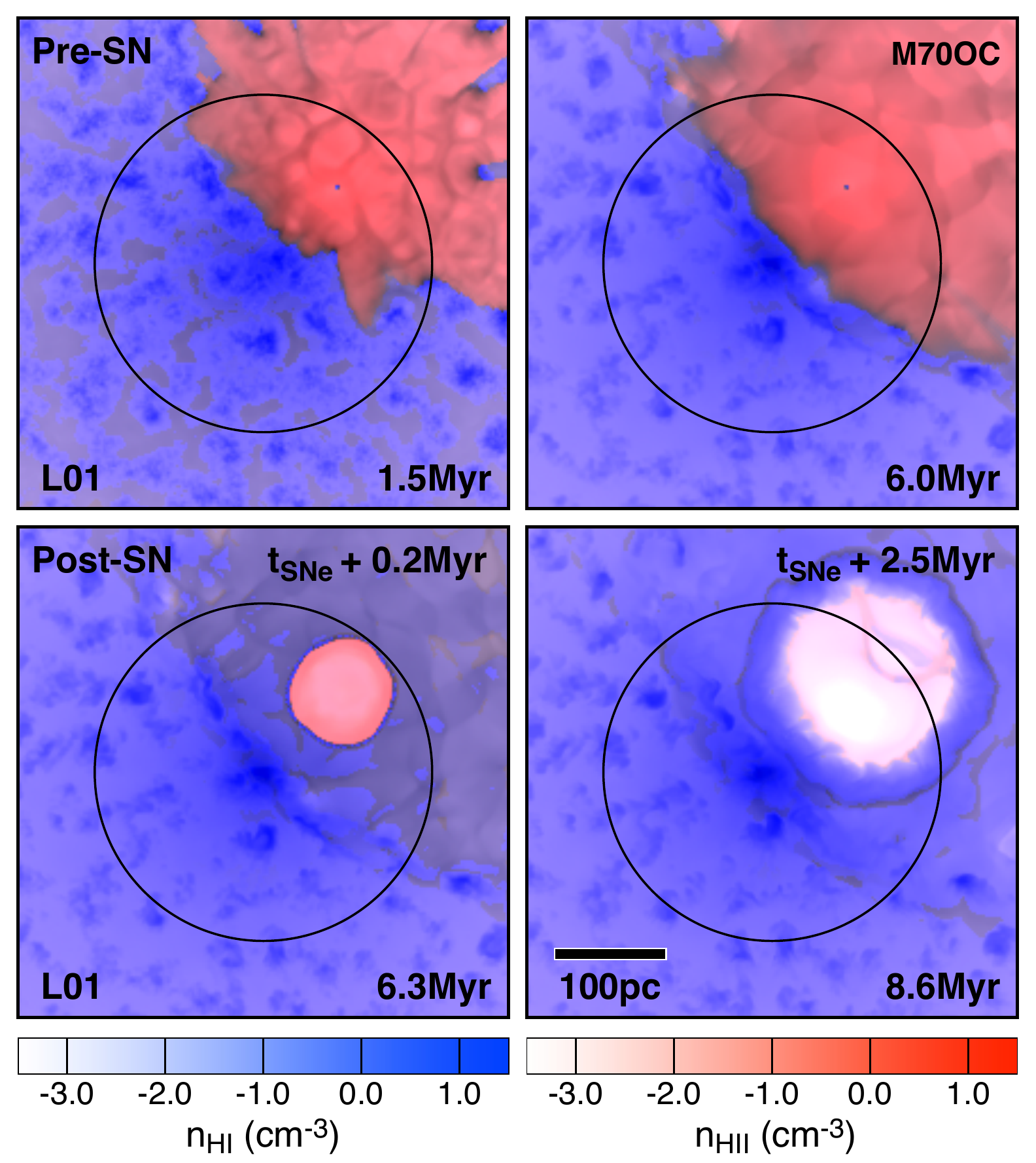}
  \caption{The development of the pre-ionization and SN phases for
  the M60, M65 and M70 cases for four different epochs; the top three figures 
  are for a central source and the lower three figures are the corresponding
  off-centred cases. The snapshots shows the distribution of
  ionised and neutral gas before, during and after the SN event at 6.0 Myr.
  This figure is a colour composite of the information contained in Figs.~\ref{f:M60CCcol}
  to \ref{f:M70OCcol}. Each image is a density slice through the centre of
  the halo. The red shows the distribution of fully ionised gas; the
  blue is the neutral gas; mauve corresponds to partially ionised gas. 
   }
   \label{f:colour}
\end{figure*}

Each of the models (M60, M65, M70) are discussed in more detail below.
In Figs.~\ref{f:M60CCcol} to \ref{f:M70OCcol}, we present time-dependent 2D slices of the gas distribution before
and after the SN explosion. These are not summed along the third dimension and so 
therefore do not illustrate the full extent of the ionized gas (cf. Fig.~\ref{f:colour}). 
Unless otherwise stated, all computations are performed
over a $648^3$ grid with two levels of the hierarchy (L00, L01) illustrated in Fig.~\ref{f:grid} 
(see \S\ref{s:hydro}). In a subset of cases, it was necessary to investigate a 3-level hierarchy 
(L00, L01, L02) over a $1944^3$ grid which we clarify below. 

It is worthwhile to pay particular attention to Fig.~\ref{f:retention}. Here
we present 1D plots of the evolution of the gas mass (A-F) and energy (G-L) retention fraction over much
longer timescales than shown in the half-tone figures (up to 25 Myr). In each plot, we present two 
sets of curves for retention within $r\leq$\rvir\ (upper) and $r\leq$\rs\ (lower).

The left panels (A, C, E, G, I, K) show all combinations $-$ clumpy (solid curves) vs. smooth (dashed curves), off-centred (light curves) vs. centred (heavy curves), adiabatic (red curves)
vs. cooling (black curves).
These plots are particularly illustrative and allow us to discuss the role of internal 
and kinetic energy in the developing explosion. For all models, the adiabatic cases (red curves)
confirm earlier work in that there is little or no gas retention \citep{maclow99}.
In the right panels (B, D, F, H, J, L), we show only the cooling clumpy models for the centred and off-centred cases; the gas retention is clearly more favourable for the larger halo masses.

One important insight emerges which explains why we have arranged the figures in this way. 
In each row, we compare two distinct halo masses (e.g. M55 vs.
M60 in the top row) to illustrate the impact of pre-ionization due to the supernova progenitor.
The left panels (e.g. A) have no pre-ionization in contrast to the right panels (e.g. B) which include this effect.
Broadly speaking, the results
on retention are similar. The same holds true when we compare M60 models without
pre-ionization to M65 models with pre-ionization, and M65 models without pre-ionization to M70
models with pre-ionization. These are important limiting
cases when one considers enrichment in halos
that do {\it not} have an important pre-ionization phase, e.g. for star formation histories
limited to low-mass
and/or Type Ia supernovae.

\subsection{M60 models}

The time-dependent behaviours of the M60 central and off-centred explosions are illustrated in 
Figs.~\ref{f:M60CCcol} \& \ref{f:M60OCcol} respectively. The evolutionary phases of an
explosive wind are well understood analytically \citep{weaver77} although we see additional
substructure due to the complexity of the initial medium.
The supernova ignites within a fully ionised, diffuse nebula confined by a neutral
gas shell. A dense radiative shell quickly forms and drives a shock into the neutral
shell. The momentum-driven shell does not stall and essentially evacuates all
gas within the halo after 15 Myr. In the off-centred case, one hemisphere is initially
evacuated within 15 Myr, with the opposite hemisphere losing most but not all of
its gas. At late times, there is some re-accretion from the far side leading to $<$10\% 
gas retention within the core regions.
The energy input from the pre-ionization phase, in combination with the SN
explosive energy, conspire to doom the M60 halos to invisibility.

\smallskip
\noindent{\it Numerical artefacts.} In order to examine the impact of artefacts, 
this case is treated with a 3-level grid. No such anomalies are seen in the 
pre-ionization phase. In the SN event, numerical artefacts evident in the first
600 kyr near the origin (Fig.~\ref{f:M60CCH}) are 
caused by the carbuncle instability (cf. Sutherland 2010). This effect is amplified 
by strong cooling in the early stages ($<125$~kyr) and leads to the `clover leaf'
pattern at 6.3 Myr inside $r_s$ (L01). Our algorithm stabilises the
instability when the radiative thin shell becomes only partially
resolved ($< 3$ cells thick) at later times, such that any structure that evolves in the
interaction between the spherical shell and the cartesian grid remains at
all later stages.  The instability is strongest where the radiative
shock front becomes orthogonal to the underlying grid, in this case
along the cartesian axes, giving rise to the boxy structures with apices
along the $x-$ and $y-$ axes seen here at 16 Myr (L00).
After the local density has dropped so that the blast--wave is no longer
radiative, the structure then simply expands with the now adiabatic
blast.  The additional L2 level is needed to better resolve the early blast, 
and the remaining visible non-spherical
structure has radius errors of approximately $10-15$\%. 

\smallskip
\noindent{\it Gas retention and energetics.}
We now discuss the issue of gas retention in the M60 models in the context 
of the energy (internal+kinetic) of the disturbed gas. 
Table~\ref{t:energy_retention} shows the amount of energy retained at critical radii,
$r= r_s$ and $r = r_{\rm vir}$, immediately before and after the supernova explodes,
and at 5~Myr and 25~Myr after the supernova event. We show density slices
through our clumpy M60 models over this time frame in Figs.~\ref{f:M60CCH} and \ref{f:M60OCH}.

We recall that during the pre-ionization phase, radiation leaks out from the centre
in the first few Myr leading to a fully developed, low density nebula by 6 Myr (see \S\ref{s:preion}).
In Fig.~\ref{f:M60CCH}, the central SN explodes into a warm low-density nebula, superheats the gas 
through shock heating, and removes essentially all of the gas from the dark matter halo out to 
\rvir\ after 25 Myr.  This is borne out in Table~\ref{t:mass_retention}. 

For the off-centred case in Fig.~\ref{f:M60OCH}, the progenitor supernova is placed at the half 
gas-mass radius to reflect the fact that most stars form off-centre. 
During the pre-ionization phase,
radiation leaks out from the centre in the first few Myr (see \S\ref{s:preion}), and
a conic `zone of avoidance' is set up diametrically opposite the progenitor. 
By 6 Myr, a low density nebula 
extends across most of the core region. 
The SN explodes into an off-centred, a warm low-density nebula, 
superheats the gas through
shock heating, and removes much of the gas asymmetrically from the dark matter halo out to 
\rvir\ after 25 Myr. The early zone of avoidance leads to far-side gas re-accreting into the core
region after the SN event has passed, but less than 10\% of the original gas remains within \rvir\
on long timescales.

\subsection{M65 models}
\label{s:m65}

We now discuss gas retention in the M65 models within the context of the mass and energy 
of the disturbed gas. The time-dependent behaviours of the M65 central and off-centred 
explosions are 
illustrated in Figs.~\ref{f:M65CCcol} \& \ref{f:M65OCcol} respectively {\it before and 
including} the explosion, and in Figs.~\ref{f:M65CCH} \& \ref{f:M65OCH} respectively 
{\it after} the supernova event. To aid comparison, the projected neutral and ionised gas in 
Figs.~\ref{f:M65CCcol} \& \ref{f:M65OCcol} are combined into a colour composite in
Fig.~\ref{f:colour}.

Tables~\ref{t:mass_retention} and
\ref{t:energy_retention} show the percentage of mass and energy retained at
$r_s$ and $r_{\rm vir}$ just before and after the supernova explosion, and
after delays of $5$ and $25$~Myr.
We see that more gas mass is retained in the core in the off-centred case
compared to the central case after 25 Myr post supernova. 
The tabulated energy retentions are percentages relative to the
binding energy of the gas bound by the dark halo. A system loses energy either
through cooling (e.g. radiative, adiabatic) or mass loss; conversely, a system gains energy 
either through heating (e.g. shocks, UV radiation) or accretion. In the off-centred model,
we see the effects of mass loss through the virial radius at the same time that the core
is accreting gas.

In Fig.~\ref{f:M65CCcol},
the supernova ignites at 6.1 Myr within a fully ionised, diffuse nebula confined by a neutral
gas shell. A dense radiative shell quickly forms and drives a shock into the neutral
shell. A lot of energy is radiated away in the time frame 0.2 to 2.5 Myr.
The momentum-driven shell stalls at 1 Myr. At this point, the shock slows down
and becomes non-radiative. The drop in temperature leads to a rapid increase in
density. The neutral shell then becomes energy-driven and more diffuse as it 
propagates outwards at about the halo escape speed and ultimately stalls at 
the virial radius. Most of the gas just survives in the M65 progenitor ionization $+$ SN model.

In Fig.~\ref{f:M65OCcol}, the off-centred HII region runs out down the halo density gradient rapidly on the side near the star, but struggles to ionise through the
core to the opposite side.  The geometry allows the ionization to affect a larger volume
in the outer halo than indicated by the integrated analysis in 
Fig.~\ref{f:recphot}.  The SN occurs again in a smoothed region and reaches a significant fraction of
the core radius by 6.5 Myr.

In these models, no more than half the gas survives the SN explosion, is enriched with the 
specific abundance yields of the discrete SN event.
Our highest resolution simulations reveal why cooling is so
effective in retaining gas compared to other factors. In the early stages, 
the super-hot metal-enriched SN ejecta exhibit strong cooling, leading to much of the
explosive energy being lost. This remains true regardless of the assumed metallicity floor 
at the time of the explosion.

In Figs.~\ref{f:M65CCH} \& \ref{f:M65OCH}, we explore the post-supernova phase over larger
radial scales (L00) and on longer timescales. The UV radiation from the supernova progenitor
escapes along some sight lines over the full spatial extent of the models. In both cases, the 
development of the supernova-driven bubble is well defined. A thick HI shell forms by 10 Myr
ahead of the thin metal-enriched shell and this material starts to reaccrete to the core by 25 Myr.
In the off-centred case, the central column densities
are higher because the shock front wraps around the core and compresses it.

\subsection{M70 models}

We now consider the retention of gas and energy in the M70 models. The HI and HII column densities before and just after the supernova are shown in 
Figs.~\ref{f:M70CCcol} and \ref{f:M70OCcol}, while Figs.~\ref{f:M70CCH} and \ref{f:M70OCH} show the HI density evolution after the supernova. 
For ease of comparsion, Figs.~\ref{f:M70CCcol} and \ref{f:M70OCcol} are also shown on the colour composite figure Fig.~\ref{f:colour}.
The M65 models walk the tightrope between survival and disruption, retaining half their baryons at
best. Tables~\ref{t:mass_retention} and \ref{t:energy_retention} show percentages of gas and energy retained just before and after the explosion, as well as after 5 and 
25~Myr. The M70 models clearly
survive a single supernova event, losing very little gas and by 25~Myr contain more gas than in their initial state as a result of accretion. As a 
result of cooling, the energies within 
both \rs\ and \rvir\ are much lower than the initial values at both 5~Myr and 25~Myr, suggesting that the gas is not flowing outwards.

For the M70 models, for the higher total gas mass, more than a single event could explode 
at a given epoch. But
this is a complication that we do not consider here because the combined
impact of the two events will depend on their masses (one is likely to be in
the low-mass limit) and locations within the halo.
In Fig.~\ref{f:M70CCcol}, the central supernova ignites at 6.0~Myr.
The photon flux is insufficient to ionise a significant part of the core volume, and the HII
region is stifled. In the off-centred case in Fig.~\ref{f:M70OCcol},
the off-centred HII region is able to reach the edge of the core, and 
the falling density slope to the outer halo ensures that 
the photoionization fronts accelerate outward on the side containing the star. 
The supernova is able to just reach the scale radius by 
6.5~Myr, but has only a small effect on the bulk of the halo. 

The post-supernova phase is shown in Figs.~\ref{f:M70CCH} and \ref{f:M70OCH}. For the central supernova model, the bubble at $t=2.5$~Myr is smaller than in the M65 case and sweeps up denser gas. By 10~Myr the bubble collapses and dense gas accretes back to the centre. In the off-centred case, the behaviour is similar, 
although a small amount of radiation reaches the edge of the grid on the side nearest to the supernova, while the gas on the opposite 
side of the halo is largely unaffected. As in the M65 off-centred case, the central densities are increased at 2.5~Myr when the shock compresses the core, however 
the effect is weaker than in the M65 model because the shock is unable to wrap around the core.

Our models consider dwarf galaxies to be isolated, but in practice the vast majority reside
in the extended halo of a host galaxy, albeit beyond the virial radius. In the M65 and M70 models, there is clear evidence of the expelled gas beginning to fall back after 20 Myr. A proper consideration of this gas requires
that we consider the environment of the halo which is beyond the scope of the current work.
We leave this to a later study.

%\subsection{Mixing of metals}
%\label{s:mix}

\subsection{Summary}

\begin{figure*}[htb!]
   \includegraphics[scale=0.35]{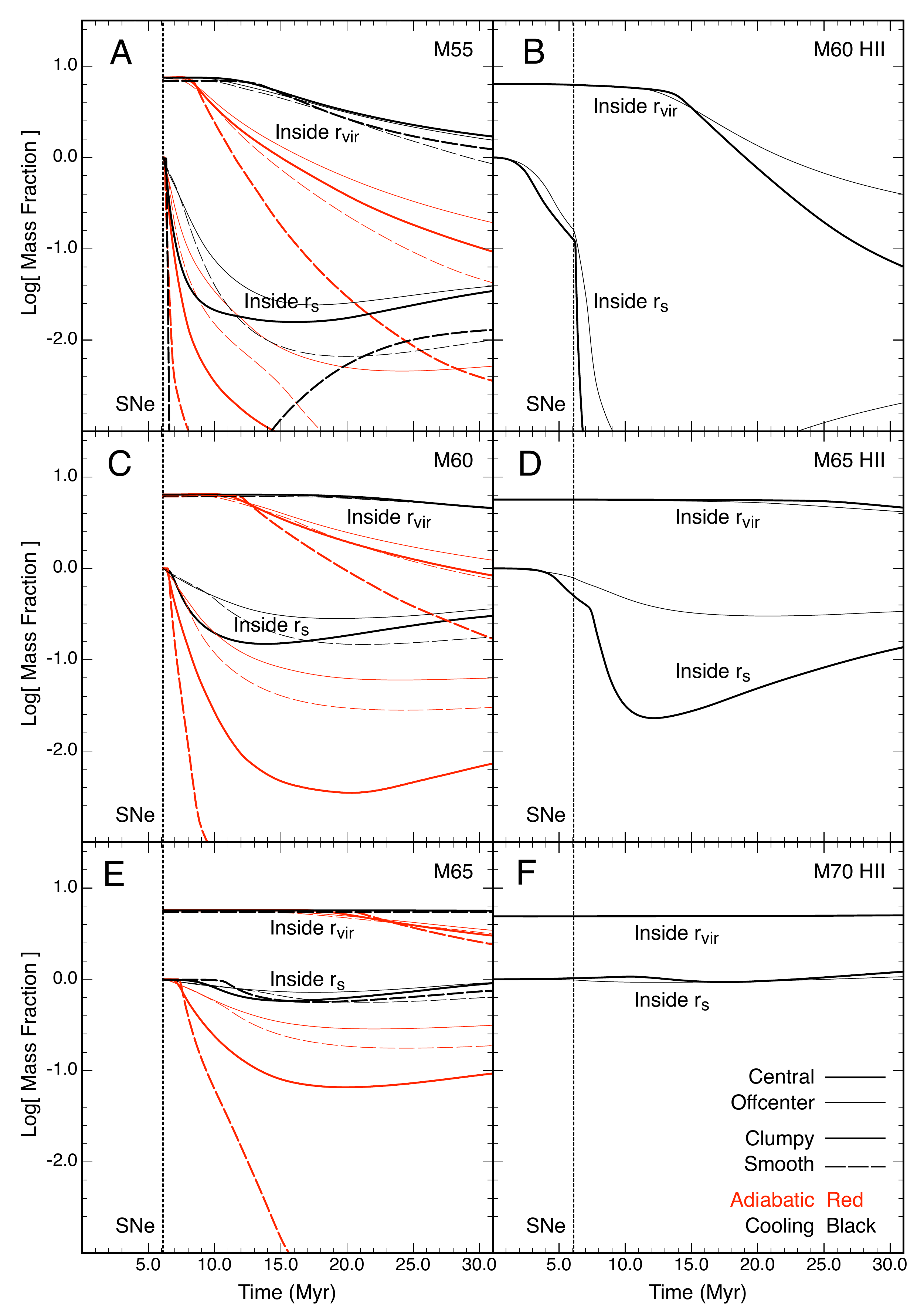} 
    \includegraphics[scale=0.35]{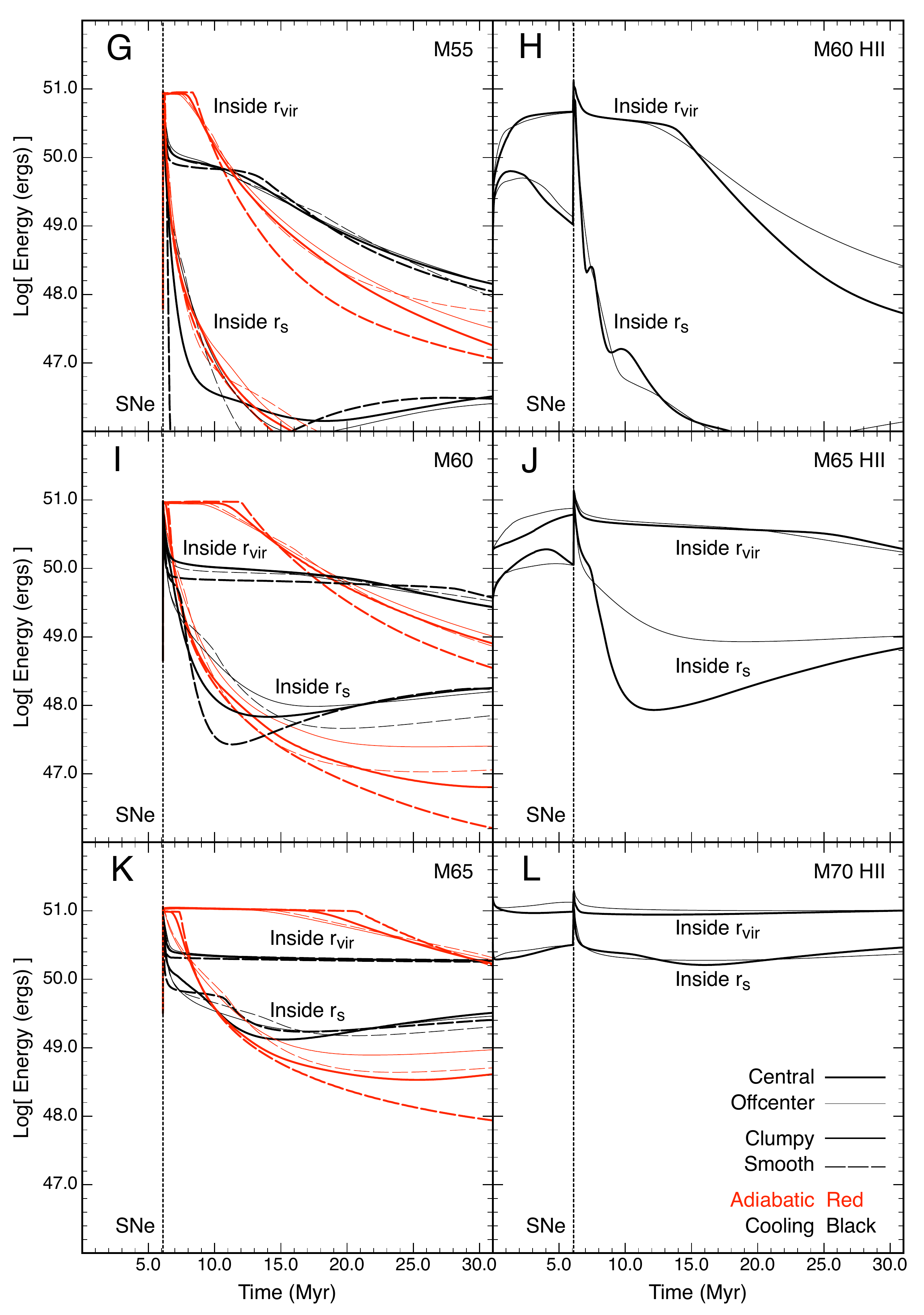} 
  \caption{Gas mass (A-F) and energy (G-L) retention inside the M55, M60, M65 and 
  M70 models. Panels A, C, E and G, I, K are for SN explosions without pre-ionization 
  (Type Ia SN); panels B, D, F and H, J, L are for SN explosions with pre-ionization (Type II SN). 
  For the former, we start the models at the explosion so there is no data prior to the event.
  The time of the SN event is shown as a vertical dashed line; the impact of 
   pre-ionization (see Table~\ref{t:mass_retention}) is clearly evident. The key to all
   curves is given in both figures.   The top set of curves in each panel is for quantities within
   a radius $r_{\rm vir}$; the bottom set is for quantities within $r_{\rm s}$.
   The mass fraction is defined relative to the original gas mass within the
   scale radii as indicated at $t=0$.
   We show both the adiabatic (red curves) and cooling models (black curves) for both clumpy and smooth media. Consistent with earlier work, the smooth adiabatic case is much lossier than the
   case for cooling halos. Note that the adiabatic case is energy conserving until gas is
   lost over a boundary.
   In the energy retention figures, note that M70's explosive energy has only a marginal impact 
   on the energy retained from the progenitor phase.
   }
   \label{f:retention}
\end{figure*}

As described in the preceding sections, the summary plots for the single supernova
event models in Fig.~ \ref{f:retention} (supported
by Figs.~\ref{f:M60CCcol} to \ref{f:M70OCcol}) reveal that:

\begin{enumerate}

\item For all M55 models and M60 adiabatic models,
essentially none of their original gas mass remains within a scale 
radius at $25$~Myr. Thus subsequent star formation does not occur.

\item For all M60 cooling models, only a few percent of the original gas mass remains 
within a scale radius at $25$~Myr. No significant star formation can occur.

\item For the M65 models, all of the cooling scenarios retain 70-80\% of the original gas 
inside the virial radius at 25 Myr. This is sufficient for ongoing star formation. 
None of the adiabatic models retain more than 20\% of the original gas.

\item In the borderline M65 cooling cases, off-centred models retain 2-3 times more gas within a scale radius 
but the supernova metals mix less efficiently with this gas.

\item For the M70 models, all of the gas survives a supernova event and its radiative progenitor. We stress 
that we have not considered clustered multi-supernova events which can occur in these models.

\item Cooling is the most important factor governing gas retention within the virial radius at late times.

\item Clumpiness is far more important in the adiabatic case than in the cooling case for gas retention.

\end{enumerate}

%\clearpage
\section{Discussion}
\label{s:discuss}

\subsection{Background}
The least massive galaxies lie at the bottom of the CDM
hierarchy. Our focus here is not their role as building blocks
in galaxy evolution, but what they can tell us about the yields of the
first stars prior to the reionization epoch, and the following
generations during and immediately after reionization 
\citep[e.g.][]{norris10,blandhawthorn10b,frebel12,ritter15}. At least some of the first galaxies must 
have resided within low mass halos. We anticipate that such systems, if they can be
identified in the present day universe, are likely to carry the most ancient chemical signatures
because of their age and because of the relatively low number of enrichment events over
cosmic time.

If we can identify such objects, it should be possible to obtain a relatively
clean signature of the primordial yields from the earliest stellar generations.
The star-formation efficiency is greatly diminished in low-mass galaxies,
thereby reducing the total number of enrichment events.
There are likely to be instances where a large fraction of the metals are
blown out of the galaxy, and this may obscure the signatures of earlier
generations of stars. This can occur when the relatively rare supernovae
that go off in these galaxies are correlated in time. But we would still
expect to find the signatures of a single enrichment event in at least
some intrinsically low-mass galaxies.

We now discuss our models in light of two recent discoveries: ultra-faint
dwarf galaxies (near field) identified in all-sky imaging surveys \citep{simon07,kirby08}, 
and very metal poor damped Ly$\alpha$ systems
(far field) identified along quasar sight lines \citep{erni2006,cooke10}. 
The two classes of objects in turn although a more extensive discussion is given 
elsewhere \citep{webster14,webster15a,webster15b}. 
We refer the reader to Figs.~\ref{f:halos} and \ref{f:mass_radius} that show slow 
evolution is expected between the formation redshift at $z=10$, the epoch of
the DLAs ($z=2-4$), and present-day ultra-faint dwarfs. The overall effect is
additional dark matter that is mostly accreted to the outer envelope of the dwarf,
rather than onto the central cusp (Fig.~\ref{f:pots}).

%As we have seen, the clumpy off-centred supernova models retain more gas
%mass than the central supernova models for a given halo mass. This is because the
%supernova shock wraps around the dense core and compresses the gas, thereby
%leading to higher central densities.  A byproduct of the supernova being off-centred is that
%most of the metals escape over at least half the available solid angle. The ejecta
%moving around the core does not mix easily with dense gas in the nuclear regions. 
%This is in contrast to the centred
%supernova models where the metals mix much more easily with the surrounding
%dense gas. However, the core regions are now more easily evacuated, and the 
%gas density (and the projected column density) is reduced. In our `one-shot' models,
%the location and mass of the supernova progenitor has important consequences for the
%projected properties in terms of the overall stellar abundances and their scatter, 
%the degree of gas/metal mixing
%and the projected gas columns at high redshift.

\subsection{Ultra-faint dwarfs}

In recent years, the Sloan Digital Sky Survey has doubled the number
of known dwarf galaxies that orbit the Galaxy (for a review, see \citet{tolstoy09}), 
particularly at the low mass end. These newly discovered dark-matter 
dominated galaxies are uniformly very old and cover a wide range in luminosity 
down to $\sim 10^3$\Lsun. Unlike globular clusters,
they show a wide range of eccentricities which may indicate that some are 
undergoing tidal disruption \citep{martin08}. Objects less luminous than $10^5$\Lsun\ 
are referred to as ultra-faint dwarfs (UFDs). 

There has been extensive discussion concerning the integrated masses of UFDs
as observed today \citep{simon07,strigari08,martin08,bovill09,wolf10,frebel14}.
Some studies quote a characteristic UFD mass within 300~pc, i.e. M$_{300} \approx 10^7$ M$_\odot$ \citep{strigari08, walker09} although this is an arbitrary physical
scale that is not easily related to well known scaling relations. The characteristic virial 
mass \mvir\ today is highly uncertain with estimates ranging
from $10^8$ to $10^9$ M$_\odot$. 
For a dark halo described by the NFW profile, the mass within a fixed radius gives
the mass within any other radius, assuming this is an accurate description of the
underlying gravitational potential. For our work, we need to know the virial mass when 
the UFD first formed. The halo parameters in cosmological simulations have a
strong dependence with redshift making any estimates rather uncertain \citep{power03,duffy08}. We have adopted \mvir\ $\lesssim$ $10^7$\msun\ because the 
anticipated total population of UFDs may be too large to support larger progenitor
masses.

The simulated `observables' from our  models are broadly consistent with the most metal-poor UFDs \citep{webster14,
webster15a}. The data for these systems are summarised by \citet[][Fig. 1]{frebel12}.
The metallicity range for our models is -1.0 $<$ [Fe/H] $<$ -4.0 with a mean
[$\alpha$/Fe] enhanced by a factor of a few with respect to the Solar value depending
on the initial mass function at the time of formation \citep{woosley95}. We have 
begun to consider the $r$ and $s$ process signatures in the gas, work that is to
be presented in later papers. The internal kinematics of the 
baryons are less than 4 km s$^{-1}$ within the inner scale radius, consistent
with observations \citep{walker09}.

\citet{brown14} find that the star formation history of UFDs is marked
by one or two bursts of star formation before truncation by a global event.
By construction, such a history is consistent with our models.
UFDs have stellar populations with mean metallicities in the range
-2.5 $<$ [Fe/H] $<$ -2.0 \citep{kirby13a}.
Within the framework of our models, this requires (i) $1-2$ SN events in
a low mass halo, specifically the M65 event-centred models (\S\ref{s:m65});
(ii) $3-10$ SN events in a starburst within a dark matter halo
an order of magnitude more massive (e.g. M70). These cases can give rise to
measurably distinct chemical histories, star formation histories and star cluster
populations. For example, [$\alpha$/Fe] is enhanced in starbursts
relative to quiescent star forming regions. Direct evidence for this comes from 
x-ray observations of starburst winds \citep{martin2002}.

At the time of publication, we note that \citet{onorbe15} have presented a new
model for UFDs where the baryons reside within dark matter 
halos that are orders of magnitude more massive than we consider
(\mvir $\approx$ $3\times 10^9$\msun). Given that all of the baryons 
are retained within our M70 models, for the same initial gas fraction, we cannot
easily distinguish their model from ours in terms of the predicted chemical 
abundances.

In summary, if UFD masses are largely unchanged since their 
early formation, we would expect these halos to show significant scatter in [Fe/H]
at low metallicity, and for some of the oldest stars to constrain the yields of 
early stellar generations \citep[e.g.][]{frebel12}.

\subsection{Very metal poor DLAs}
An interesting development is the recent discovery of very
metal-poor damped Ly$\alpha$ systems (DLA) along QSO sight lines
\citep{pettini08, penprase10}. If these are protogalactic
structures that have recently formed from the IGM, there is the
prospect of identifying the chemical imprint of early generations of
stars \citep[e.g.][]{pettini02}. Interestingly, two of these systems with
[Fe/H] $\sim$ -3 may bear the hallmarks of early stellar enrichment 
\citep{erni2006,cooke10,cooke15}.

First, \citet{erni2006} identify a DLA towards the QSO Q0913+072 ($z=2.785$)
with an iron abundance characteristic of the IGM at that redshift. The C, N, O,
Al, Si abundances show an odd-even effect reminiscent of the most
metal-poor stars in the Galactic halo. This pattern is created in models where
the neutron flux is low \citep[e.g.][]{heger02}, presumably due to the 
low overall metal abundance. A more striking signature is the strong [N/H] depletion
which \citet{pettini02} has argued is further evidence for a system 
that has recently formed from the IGM. \citet{erni2006} argue that the abundances 
appear to agree with 10$-$50 M$_\odot$ zero metallicity Pop III models.

Secondly, \citet{cooke15} identify half a dozen DLA objects that
resemble the properties of UFDs with mean [Fe/H] in the range
-2.5 $<$ [Fe/H] $<$ -2.0 and line-of-sight dispersion velocities of 1 $<$ $\sigma_{\rm LOS}$
$<$ 5 km s$^{-1}$ corrected for thermal and turbulent broadening. 
These physical parameters are in line with our M65 and M70 models \citep{webster15b}.

Finally, we examined whether the line-of-sight properties of our models would quality them as DLAs.
For all very metal poor DLAS, \citet{cooke15} infer the total mass of
neutral gas to be $10^4 \lesssim M_{\rm gas} \lesssim 10^7$ M$_\odot$ within a linear scale
of $\lesssim 30-1300$ pc although their UFD candidates are at the low end of these
ranges. 
There are no sight lines in the M55
models, and few sight lines in the M60 models, that reach the DLA threshold. But
the M65 and M70 models fare better with a significant fraction of sight lines within a radius of $\sim r_s$
being classified as a DLA-type spectrum. Furthermore, the inferred gas metallicity of [Fe/H] $\sim$ -3
falls well within our predicted range. This gas is mostly self-shielded from any
external radiation and therefore does not require an ionization correction.

\section{Conclusions}
\label{s:conclude}

To investigate the issue of gas retention in the lowest mass dark--matter
halos, we performed a series of experiments in the form of
high--resolution radiative hydrodynamic simulations dominated by the
gravitational potential of the dark matter. We find that the broad properties
of UFDs and very metal poor DLAs are consistent with the host halos 
having intrinsically low mass. Our calculations are based on the critical
period a few Myr before the SN explodes, and up to 25 Myr after the event.
These cases are explored over much longer timescales ($\sim 600$ Myr) elsewhere
which allows us to consider the effect of subsequent SN events and the homogenising
effects of re-accreting gas dislodged by earlier events
\citep{webster14,webster15a,webster15b}.

In summary, dark matter halos with virial masses as little as $3\times 10^6$\msun\ 
are able to retain some gas, a limit that is an order of magnitude lower than found previously
\citep{maclow99}.
The effects of cooling, the clumpiness of the medium, and the position of the explosion were 
studied. Consistent with earlier work, the baryons do {\it not} survive in 
smooth adiabatic or smooth cooling models in the event of a supernova.
Models with radiative cooling, 
clumpy media and off--centred explosions are most favourable to gas retention.
Our highest resolution simulations reveal why cooling is so
effective in retaining gas compared to any other factor. In the early stages, 
the super-hot metal-enriched SN ejecta exhibit strong cooling, leading to much of the
explosive energy being lost.
This remains true regardless of the assumed metallicity floor at the time of the explosion.

An added complication is to consider asymmetric explosions in 
core-collapse supernovae \citep{grefenstette14} rather than the
isotropic explosions considered here. But we believe these do {\it not} lead to 
radically different outcomes when compared to the off-centred, isotropic supernova models. 
To first order, the overall mixing is dictated by geometry such that an energetic pulse
experiencing {\it any} asymmetry leads to broadly similar metallicity spreads. If confirmed
by later work, this principle will greatly reduce the parameter space of future simulations.
It remains to be seen if asymmetric vs. symmetric explosions lead to the ejection
of different abundance yields.
 
We recognise at the low mass limit explored in our models, the predictions
are subject to stochastic effects due to the rarity of supernova events.
Any assumed initial mass function will be poorly sampled in
this limit. In rare instances, simultaneous supernova events are
statistically possible but these are more likely in higher mass halos than
considered here.  Conversely, it is conceivable that low-mass star formation can proceed at the adopted 
metallicity threshold of
[Fe/H]=-4 \citep{bromm11} such that no prior supernova activity is required beyond
the enrichment from the first stars.
This leads to a small fraction of stars extending down to the metallicity floor, a signature
observed in only a few cases, e.g. Bootes~I \citep{gilmore13} and Segue~I \citep{norris10}.

This work has identified new lines of enquiry. In particular, the central gas densities 
are chosen to limit the gas mass inside one scale
radius to about 10\% of the dark matter mass on the same scale (\S\ref{s:problem}). 
Such a highly concentrated baryon fraction is to
be expected because the star formation efficiency is likely to be very low
before the onset of metal (and therefore dust) production \citep{krumholz12}.
In future papers, we will consider higher baryon fractions including those that dominate over
the dark matter core. This requires us to include self gravity into the {\it Fyris} code
through the use of fast fourier transforms over the refined adaptive grid. In these models, 
we treat the pre-supernova phase with higher resolution grids. We have already begun to
implement self-consistent cooling arising from variable enrichment over the gas
distribution. Our future models will include the formation of dust and molecules.
A complete understanding of UFDs will need to include the details of the subhalo orbit within
the evolving Galactic potential and the coronal gas distribution responsible for the gas stripping 
\citep{nichols12,nichols15}. 

 %What causes bursts? see QBH08's delay difference equation

\acknowledgments  JBH is supported by an Australian Laureate Fellowship from the
Australian Research Council (ARC). JBH is indebted to Merton College, 
Oxford for a Visiting Research Fellowship and to the Leverhulme Trust for a Visiting 
Professorship to Oxford. DW is funded by an Australian Postgraduate Award.
We are grateful to the Kavli Institute at UC Santa Barbara for hosting us in the final
stages of this work. We are indebted to Julio Navarro, Anna Frebel, Josh
Simon, Marla Geha, Evan Kirby and Charlie Conroy for insightful comments.
We have benefited from numerous incisive comments and insights from an
anonymous referee.

\bibliographystyle{apj}
\bibliography{refs}

\appendix

\section{Appendix A -- Abundances} 

The host halo gas metallicity was taken to be [Fe/H] = -4.0. The abundances, by number relative to hydrogen = 1.0,  are shown in Table
\ref{t:abund}. The solar abundances are based on
\cite{asplund2005}, with a mild alpha element enhancement pattern:
Carbon ($\times1.5$), Oxygen($\times5.0$), Magnesium($\times2.5$),
Silicon($\times2.5$), and Calcium($\times2.5$) (K13). Helium evolution follows \citep{dopita06}, $n_{He}/n_H$ $=$ 0.0737+0.0024 Z/Z$_\odot$.

{\small
\begin{table}[htdp]
\caption{Gas Phase Abundances, \\
$[{\rm Fe/H}] = -4.0$, Neutral: $\mu = 1.21$, Ionised: $\mu = 0.59$}
\begin{center}
\begin{tabular}{r c c c c }
\hline
       &SN25 Z=0&SN25 Z=0&`Primordial'&Solar\\
Element&Mass Yields&Abundances& Abundances&Ratios\\
\hline
\hline
H (X)&10.60&0.00&0.00&0.00\\
He (Y)&8.03&-0.72&-1.13$^*$&-1.01\\
Metals(Z)&4.41&-1.62&-6.27&-3.09\\
\hline
C&2.94E-01&-2.63&-7.28&-4.10\\
N&5.91E-04&-5.40&-10.05&-4.79\\
O&2.79E+00&-1.78&-6.43&-3.34\\
Ne&5.33E-01&-2.60&-7.25&-3.91\\
Na&1.03E-03&-5.37&-10.02&-5.75\\
Mg&1.20E-01&-3.33&-7.98&-4.42\\
Al&8.08E-04&-5.55&-10.19&-5.61\\
Si&3.51E-01&-2.92&-7.57&-4.49\\
S&1.86E-01&-3.26&-7.91&-4.79\\
Ar&3.14E-02&-4.13&-8.78&-5.20\\
Ca&2.48E-02&-4.23&-8.88&-5.64\\
Fe&7.38E-02&-3.90&-8.55&-4.55\\
Ni&4.42E-04&-6.15&-10.79&-5.68\\
\hline
\multicolumn{5}{l}{\footnotesize * Helium from BB NS, metals scaled to [Fe/H] = -4.0.}
\end{tabular}
\end{center}
\label{t:abund}
\end{table}
} %small

The supernova ejecta composition was modelled on the well studied M25 supernova, cf. \cite{kobayashi2006, nomoto2006}. 
Using the Kobayashi compilation yields $2.79~M_\odot$ of oxygen in the $\sim4.4~M_\odot$ of total metals ejected, with $0.072 M_\odot$ of Fe.

Table \ref{t:abund} shows the SNe mass yields, and the abundances by number, for the ejecta and ambient medium,  with a final column showing solar abundances ratios for reference.

\section{Appendix B -- Fractal Interstellar Medium}

\subsection{Log--normal density distribution} \label{s:lognormal}

We use a log--normal distribution to describe the single--point
statistics of the density field of our non--uniform ISM. The log--normal
distribution is a skewed continuous probability distribution.  Unlike
the normal distribution, it has a non-zero skewness, variable kurtosis,
and in general the mode, median and mean are unequal.  

The log--normal
distribution appears to be a nearly universal property of isothermal
turbulent media in experimental, numerical and analytical studies (e.g. 
\citet{nordlund99a, warhaft00, pumir94}).   
Moreover, it is encouraging  that the log--normal
distribution is the limiting distribution for the {\em product} of
random increments, in the same way that the normal distribution plays
that role for {\em additive} random increments. It is thus compatible,
at least conceptually, with a generic cascading process consisting of
repeated folding and stretching.

With a log--normal distribution,  which is on average isotropic, the natural logarithm of the ISM
density field is a Gaussian which has a mean $m$ and variance $s^2$. The
probability density function for the log-normal distribution of the mass
density $\rho$  is, 
\begin{equation} P(\rho) = \frac{1}{s \sqrt{2 \pi}
\, \rho} \exp \left \lbrack \frac{-(\ln \rho - m)^2}{2s^2}\right \rbrack
\, . \end{equation} 
The mean $\mu$ and variance $\sigma^2$  of the
density are given by 
\begin{eqnarray} \mu & = & \exp[ m + s^2/2] \,. \\
\sigma_F^2 & = & \mu^2 \, (\exp[ s^2]-1) \, . \label{e:log_normal_pars}
\end{eqnarray}

In these simulations we adopt $\mu = 1.0$, $\sigma_F^2 = 5.0$, as our
standard log--normal distribution.  With these
parameters, densities below the mean comprise one quarter of
the mass, and occupy three quarters of the volume, and the mean is
approximately 20 times the mode.  See \cite{sutherland2007}  for more details.

\subsection{Power-law density structure}

The two--point structure of a homogeneous turbulent medium may be
described in Fourier space.   We denote the Fourier transform of the
density $\rho(\r)$ by $F(\k)$ (where $\k$ is the wavenumber vector). The
isotropic power spectrum $D(k)$ is the integral over solid angle in
Fourier space of the spectral density $F(\k) F^*(\k)$. In three
dimensions: 
\begin{equation} D(k) = \int  k^2 F(\k) F^*(\k) \> d \Omega . \label{e:Ek} 
\end{equation}. 
Even if the spectral density is
anisotropic, the angular integral averages the spectral density into a
one dimensional function of $\k$ only.
For a power-law dependence on $\k$, $D(\k) \propto \k^{-\beta}$ and
$\beta = 5/3$, the spectrum is referred to as Kolmogorov turbulence. 

In order to simultaneously achieve log--normal single-point statistics
and a power--law self--similar structure, we have implemented the
practical method developed by atmospheric scientists, for constructing
two and three dimensional  terrestrial cloud models, which are used in
radiative transfer calculations \citep{lewis02}.

The remaining choice in this procedure is to select the range of wave
numbers over which to generate the fractal, in particular the minimum
wave number $\k_{\rm min}$, which determines the largest structure scale
in the resulting fractal with respect to the spatial grid.   The main
domain of interest on the intermediate level in the simulations covers
at least two dark matter scale radii in each model.  We choose $\k_{\rm
min} = 8$ over this domain, corresponding to approximately two cloud
structures per scale radius.  On the upper level, the domain covers
three times the extent, the fractal is extended in a $\k_{\rm min} = 24$
model.  The lowest level is initially populated with a simple resampled
version of the intermediate level density field.  Each level of the
simulations is sampled in a $216^3$ cells and so even the coarse
$\k_{\rm min} = 24$ upper level has well resolved small clouds.

\subsection{Equilibrium turbulent distribution} \label{s:turb_pot}

Here we are considering purely spherical distributions, neglecting
rotational formation of disk structures. The reasons are two--fold. 
First, flattening  due to rotation will provide a natural axis for the
escape of the SN energy and potentially greatly reduce the efficiency of
the supernova in evacuating the region of potentially star--forming gas.
We are looking for a conservative estimate to the lowest mass halo that
retains gas, and so wish to retain spherical symmetry, in order to
maximises the potential interaction between the host ISM and the SNe. 
Secondly it is less clear that these very small halos include
significant angular momentum, and so a non- rotating model is the
simplest applicable one 

Taking the mean global temperature of the ISM to be $\tilde T$ and the
line-of-sight turbulent velocity dispersion to be $\sigma_t$, the mean
density of gas in the potential, $\Phi$, is given by \citep
{sutherland2007}: 
\begin{equation} \frac {\rho(r)}{\rho(0)} =
\exp  \left[ - \frac {\sigma_D^2}{\sigma_g^2} \Phi(r) \right]^\beta \, .
\label{e:warm_gas} 
\end{equation} 
where we take $e = 0$, and 
$\sigma_g^2 = \sigma_t^2 + k \tilde T / \mu m$ and $\sigma_D$ is the velocity
dispersion of the dark matter, assuming also that $\tilde T$ is the mean
ISM temperature,  and $\sigma_t$ is constant throughout the ISM.  

This mean global density distribution is multiplied by the fractal generated
above, giving a clumpy distribution with the same mean global
properties, and which is stable enough over the timescale of the
simulations to remain static in the absence of a supernova explosion.

\begin{figure*}
\plotone{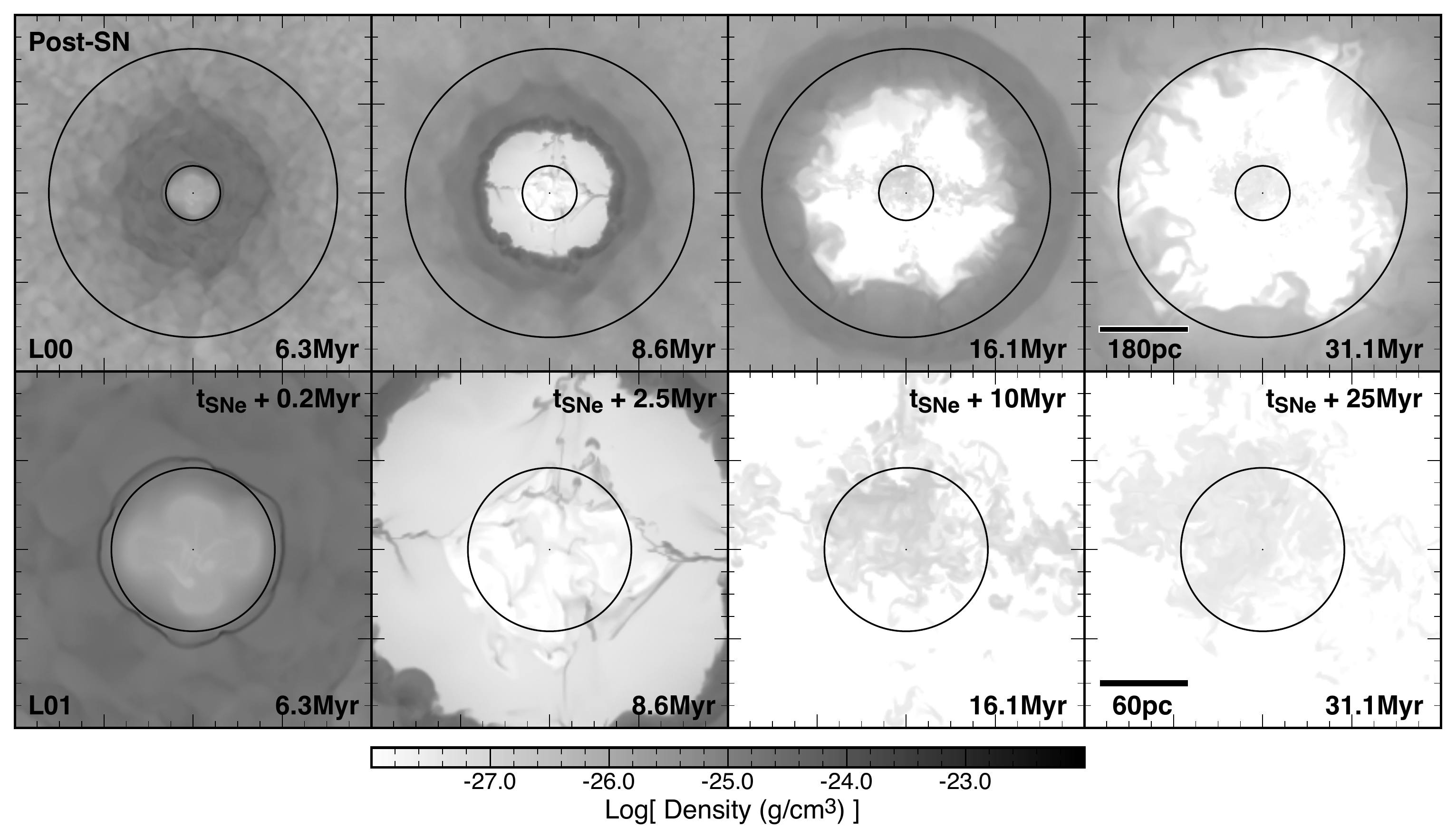}
\caption{ \label{f:M60CCH}
M60 Central Clumpy explosion commencing at the end of the progenitor ionization (pre-ionization) phase.  
The SN explodes into a warm low-density nebula
pre-heated by the progenitor. The SN wind superheats the gas through
shock heating, and removes essentially all of the gas from the dark matter halo out to \rvir\ after 30 Myr. 
For the uppermost 4 panels (L00), the inner circle has radius \rs; the outer circle has radius \rvir. Each panel has been magnified in the next 4 panels (L01); the circle has radius \rs. 
All images are central plane density slices at $z = 0.0$.  Log of the normalised density variable in units of $1.0\times 10^{-24}$ g~cm$^{-3}$: greyscale minimum $ = -3.0$ (white), maximum $ = 2.0$ (black).  A scale bar in parsecs is shown. 
}
\end{figure*}

\begin{figure*}
\plotone{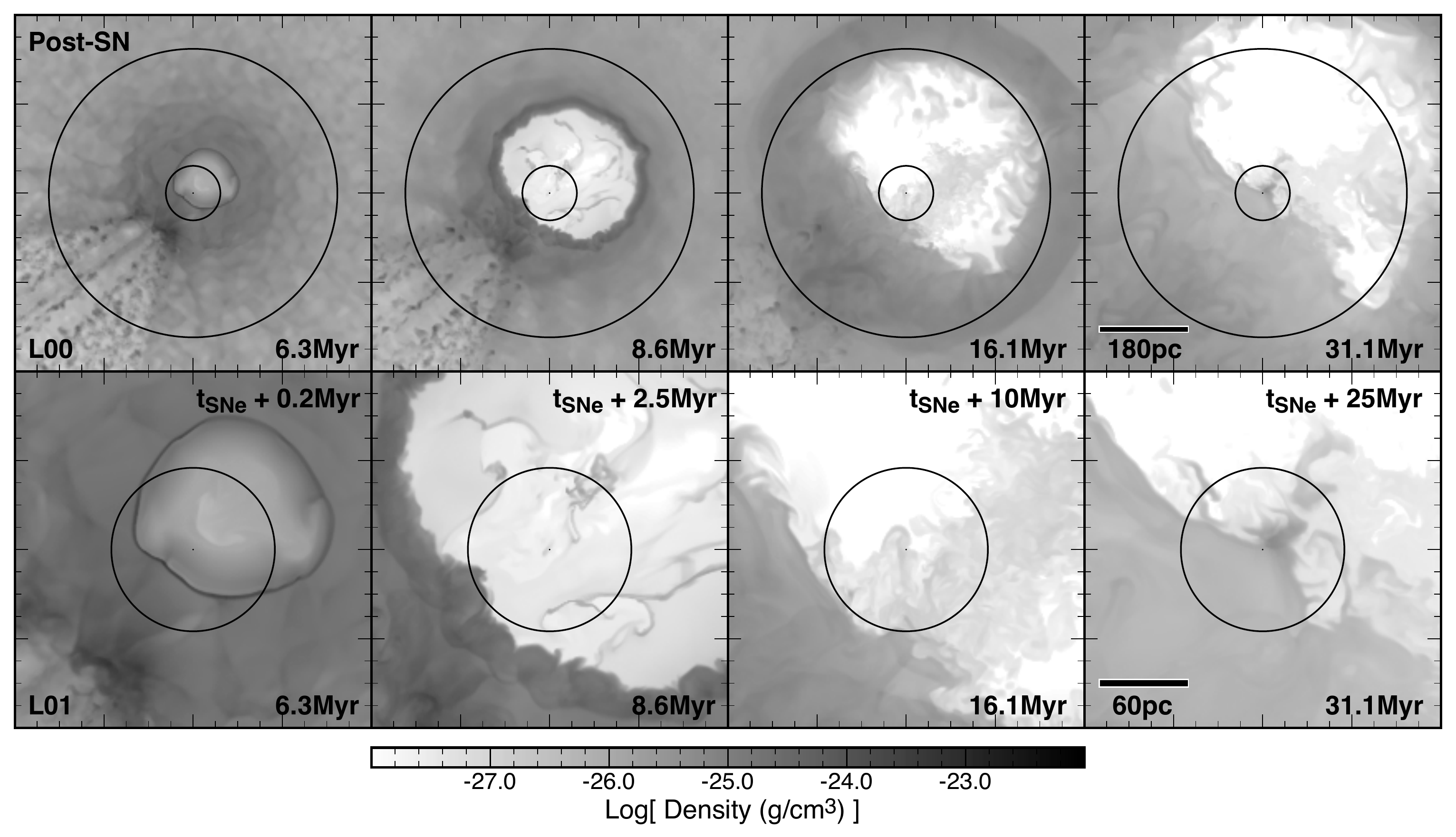}
\caption{ \label{f:M60OCH}
M60 Off-centred Clumpy explosion commencing at the end of the progenitor ionization phase; see Fig.~\ref{f:M60CCH} for details. The star is placed at the half 
gas-mass radius to reflect the fact that most stars form off-centre. 
The conic `zone of avoidance' sets up in a region that is diametrically opposite the progenitor star; 
a low density nebula extends across most of the core region before the supernova ignites.
The SN explodes into an off-centred, a warm low-density nebula, superheats the gas through
shock heating, and removes much of the gas asymmetrically from the dark matter halo out to 
\rvir\ after 30 Myr. The early zone of avoidance leads to far-side gas re-accreting into the core
region after the SN event has passed, but less than 10\% of the original gas remains within \rvir\
on long timescales.
}
\end{figure*}

\begin{figure*}
\plotone{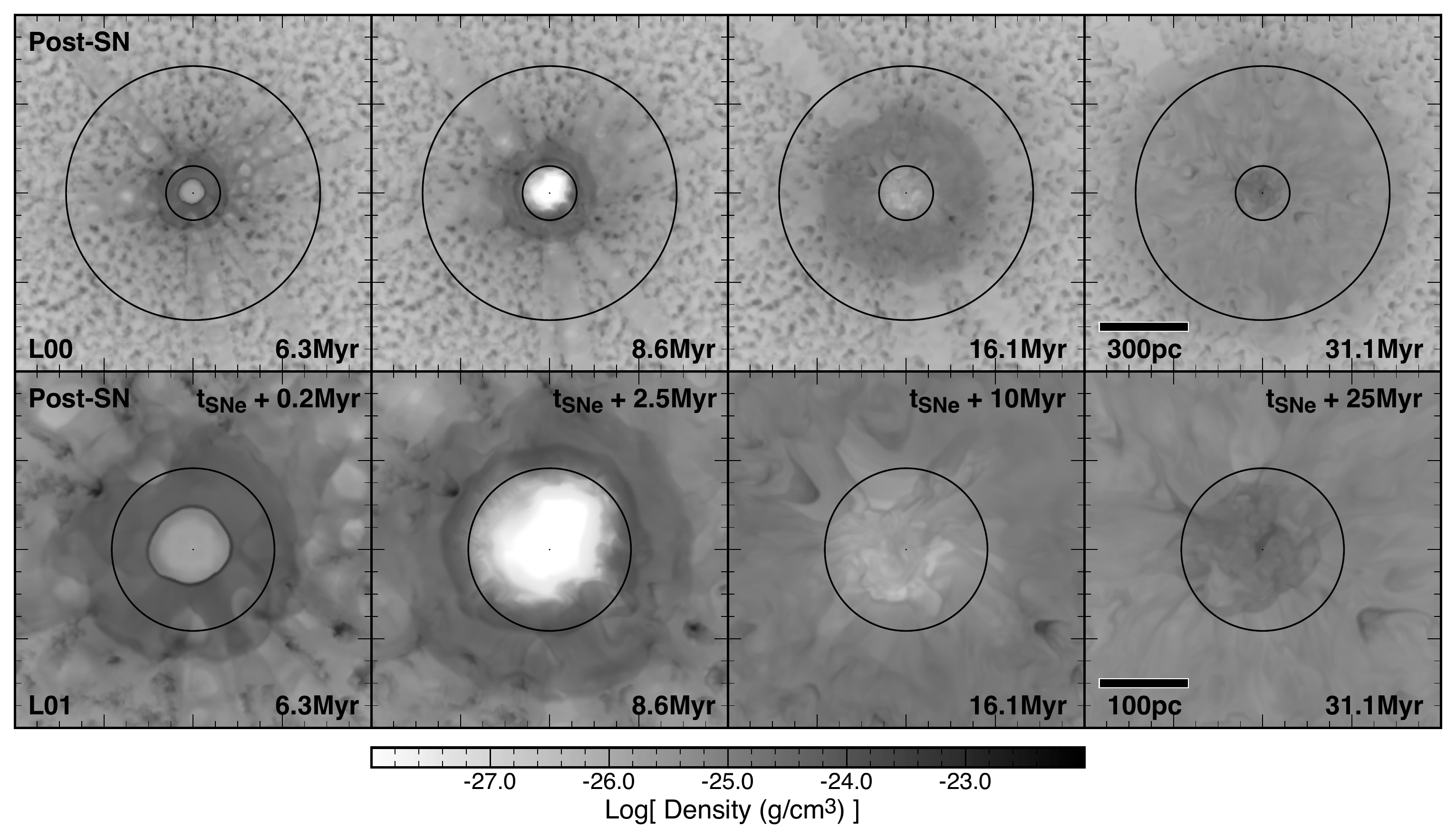}
\caption{ \label{f:M65CCH}
M65 Central Clumpy explosion commencing at the end of the progenitor ionization phase. For the uppermost 4 panels (L00), the inner circle has radius \rs; the outer circle has radius \rvir. Each panel has been magnified in the next 4 panels (L01); the circle has radius \rs. Unlike the M60 case, little radiation leaks out from the centre for the first 6 Myr. The ionization is almost entirely restricted to the core region ($r<$\rs). 
The SN explodes into a warm, moderately dense nebula, superheats the gas through shock heating, but the gas cools as the nebula expands, which checks its progress.
A diffuse ionised nebula forms that extends across the halo but is still bound by it. The core region
cools and condenses, leading to re-accretion after the bubble collapses starting as early as 3 Myr after 
the SN explosion. All images are central plane density slices at $z = 0.0$.  Log of the normalised density variable in units of $1.0\times 10^{-24}$ g~cm$^{-3}$: greyscale minimum $ = -3.0$ (white), maximum $ = 2.0$ (black).  A scale bar in parsecs is shown.
}
\end{figure*}

\begin{figure*}
\plotone{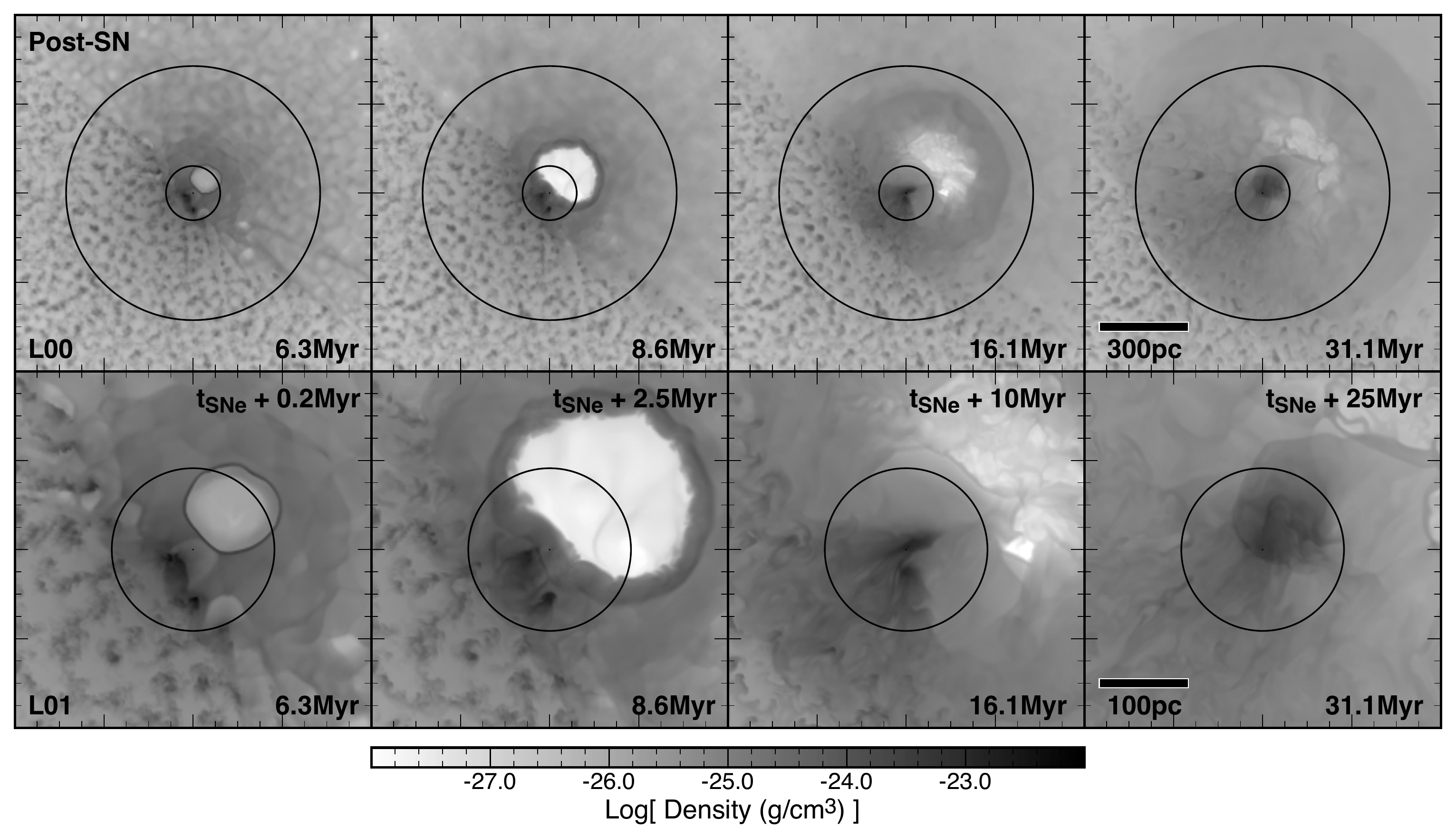}
\caption{ \label{f:M65OCH}
M65 Off-centred Clumpy explosion commencing at the end of the progenitor ionization phase; see Fig.~\ref{f:M65CCH} for details.
The star is placed at the half gas-mass radius to reflect the fact that most stars form off-centre. 
The SN explodes into an off-centred, warm moderately dense nebula and
superheats the gas through shock heating. This sends a blast wave into the surrounding nebula but
fails to maintain the hot cavity that starts to collapse after 6 Myr. The early zone of avoidance leads to 
far-side gas re-accreting into the core region after the SN event has passed. On 30 Myr timescales, 
there is essentially no gas loss from the dark halo.}
\end{figure*}

\begin{figure*}
\plotone{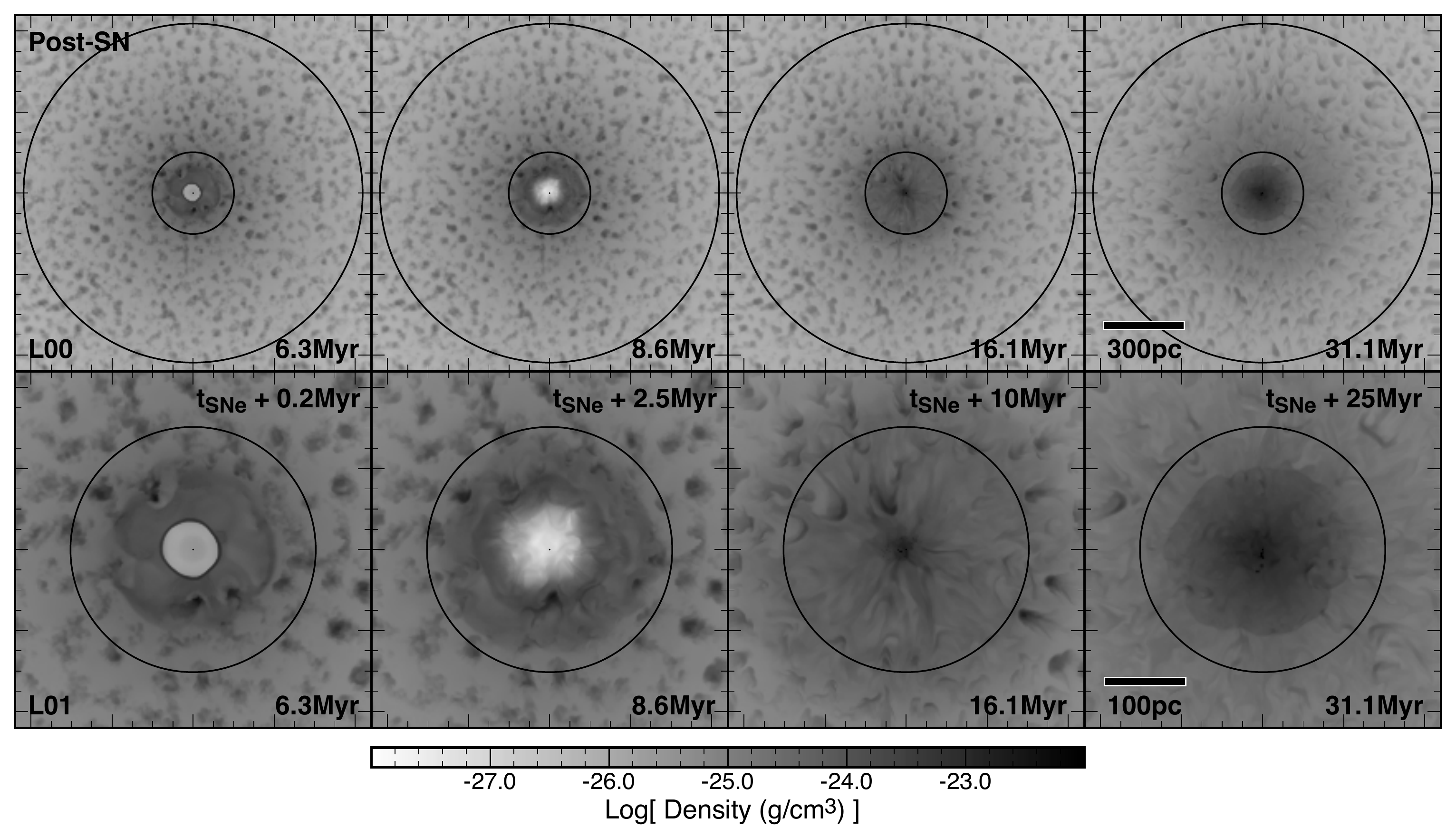}
\caption{ \label{f:M70CCH}
M70 Central Clumpy explosion commencing at the end of the progenitor ionization phase. 
For the uppermost 4 panels (L00), the inner circle has radius \rs; the outer circle has radius \rvir. Each panel has been magnified in the next 4 panels (L01); the circle has 
radius \rs. Much like the M65 case, little radiation leaks out from the centre for the first 6 Myr. The ionization is almost entirely restricted to the core region ($r<$\rs).
The SN explodes into a warm, moderately dense nebula, superheats the gas through shock heating, but the gas cools as the nebula expands, which checks its progress.
A diffuse ionised nebula forms that extends across the halo but is still bound by it. The core regions
cool and condense, leading to re-accretion after the bubble collapses starting as early as 3 Myr after 
the SN explosion. All images are central plane density slices at $z = 0.0$.  Log of the normalised density variable in units of $1.0\times 10^{-24}$ g~cm$^{-3}$: greyscale minimum $ = -3.0$ (white), maximum $ = 2.0$ (black).  A scale bar in parsecs is shown. 
}
\end{figure*}

\begin{figure*}
\plotone{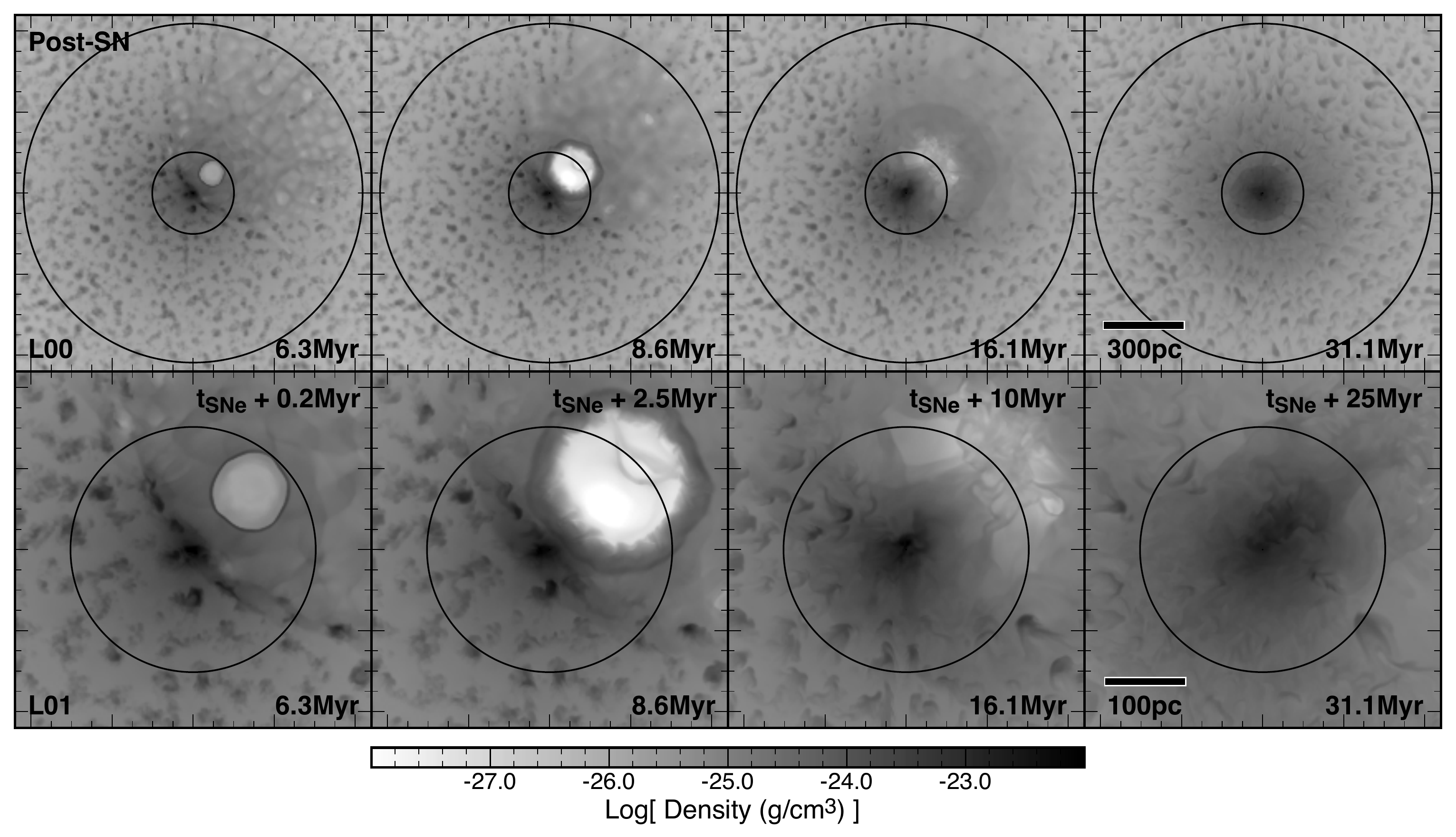}
\caption{ \label{f:M70OCH}
M70 Off-centred Clumpy explosion commencing at the end of the progenitor ionization phase; see Fig.~\ref{f:M70CCH} for details. After the progenitor phase, only about half
the gas (i.e. one hemisphere) is ionised due to shadowing from the core region which becomes
denser due to compression by the overpressured ionised volume.
The star is placed at the half gas-mass radius to reflect the fact that most stars form off-centre. 
The SN explodes into an off-centred, warm moderately dense nebula and
superheats the gas through shock heating. This sends a blast wave into the surrounding nebula but
fails to maintain the hot cavity that starts to collapse after 6 Myr. The early zone of avoidance leads to 
far-side gas re-accreting into the core region after the SN event has passed. On 30 Myr timescales, 
there is essentially no gas loss from the dark halo.}
\end{figure*}

\end{document}